\RequirePackage{pdf14}

\documentclass[letterpaper]{article}
\usepackage[usenames,dvipsnames,table]{xcolor}
\usepackage{graphicx}
\usepackage{caption}
\usepackage{subcaption}
\usepackage{array,setspace,mathrsfs,amsthm,amsfonts,amsmath,yfonts,dsfont,bbm,colonequals,mathtools}
\usepackage{./auxiliary/tikz-cd}

\usepackage{./auxiliary/jheppub}
\usepackage{afterpage}
\usepackage{xspace}
\usepackage{stmaryrd}
\usepackage[normalem]{ulem}
\usepackage{tikz}
\usetikzlibrary{decorations.markings}

\newcommand*\pFqskip{8mu}
\catcode`,\active
\newcommand*\pFq{\begingroup
        \catcode`\,\active
        \def ,{\mskip\pFqskip\relax}%
        \dopFq
}
\catcode`\,12
\def\dopFq#1#2#3#4#5{%
        {}_{#1}F_{#2}\biggl[\genfrac..{0pt}{}{#3}{#4};#5\biggr]%
        \endgroup
}

\newcommand\eea{\end{eqnarray}}
\newcommand\ee{\end{equation}}

\newcommand{\includegraphicsif}[2]{\IfFileExists{#2}{\includegraphics[#1]{#2}}{\includegraphics[scale=.5]{./figures/dummy.pdf}}} 

\newcommand\mf{\mathfrak}

\newcommand\nn{\nonumber}
\newcommand\ph{\phantom}
\newcommand\wt{\widetilde}

\newcommand\eg{{\it e.g.}}
\newcommand\ie{{\it i.e.}}

\newtheorem*{result*}{Result}

\newtheorem*{fact*}{Fact}
\newtheorem{conj}{Conjecture}
\newtheorem*{conj*}{Conjecture}
\newtheorem*{spec*}{Speculation}

\newtheorem*{cor*}{Corollary}

\newcommand\Cb{\mathbb{C}}

\newcommand\Rb{\mathbb{R}}

\newcommand\Zb{\mathbb{Z}}

\renewcommand\AA{\mathcal{A}}
\renewcommand\SS{\mathcal{S}}

\newcommand\CC{\mathcal{C}}

\newcommand\JJ{\mathcal{J}}

\newcommand\MM{\mathcal{M}}
\newcommand\NN{\mathcal{N}}
\newcommand\OO{\mathcal{O}}

\newcommand\QQ{\mathcal{Q}}
\newcommand\RR{\mathcal{R}}

\newcommand\UU{\mathcal{U}}

\newcommand\ZZ{\mathcal{Z}}

\newcommand\elem{\in}

\newcommand\qq{\mathbbmtt{Q}}
\newcommand{\q}[1]{\mathbbmtt{Q}\,_{#1}}

\newcommand\uf{{\mf{u}}}
\newcommand\sof{{\mf{so}}}
\newcommand\suf{{\mf{su}}}
\newcommand\spf{{\mf{sp}}}
\newcommand\ospf{{\mf{osp}}}

\newcommand\psuf{{\mf{psu}}}
\newcommand\slf{{\mf{sl}}}

\newcommand\gf{{\mf{g}}}

\newcommand\sgn{\mathrm{sgn}}

\newcommand\restr[2]{{
  \left.\kern-\nulldelimiterspace 
  #1 
  \vphantom{\big|} 
  \right|_{#2} 
  }}
\newcommand\fverb{\setbox\fverbbox=\hbox\bgroup\verb}
\newcommand\fverbdo{\egroup\medskip\noindent%
			\fbox{\unhbox\fverbbox}\ }
\newcommand\fverbit{\egroup\item[\fbox{\unhbox\fverbbox}]}
\newbox\fverbbox

\title{Deformation quantization and superconformal symmetry in three dimensions}
\author[1]{Christopher Beem,\!}
\author[2]{Wolfger Peelaers,\!}
\author[3]{Leonardo Rastelli\ph{,}\!}

\emailAdd{beem@sns.ias.edu}
\emailAdd{wolfger.peelaers@rutgers.edu}
\emailAdd{leonardo.rastelli@stonybrook.edu}

\affiliation[1]{School of Natural Sciences, Institute for Advanced Study, Einstein Drive, Princeton, NJ 08540, USA}
\affiliation[2]{New High Energy Theory Center, Rutgers University, Piscataway, NJ 08854, USA}
\affiliation[3]{C.~N.~Yang Institute for Theoretical Physics, Stony Brook University, Stony Brook, NY 11794, USA}

\bigskip
\abstract{
We investigate the structure of certain protected operator algebras that arise in three-dimensional $\NN=4$ superconformal field theories. We find that these algebras can be understood as a quantization of (either of) the half-BPS chiral ring(s). An important feature of this quantization is that it has a preferred basis in which the structure constants of the quantum algebra are equal to the OPE coefficients of the underlying superconformal theory. We identify several nontrivial conditions that the quantum algebra must satisfy in this basis. We consider examples of theories for which the moduli space of vacua is either the minimal nilpotent orbit of a simple Lie algebra or a Kleinian singularity. For minimal nilpotent orbits, the quantum algebras (and their preferred bases) can be uniquely determined. These algebras are related to higher spin algebras. For Kleinian singularities the algebras can be characterized abstractly -- they are spherical subalgebras of symplectic reflection algebras -- but the preferred basis is not easily determined. We find evidence in these examples that for a given choice of quantum algebra (defined up to a certain gauge equivalence), there is at most one choice of canonical basis. We conjecture that this is the case for general $\NN=4$ SCFTs.
}


\arxivnumber{1601.05378}

\begin{document}
\setcounter{tocdepth}{2}
\maketitle
\setcounter{page}{1}


\section{Introduction}
\label{sec:intro}

An important feature of conformal field theories --- and one that has received increased attention in recent years --- is that they admit a nonperturbative algebraic formulation in terms of the operator product expansion (OPE) of local operators. This formulation makes no explicit reference to elementary fields or path integrals. Instead, the basic data is taken to be the collection of local operators --- organized into representations of the conformal algebra and any additional symmetry algebras --- and the OPE coefficients that serve as structure constants for the operator algebra. This information is often referred to as the CFT data of the theory, and any correlation function of finitely many local operators is determined algorithmically in terms of it.

The CFT data is strongly constrained by the requirement that the OPE be associative. This associativity condition is normally presented as the requirement that crossing symmetry hold for four-point functions. The project of extracting useful results from crossing symmetry is known as the conformal bootstrap (see, \eg, \cite{Rychkov:2016iqz} and references therein). The difficulty of the conformal bootstrap approach stems from the very complicated form of the constraints of crossing symmetry. Indeed, outside of rational theories in two dimensions, the list of local operators always involves an infinite number of representations of the symmetry algebra, and the crossing symmetry equations amount to an infinite number of coupled functional equations depending on an infinite amount of the CFT data. Starting with \cite{Rattazzi:2008pe}, spectacular progress has been made extracting bounds, and sometimes precise estimates, for certain operator dimensions and OPE coefficients using numerical methods (see, \eg, the paradigmatic application of these methods to the critical Ising model in three dimensions \cite{ElShowk:2012ht, El-Showk:2014dwa, Kos:2014bka}). Furthermore, the crossing equations become more tractable in the Lorentzian lightcone limit, making it possible to extract interesting asymptotic results for operators with large spin \cite{Alday:2007mf,Fitzpatrick:2012yx, Komargodski:2012ek,Alday:2013cwa}. Nevertheless, it remains the case that the OPE algebra as a whole is too difficult to treat analytically in any holistic fashion.

In supersymmetric theories, there is a well-established mechanism to extract a more tractable algebraic structure from the full OPE algebra dating back to \cite{Lerche:1989uy}. The trick is to identify a supersymmetry transformation and treat it as a differential acting on the space of local operators, and then pass to cohomology with respect to that differential. The OPE algebra descends to an algebra defined for cohomology classes of local operators, with certain OPE coefficients of the full theory appearing as structure constants of this cohomological algebra. Generally the cohomological algebra is a simpler object than the full OPE algebra. The observables that survive the passage are independent of various arguments --- \eg, coupling constants or operator positions --- and this is usually enough to render the OPE algebra much more tractable.

The simplest version of this procedure takes place when the supercharge in question is the scalar supercharge in a topologically twisted theory. In such a case, the cohomology class of a local operator is completely independent of its position in spacetime. The resulting operator algebra is a kind of $d$-dimensional topological algebra, which is more or less a commutative associative algebra.\footnote{In fact, there is more structure in this truncation than just a commutative associative algebra. Topological descent allows one to define additional operations on local operators. For the case of topologically twisting, this leads to what mathematicians call an $E_d$ algebra \cite{ben_zvi_private}.} While these topological algebras are interesting for many purposes --- especially since they can be defined in non-conformal theories as well --- they don't participate in a powerful truncation of the bootstrap problem, in the sense that imposing associativity doesn't lead to strong constraints on the CFT data of the theory.

By considering more general nilpotent supersymmetries, it is possible to isolate more intricate algebraic structures for which the reduced bootstrap problem is still very interesting. A prime example of this phenomenon was presented in \cite{Beem:2013sza}, where it was shown that in four-dimensional $\NN\geqslant2$ SCFTs there is a cohomological truncation that has the structure of a two-dimensional chiral algebra (a similar truncation appears in six-dimensional theories with $(2,0)$ superconformal symmetry \cite{Beem:2014kka}). Aside from being a surprising connection between two- and four- (or six-)dimensional physics, the appearance of chiral algebras in the context of higher dimensional SCFTs is exciting because chiral algebras are strongly and \emph{tractably} constrained by associativity. This means that a wealth of CFT data can potentially be recovered from limited input by solving the chiral algebra bootstrap problem for the theory in question, see for example \cite{Beem:2014rza,Lemos:2014lua}. Indeed, this ``mini-bootstrap'' problem associated to chiral algebras provides an essential starting point for the more laborious numerical bootstrap analysis of four- and six-dimensional theories with extended supersymmetry \cite{Beem:2014zpa,Beem:2015aoa}.

In this paper we study an analogous algebraic structure that arises in three-dimensional theories with $\NN\geqslant4$ superconformal symmetry. It was observed already in \cite{Beem:2013sza} that with this much supersymmetry one can identify a cohomological reduction of the OPE algebra that takes the form of a one-dimensional topological algebra. Here a one-dimensional topological algebra means a not-necessarily-commutative, associative algebra that additionally has an evaluation map (\ie, one can take expectation values by evaluating the corresponding correlation functions in the full SCFT). In this paper, we will refer to this algebra as the \emph{protected associative algebra} of an $\NN=4$ SCFT. This structure was developed and exploited successfully in \cite{Chester:2014mea} to determine a large number of OPE coefficients in several $\NN=8$ superconformal field theories and to derive a number of relations amongst OPE coefficients that must hold in any $\NN=8$ theory.\footnote{See also \cite{Liendo:2015cgi} for a discussion in $\NN=6$ theories.} These analytic results were then used to improve the numerical analysis of \cite{Chester:2014fya}. Our aim is to understand the general properties of these associative algebras and to investigate the corresponding bootstrap problem.

Our first observation is that the operators that participate in the protected associative algebra are in one-to-one correspondence with the highest weight states of half-BPS representations of the superconformal algebra. These are the Higgs or Coulomb branch chiral ring operators; there is a separate truncation for each. The multiplication of these operators in the associative algebra is no longer simply chiral ring multiplication, but rather is a noncommutative deformation thereof. A key observation is that the leading order deformation is determined by the Poisson bracket of the appropriate chiral ring, so we are dealing with deformation quantization of the chiral ring. It is worth noting that quantizations of the chiral ring of supersymmetric gauge theories have appeared in a variety of contexts in recent years \cite{Gaiotto:2010be,Yagi:2014toa,Bullimore:2015lsa,Braverman:2016wma}. The connection between supersymmetric field theory and quantization goes back further still \cite{Cattaneo:1999fm,Gukov:2008ve,Gukov:2010sw}. A connection between these two sets of references is provided by \cite{Nekrasov:2010ka}. While it is interesting to ask whether the present work can be tied into this extant circle of ideas, the appearance of the algebraic structure of interest in those references has a somewhat different flavor to ours. We briefly comment on this in the conclusions.

Deformation quantization of hyperk\"ahler cones has been a subject of considerable study by mathematicians, particularly in the context of geometric representation theory \cite{Braden:2014cb,Braden:2014iea}.\footnote{In the literature, the problem is usually described as deformation quantization of symplectic singularities, and the essential features of the problem can be understood in a single complex structure. As we will see, superconformal operator algebra suggests that taking the full hyperk\"ahler structure of these singularities more seriously may be worthwhile.} At first blush, this appears to be a major boon, since there is a classification theorem for these algebras that suggests a finite-dimensional space of solutions to our bootstrap problem. However, we will see that this classification provides only the starting point for us, because the algebras classified in the theorem are defined up to an infinite-dimensional group of equivalence relations --- referred to as gauge equivalence --- that do not define equivalences in the physical context. Thus, while a first approximation of our problem suggests a finite-dimensional classification of its solutions, the next approximation involves an infinite-dimensional space of solutions!

A more careful consideration of the problem reveals that the situation is somewhere in between. Certain selection rules and unitarity conditions in the parent SCFT imply a number of properties for structure constants of the protected associative algebra that are not traditionally imposed in the mathematics literature and that are not preserved by gauge transformations. This suggests that these extra conditions may play the role of gauge fixing conditions. It is not clear whether these gauge fixing conditions should be perfect, or if they should even be feasible for any given point in the space of algebras-modulo-gauge-equivalence.

We develop our intuition in a few examples. Because the deformation quantization problem is naturally formulated using the data of the Higgs or Coulomb branch of the SCFT in question, our examples are organized according to a choice of hyperk\"ahler cone. The first class of examples are theories that have the minimal nilpotent orbit of a complex, simple Lie algebra $\gf$ as a branch of their moduli space. The quantum algebra for these theories has the special property that the $\gf$-symmetry is already enough to completely fix the gauge redundancy, meaning that for these cases the classification theorem can be applied directly to the bootstrap problem. When $\gf=\slf_2$ this means that the bootstrap problem has at most a one-dimensional space of solutions, while for $\gf\neq\slf_2$ the bootstrap problem has (at most) a unique solution. Interestingly, these algebras have already made an appearance in the physics literature; they are (generalized) higher spin algebras.

A more difficult class of examples to analyze are the $A$-type Kleinian singularities $\Cb^2/\Zb_{n\geqslant3}$.\footnote{The case $\Cb^2/\Zb_2$ is special because it has enhanced symmetry. In fact it is the same as the minimal nilpotent orbit of $\slf_2$, and so it falls into the previous category.} In these cases it is easy to characterize the quantum algebra up to gauge equivalence, but the space of gauge transformations is genuinely infinite dimensional so the power of our gauge fixing conditions comes into question. We have not been able to solve the complete set of gauge fixing conditions exactly, but we can systematically solve them for operators of increasing dimension up to some cutoff. Our results suggest that imposing selection rules without unitarity is enough to cut down the infinite dimensional equivalence classes to an affine finite dimensional space. What's more, we find convincing evidence in the examples we analyze that the additional requirements of unitarity are sufficient to completely fix the gauge freedom for allowed algebras (and also rule out other algebras). This leads us to make the following ambitious conjecture:
\begin{conj}\label{conj:main}
The superconformal selection rules and unitarity conditions given in Section \ref{sec:algebra_properties} of this paper, when solvable, are perfect gauge fixing conditions for the deformation quantization of hyperk\"ahler cones.
\end{conj}
\noindent From a mathematical point of view, this would be a surprising result and should provide significant motivation to investigate the canonical bases that satisfy these conditions. From the physics point of view, this would provide a succinct characterization of the data necessary to completely determine the protected associative algebra via the bootstrap.

The organization of the rest of the paper is as follows. In Section \ref{sec:superalgebra} we recall the details of the $\ospf(4|4,\Rb)$ superconformal algebra and describe the cohomological reduction that leads to the protected associative algebra. We emphasize the precise relationship between the structure constants of the associative algebra and the OPE coefficients of the full SCFT. In Section \ref{sec:algebra_properties} we describe the properties that the associative algebra must possess. These conditions make it clear that the associative algebra will be a special instance of deformation quantization of the (Higgs or Coulomb branch) chiral ring. In Section \ref{sec:math_phys} we review the basic structure of deformation quantization as an algebra problem as well as describing simplifications and mathematical results relevant to the case of coordinate algebras of hyperk\"ahler cones. We further point out how the bootstrap problem for the protected associative algebra relates to and expands upon the more standard mathematical problem. In Section \ref{sec:nilpotent} we consider our first class of examples: theories that have minimal nilpotent orbits as branches of their moduli spaces of vacua. We find an intriguing connection between these algebras and higher spin algebras. In Section \ref{sec:kleinian} we consider the examples of $A$-type Kleinian singularities. Our results lead us to Conjecture \ref{conj:main}. We conclude by pointing out a number of open questions and interesting directions in which to extend the present work. In two  appendices we provide background about the algebraic geometry of hyperk\"ahler cones and details of the calculations of moment map two-point functions using supersymmetric localization.

\section{Superconformal algebra and cohomology}
\label{sec:superalgebra}

The cohomological truncation of \cite{Beem:2013sza,Chester:2014mea} follows from the structure of the three-dimensional $\NN=4$ superconformal algebra. Here we recall the structure of this superconformal algebra and how it gives rise to the truncation of interest. Most of the content of the first two subsections was already presented in \cite{Chester:2014mea}, and greater detail about the superconformal algebra can be found in \cite{Dolan:2008vc}. We include the full discussion for completeness and to establish notation.

\subsection{The superconformal algebra and its half BPS representations}
\label{subsec:superconformal_algebra}

The superconformal algebra in question is $\ospf(4|4,\Rb)$. This is a lie superalgebra whose even part is given by
\begin{equation}
\label{eq:even_superconformal_algebra}
\sof(4)_R \times \spf(4,\Rb) \cong (\suf(2)_H\oplus\suf(2)_C) \times \sof(3,2)~.
\end{equation}
The first factor is the $R$-symmetry and the second is the three-dimensional conformal algebra in Lorentzian signature. The subscripts $C$ and $H$ indicate that the corresponding $R$-symmetries act as isometries of the Coulomb and Higgs branches of the moduli space of vacua, respectively.\footnote{Recall that in three-dimensional $\NN=4$ superconformal field theories, both Coulomb and Higgs branches are hyperk\"ahler cones and admit $\suf(2)_C$ and $\suf(2)_H$ isometries, respectively, that rotate their complex structures as a triplet. See appendix \ref{app:HK_cones} for additional details.} While Lagrangian constructions treat the Coulomb and Higgs branches differently, our focus is purely on superconformal fixed points and as such there is no meaningful difference between these two factors.

The odd part of this superalgebra is spanned by Poincar\'e supercharges $\{Q^{a\tilde a}_{\alpha}\}$ and special conformal supercharges $\{S_{a\tilde a}^{\alpha}\}$. These all transform in the $({\bf 2},{\bf 2},{\bf 2})$ of $\suf(2)_H\times \suf(2)_C\times \sof(2,1)_{L}$. When quantizing on the two-sphere, hermitian conjugation is defined for the supercharges according to $\left(Q^{a\tilde a}_{ \alpha }\right)^\dagger = S_{a\tilde a}^{\alpha}$.\footnote{Note that relative to \cite{Chester:2014mea} our supercharges are rescaled by a factor of $\sqrt{2}$, and for $S$ also a factor of $i$: $Q_{\rm there} = \sqrt{2} Q_{\rm here}$ and $S_{\rm there} = i\sqrt{2} S_{\rm here}$.}
Anticommutators of these odd generators take the form\footnote{Here we are adopting gamma-matrix conventions of \cite{Dolan:2008vc}. In particular $P_{\alpha\beta} = \left(\begin{smallmatrix} P_0 + P_2 & P_1\\
P_1 & P_0-P_2 \end{smallmatrix}\right),$ $K^{\alpha\beta} = \left(\begin{smallmatrix} -K_0 + K_2 & K_1\\
K_1 & -K_0-K_2 \end{smallmatrix}\right)$ and $M_\alpha^{\ \beta} = i \left(\begin{smallmatrix} M_{02} & M_{01} - M_{12}\\
M_{01} + M_{12} & -M_{02} \end{smallmatrix}\right)$.}
\begin{equation}
\begin{split}
\{Q^{a\tilde a}_{\alpha } , Q^{b\tilde b}_{\beta } \} &= 2 \epsilon^{ab}\epsilon^{\tilde a \tilde b} P_{\alpha \beta}~,\\
\{S_{a\tilde a}^{\alpha } , S_{b\tilde b}^{\beta } \} &= 2 \epsilon_{ab}\epsilon_{\tilde a \tilde b} K^{\alpha \beta}~,\label{3dalgebra1}\\
\{Q^{a\tilde a}_{\alpha } , S_{b\tilde b}^{\beta } \} &= 2 \delta^a_b \delta^{\tilde a}_{\tilde b} \left(M_\alpha^{\ \beta} + \delta_\alpha^\beta D\right) - 2\delta_\alpha^\beta(R^{a}_{\ b} \delta^{\tilde a}_{\tilde b}  +  \delta^a_b \tilde R^{\tilde a}_{\ \tilde b}  )~,
\end{split}
\end{equation}
where the generators of $\sof(2,1)\cong\slf(2,\mathbb R)$, $\suf(2)_H$, and $\suf(2)_C$ are respectively defined as
\begin{equation}\
M_\alpha^{\ \beta} = \begin{pmatrix}
J_3 & \ph{-}J_+ \\
J_- & -J_3
\end{pmatrix}~, \qquad 
R^a_{\ b} = \begin{pmatrix}
R\ph{{}_{-}} & \ph{-}R_+ \\
R_- & -R
\end{pmatrix}~, \qquad 
\tilde R^{\tilde a}_{\ \tilde b} = \begin{pmatrix}
\tilde R\ph{{}_{-}} & \ph{-}\tilde R_+ \\
\tilde R_- & -\tilde R
\end{pmatrix}~.
\end{equation}

Unitary highest weight representations of $\ospf(4|4,\Rb)$ have been classified in \cite{Dolan:2008vc}. Of particular interest to us are the half-BPS multiplets, denoted there as $(2,B,\pm)$. The shortening conditions for these multiplets are such that their superconformal primaries are annihilated by four supercharges and obey certain restrictions on their quantum numbers: 
\begin{equation}
\begin{split}
(2,B,+):\qquad Q_\alpha^{1\tilde a}|\psi_{\rm scp}\rangle=0~,\quad \Delta &= r~,\quad \tilde r=0~,\\
(2,B,-):\qquad Q_\alpha^{a\tilde 1}|\psi_{\rm scp}\rangle=0~,\quad \Delta &= \tilde r~,\quad  r=0~,
\end{split}
\end{equation}
where $\Delta$, $r$, and $\tilde r$ are the respective eigenvalues of $D$, $R$, and  $\tilde R$. The superconformal primaries in these multiplets are Higgs or Coulomb branch chiral ring operators, \ie, their expectation values parameterize the Higgs and Coulomb branches of vacua, respectively. For ease of discussion, we will refer to $(2,B,+)$ multiplets as \emph{Higgs branch multiplets} and $(2,B,-)$ multiplets as \emph{Coulomb branch multiplets} in the remainder of the paper. We will refer to the full $\suf(2)_{H/C}$ representations of the superconformal primaries in these multiplets as \emph{Higgs branch operators} and \emph{Coulomb branch operators}, respectively. Finally, we will refer to the highest weight states in those representations (for a given choice of Cartan decomposition) as \emph{Higgs branch chiral ring operators} and \emph{Coulomb branch chiral ring operators}, respectively.

The structure of conformal primaries in a Higgs branch multiplet with $r>1$ is displayed in \eqref{(2,B,+)multiplet}, with operators labelled according to their quantum numbers $(\Delta;j_3;r,\tilde r)$. For $r=1$ the multiplet is smaller and is illustrated in \eqref{Higgsmultiplet_current}. This multiplet contains a conserved current as a level-two superconformal descendant. If an $\NN=4$ SCFT possesses additional global symmetries that act upon the Higgs branch, then the corresponding conserved currents will lie in these multiplets. For $r=\frac12$ the multiplet is smaller still as illustrated in \eqref{eq:Higgsmultiplet_free}. This multiplet contains a free complex scalar and complex fermion, which are the elementary fields that constitute a free half-hypermultiplet.

The Coulomb branch multiplets take an analogous form upon exchanging $\suf(2)_H$ and $\suf(2)_C$. For $\tilde r = 1/2$ one finds a free twisted half-hypermultiplet and for $\tilde r=1$ one finds a conserved current multiplet that acts on the Coulomb branch. More generally in a Lagrangian theory the Coulomb branch operators will come from dressed monopole operators in the UV. From the point of view of the conformal fixed point, the Coulomb branch is on exactly equal footing with the Higgs branch, so without loss of generality we will adopt the language of the Higgs branch henceforth. The corresponding constructions for the Coulomb branch can be obtained by appropriate linguistic substitutions.

\begin{equation}\label{(2,B,+)multiplet}
\begin{tikzcd}[row sep = scriptsize]
(r;0;r,0)  \arrow{dr}	& {} & {} \\
{} &  \arrow{dl}(r+\frac{1}{2};\frac{1}{2};r-\frac{1}{2},\frac{1}{2}) \arrow{dr} & {}\\
(r+1;1;r-1,0)\arrow{dr} & {} & \arrow{dl} (r+1;0;r-1,1) \\
{} & \arrow{dl} (r+\frac{3}{2};\frac{1}{2};r-\frac{3}{2},\frac{1}{2})   &{}\\
(r+2;0;r-2,0) &{} &{}
\end{tikzcd}
\end{equation}
\begin{equation}\label{Higgsmultiplet_current}
\begin{tikzcd}[row sep = scriptsize]
(1;0;1,0)  \arrow{dr}	& {} & {} \\
{} &  \arrow{dl}(\frac{3}{2};\frac{1}{2};\frac{1}{2},\frac{1}{2}) \arrow{dr} & {}\\
(2;1;0,0)_{\text{cons.}} & {} & (2;0;0,1)
\end{tikzcd}
\end{equation}
\begin{equation}\label{eq:Higgsmultiplet_free}
\begin{tikzcd}[row sep = scriptsize]
(\frac{1}{2};0;\frac{1}{2},0)_{\text{free}}  \arrow{dr}	& {} & {} \\
{} &  (1;\frac{1}{2};0,\frac{1}{2})_{\text{free}}  & {}
\end{tikzcd}
\end{equation}

\subsection{Protected associative algebra in cohomology}
\label{subsec:noncommutative_algebra_cohomology}

We are interested in the cohomological truncation of the operator algebra described in \cite{Beem:2013sza,Chester:2014mea}. This truncation works by restricting operators to lie on a straight line in spacetime and treating them at the level of cohomology with respect to a particular nilpotent supercharge. To this end, consider a line $\Rb_{\rm top}\subset\Rb^{2,1}$, which we will take to be the line $x_0=x_2=0$. This line is mapped to itself by either of the following subalgebras of $\ospf(4|4,\Rb)$:
\begin{equation}
\label{preserved subalgerbas}
\psuf(1,1|2)_H \oplus \uf(1)_C \oplus \sof(1,1)\;, \qquad \text{or} \qquad \psuf(1,1|2)_C \oplus \uf(1)_H \oplus \sof(1,1)~,
\end{equation}
where $\suf(1,1)\cong \sof(1,2)$ is the one-dimensional conformal algebra acting on the chosen line, and $\sof(1,1)$ is generated by the rotation (boost) generator in the plane orthogonal to $\Rb_{\rm top}$. The algebra $\psuf(1,1|2)_H$ is actually centrally extended by the (anti-)diagonal combination of $\uf(1)_C \oplus \sof(1,1),$ and similarly for $\psuf(1,1|2)_C$. The existence of a geometrically acting $\suf(1,1|2)$ subalgebra allows one to define the aforementioned cohomology. For concreteness, we choose to work with the first preserved subalgebra in \eqref{preserved subalgerbas}.

For our choice of line, the bosonic generators of this subalgebra are $L_{-1} \equiv P_{+-}$, $L_{0} \equiv D$, $L_{+1} \equiv K^{+-}$, and $R^{a}_{\ b}$, and as supercharges of $\suf(1,1|2)_H$ we take
\begin{equation}
\begin{split}
\QQ^a &\equiv Q^{a\tilde 1}_{+}\\
\tilde{\QQ}^a &\equiv Q^{a\tilde 2}_{-}\\
\SS_b  &\equiv S_{b\tilde 1}^+,\\
\tilde{\SS}_b &\equiv S_{b\tilde 2}^-.
\end{split}
\end{equation}
The algebra is centrally extended by $\mathcal Z = 2(\tilde R-J_3)$; the non-vanishing anti-commutation relations among the supercharges are given by
\begin{equation}
\begin{split}
\{ \QQ^a, \tilde{\QQ}^b\} &= 2 \epsilon^{ab} P_{+-}~,\\ 
\{ \SS_a, \tilde{\SS}_b\} &= 2 \epsilon_{ab} K^{+-}~,\\
\{ \QQ^a, {\SS}_b\}&= 2 \delta^a_b \left(J_3 +  D-\tilde R\right) - 2R^{a}_{\ b}~,\\
\{ \tilde{\QQ}^a, \tilde{\SS}_b\}&= 2  \delta^a_b \left(-J_3 +  D + \tilde R\right) - 2R^{a}_{\ b}~.
\end{split}
\end{equation}

We next need to choose the nilpotent supercharge(s) that will define our cohomology. In a slight departure from the presentation in \cite{Beem:2013sza,Chester:2014mea}, we take these supercharges to depend on an arbitrary nonzero complex number $\zeta\in\Cb^\star$:
\begin{equation}
\label{eq:qq_defs}
\q{1} \equiv \QQ^1 -  \zeta \tilde{\SS}_1~, \qquad\qquad  \q{2} \equiv \SS_1 + \tfrac{1}{\zeta}\tilde{\QQ}^1~.
\end{equation}
Note that $\zeta$ can be set to any value by the action of an inner automorphism of the superconformal algebra, so we can set it to our favorite constant value without loss of generality. Nevertheless, we will see below that for book keeping purposes it is useful to keep $\zeta$ undetermined. 

These supercharges have the property that a particular diagonal combination of $\suf(1,1)$ and $\suf(2)_H$ is $\q{i}$-exact:
\begin{alignat}{4}
\label{DRexact}
&\ph{\tfrac{1}{4}}\{\q{1}\,,\, \tfrac{1}{2}\tilde{\QQ}^2 \} &~~~=~~~& \ph{\tfrac{1}{4}}\{\q{2}\,,\, -\tfrac{\zeta}{2}{\QQ}^2 \}  &~~~=~~~& P_{+-} + \zeta R_-     &~~~\equiv~~~& \widehat{L}_{-1}~, \nn\\
&\ph{\tfrac{1}{4}}\{\q{1}\,,\, \tfrac{1}{2\zeta}{\SS}_2  \} &~~~=~~~& \ph{\tfrac{1}{4}}\{\q{2}\,,\, \tfrac{1}{2}\tilde{\SS}_2 \} &~~~=~~~& K^{+-} - \zeta^{-1}R_+ &~~~\equiv~~~& \widehat{L}_{+1}~, \\
&\tfrac{1}{4}\{\q{1}\,,~\SS_1-\zeta^{-1}\tilde{\QQ}^1\} &~~~=~~~& \tfrac{1}{4}\{ \q{2}\,,~ \QQ^1 + \zeta \tilde{\SS}_1\} &~~~=~~~& D - R   &~~~\equiv~~~& \widehat{L}_{0}~.\nn 
\end{alignat}
The generators $\widehat{L}_{0,\pm1}$ obey $\slf(2)$ commutation relations, and we will refer to this subalgebra as $\widehat{\slf(2)}$. Also note that
\begin{equation}
\label{JRtildeexact}
\{ \q{1},\q{2} \} = 4(J_3 - \tilde R) = -2 \ZZ~.
\end{equation}

The nontrivial cohomology classes of local operators with respect to the $\q{i}$ can be characterized in two steps. We first characterize the operators that give rise to nontrivial $\q{i}$ cohomology classes when inserted at the origin. We can then employ the $\q{i}$-exact translation operator $\widehat{L}_{-1}$ to transport operators to other points on $\Rb_{\rm top}$. 
\begin{description}
\item[Cohomology at the origin.] When considering local operators situated at the origin, we may restrict ourselves without loss of generality to definite eigenspaces of the operators $\widehat{L}_{0}$ and $\ZZ$. It follows from \eqref{DRexact} and \eqref{JRtildeexact} that operators inserted at the origin that are $\q{i}$-closed but not $\q{i}$-exact must satisfy
\begin{equation}
\label{conditionsatorigin}
D=R \qquad \text{and} \qquad \tilde R = J_3~.
\end{equation}
In fact, a simple cohomological argument parallel to the one in \cite{Beem:2013sza} demonstrates that these are also sufficient conditions. Note that the second equality follows from the first by unitarity. Indeed, 
\begin{align}
\{Q^{1\tilde 1}_{ + } , S_{1\tilde 1}^{ + } \} &= 2  \left( D- R  +J_3 - \tilde R \right) \geqslant 0~,\\
\{Q^{1\tilde 2}_{ - } , S_{1\tilde 2}^{ - } \} &= 2  \left( D - R - (J_3 -  \tilde R) \right) \geqslant 0~,
\end{align}
and thus $D- R  \geqslant |J_3 - \tilde R|$. It further follows that operators at the origin must be in their $\suf(2)_H$ highest weight states to define a nontrivial cohomology class.

A complete classification of the operators satisfying \eqref{conditionsatorigin} can be obtained from the detailed accounting of $\ospf(4|4,\Rb)$ superconformal multiplets in \cite{Dolan:2008vc}. The result is that only the $\suf(2)_H$ highest weight state of the superconformal primaries of Higgs branch multiplets --- \ie, Higgs branch chiral ring operators --- are allowed.

\item[Full cohomology.] The nilpotent supercharges $\q{i}$ include special conformal supercharges, so they do not commute with any ordinary translation operators. They do commute with $\widehat{L}_{-1}$, which can then be used to transport operators along $\Rb_{\rm top}$ without removing them from the kernel of the $\q{i}$. Since $\widehat{L}_{-1}$ is $\q{i}$-exact, the operators transported in such a fashion will be independent of their position at the level of cohomology. Concretely, let $\OO_k^{(11\ldots 1)}(0)$ be a dimension $k/2$ chiral ring operator. Exponentiating the action of the translation operator $\widehat{L}_{-1}$ results in
\begin{equation}\label{eq:twisted_translated_ops}
\OO_k(s) = u_{a_1}(s)u_{a_2}(s)\ldots u_{a_k}(s) \ \OO_k^{(a_1a_2\ldots a_k)}(s)\;, \qquad u_a(s) = \begin{pmatrix}
1\\
\zeta s
\end{pmatrix}~.
\end{equation}
Because the twisted translation operator is $\q{i}$-exact, the operator defined above is independent of $s$ at the level of cohomology. Nevertheless, the ordering of operators along $\Rb_{\rm top}$ remains meaningful as operators can not be moved around each other without leaving the kernel of the $\q{i}$. For the time being we will therefore retain the coordinate $s$ as a label for $\q{i}$-cohomology classes of operators, while recognizing that the dependence on $s$ is locally constant:\footnote{Because the cohomologies of the two supercharges $\q{i}$ are isomorphic and we have chosen representatives that are well-defined with respect to both, we will drop the explicit $i$ from now on. One could equally well restrict attention to a single choice of $i$ (or a linear combination).}
\begin{equation}
\OO_k^{(s)} \equiv [\OO_k(s)]_{\qq}~.
\end{equation}
\end{description}

Familiar arguments about cohomology classes of local operators imply that OPEs and correlation functions of $\q{}$-closed operators are well-defined at the level of $\q{}$-cohomology classes, so the OPE algebra of the full SCFT will reduce to a well-defined cohomological operator algebra. This algebra will necessarily still be associative, since we can perform OPEs using the associative algebra of the full SCFT and then pass to cohomology at the end. Thus the output of this machinery is an associative algebra of operators that are labelled by their position on a line, with the multiplication depending only on (at most) the ordering of the operators on the line. In addition there is an evaluation operation that comes from taking expectation values, or equivalently, picking out the coefficient of the identity operator after taking the OPE. This is what we called a ``one-dimensional topological algebra'' in the introduction.

\subsection{The free hypermultiplet}
\label{subsec:free_hyper} 

To illustrate this construction in a completely tractable example, let us analyze the simplest example of an $\NN=4$ SCFT in three dimensions: the free hypermultiplet.  The fields in this theory are dimension one-half scalars and a dimension one fermion,
\begin{equation}
q^a_{i}(x)~,\qquad \psi_{\alpha,i}^{\tilde a}(x)~.
\end{equation}
Here $a$ is an $\suf(2)_H$ doublet index, $\tilde a$ is an $\suf(2)_C$ doublet index, $i$ an $\suf(2)_F$ flavor doublet index, and $\alpha$ is a spacetime spinor index. The complete set of local operators in this theory comes from normal ordered products of (derivatives of) these fields and their complex conjugates. Since this is a free field theory, the quantum numbers of these operators behave additively under normal ordered products, so it is simple to determine which operators obey the relevant conditions \eqref{conditionsatorigin}. They are products of the scalars $q^1_{i}$, which of course are just the Higgs branch chiral ring operators.

The OPE of the scalar fields takes the following covariant form:
\begin{equation}
q^a_{i}(x)q^b_{j}(y) = \frac{\varepsilon^{ab}\epsilon_{ij}}{|x-y|} + (q^a_{i}q^b_{j})(y)+\ldots~,
\end{equation}
where the ellipsis represents conformal descendants and we use conventions that $\epsilon_{12}=\varepsilon^{12}=1$. From here we can determine the OPEs of the corresponding twisted-translated operators defined in \eqref{eq:twisted_translated_ops}. We define the cohomological local operators corresponding to the elementary scalars as
\begin{equation}
q_{i}^{(s)}\equiv [q^1_{i}(s) +  s \zeta q^2_{i}(s)  ]_{\qq}~.
\end{equation}
The OPEs are easiest to calculate by positioning one of the operators at the origin (though this is by no means essential). We find
\begin{equation}
q_{i}^{(0)}q_{j}^{(s)} = \zeta \epsilon_{ij}\  \text{sgn}(s) + (q_{i}q_{j})^{(0)}~,
\end{equation}
where $(q_{i}q_{j})^{(0)}$ is the cohomology class of the operator $(q_{i}^1 q_{j}^1)(0)$, and because we are working in cohomology we have omitted $\qq\,$-exact terms.

Let us introduce the notation $\star$ to denote the multiplication of $\qq$ cohomology classes of local operators that are ordered left to right on $\Rb_{\rm top}$, \ie,
\begin{equation}
\OO_1\star\OO_2\equiv \OO_1^{(s_1)}\OO_2^{(s_2)}~,\quad s_1<s_2~.
\end{equation}
Then the above example gives the following simple star product for the elementary scalars,
\begin{equation}
q_{i} \star q_{j} = \zeta \epsilon_{ij} + (q_{i}q_{j})~.
\end{equation}
We see immediately that this product is indeed noncommutative. It is an exercise in Wick contractions to work out the corresponding expression for the star products of more general composite chiral ring operators. The result takes an elegant form:
\begin{equation}\label{Moyal-product}
(q_{i_1}\ldots q_{i_k}) \star (q_{j_1}\ldots q_{j_l}) = (q_{i_1}\ldots q_{i_k})\exp\left[ \zeta \epsilon_{kl} \overset{\leftarrow}{\partial}_{\!q_{k}}\overset{\rightarrow}{\partial}_{\!q_l} \right] (q_{j_1}\ldots q_{j_l})~.
\end{equation}
This is the famous Moyal-Weyl-Groenewold star product, which appears when quantizing the $\Rb^2_{[x,p]}$ phase space of a single nonrelativistic particle. This observation foreshadows the general structure we will find in the coming sections.

\subsection{General form of the twisted OPE}
\label{subsec:twisted_OPE}

The twisted OPE for operators in more general (interacting) theories is determined in a simple way in terms of certain OPE coefficients for half BPS operators. Let us first introduce our conventions for these OPE coefficients. For a Higgs branch operator in the spin $\ell/2$ representation of $\suf(2)_H$, we define
\begin{equation}\label{eq:op_with_polarization}
\OO(x,y)\equiv \OO^{(a_1\cdots a_{\ell})}(x)y_{a_1}\ldots y_{a_{\ell}}~,
\end{equation}
where $y_a$ is a commuting polarization variable in the two-dimensional representation of $\suf(2)_H$. The $\suf(2)_H$ transformation properties of correlation functions and OPEs of these operators can be treated efficiently in terms of these variables. The forms of the two- and three-point functions of Higgs branch operators are determined by conformal and $\suf(2)_H$ invariance and are given by
\begin{equation}
\begin{split}
\label{eq:two_three_point_polarization}
\langle\OO_i(x_i,y_i)\OO_j(x_j,y_j)\rangle				 &=~\frac{\eta_{ij}\langle y_i,y_j\rangle^{\ell}}{|x_i-x_j|^{\ell}}~,\\
\langle\OO_i(x_i,y_i)\OO_j(x_j,y_j)\OO_k(x_k,y_k)\rangle &=~\frac{c_{ijk}\langle y_i,y_j\rangle^{\ell_{ijk}}\langle y_j,y_k\rangle^{\ell_{jki}}\langle y_i,y_k\rangle^{\ell_{kij}}}{|x_i-x_j|^{\ell_{ijk}}|x_j-x_k|^{\ell_{jki}}|x_k-x_i|^{\ell_{kij}}}~,\
\end{split}
\end{equation}
where $\langle y,\tilde{y}\rangle\equiv \epsilon^{ab}y_a \tilde{y}_b = y_{1}\tilde{y}_2-y_2\tilde{y}_1$ and $\ell_{ijk}\equiv\frac{\ell_i+\ell_j-\ell_k}{2}$. In particular, with these conventions the matrix $\eta_{ij}$ and the three-point couplings $c_{ijk}$ can be understood in terms of explicit correlators of certain components of the Higgs branch operators,
\begin{equation}
\begin{split}
\label{eq:two_three_point_normalization}
\langle\OO_i^{(1\cdots1)}(x_i)\OO_j^{(2\cdots2)}(x_j)\rangle 								&=~ \frac{\eta_{ij}}{|x_i-x_j|^{\ell}}~,\\
\langle\OO_i^{(1\cdots1)}(x_i)\OO_j^{(1\cdots12\cdots2)}(x_j)\OO_k^{(2\cdots2)}(x_k)\rangle &=~ \frac{c_{ijk}}{|x_i-x_j|^{\ell_{ijk}}|x_j-x_k|^{\ell_{jki}}|x_k-x_i|^{\ell_{kij}}}~.
\end{split}
\end{equation}
This turns out to be useful for manipulations involving twisted-translated operators. Note also that with these conventions, we have $c_{ijk}=(-1)^{\ell_{ijk}}c_{jik}$ and $c_{ij0}=\eta_{ij}$.

It is useful to have a covariant expression for the way that a Higgs branch operator appears in the OPE of two other Higgs branch operators. The form of the OPE is completely determined by \eqref{eq:two_three_point_polarization}, and is given by
\begin{equation}\label{eq:truncated_ope}
\OO_i(x,y)\OO_j(\tilde{x},\tilde{y})\sim
\frac{c_{ij}^{\ph{ij}k}\langle y,\tilde{y}\rangle^{\ell_{ijk}}}{\left|x-\tilde{x}\right|^{\ell_{ijk}}}\OO_{k}^{(a_1\cdots a_{\ell_k})}(\tilde{x}){y}_{a_1}\ldots y_{a_{\ell_{kij}}}\tilde{{y}}_{a_{\ell_{kij}+1}}\ldots \tilde{y}_{a_{\ell_k}}~.
\end{equation}
The OPE coefficients above and the previously introduced two- and three-point coefficients are related in a natural way,
\begin{equation}
c_{ijk}=c_{ij}^{\ph{ij}k^\prime}\eta_{k^\prime k}~.
\end{equation}

These OPE coefficients and two- and three-point functions are related directly to the structure constants and correlation functions that control the algebra of $\qq$ cohomology classes of local operators. To observe this coincidence, note that the twisted translated operators in \eqref{eq:twisted_translated_ops} can be written as
\begin{equation}
\OO_k(s)=\OO_k(s;y_1=1~, y_2=\zeta s)~.
\end{equation}
The terms in the OPE on the right hand side of \eqref{eq:truncated_ope} then take a simplified form
\begin{equation}\label{eq:covariant_ope}
\begin{split}
\OO_i(s)\OO_j(s^{\prime})	&\sim~ c_{ij}^{\ph{ij}k}\zeta^{\ell_{ijk}}\left(\frac{s^\prime-s}{|s^\prime-s|}\right)^{\ell_{ijk}}\OO_k(s^\prime) +\ldots~,\\
							&\sim~ c_{ij}^{\ph{ij}k}\zeta^{\ell_{ijk}}\sgn(s^\prime-s)^{\ell_{ijk}}\;\OO_k(s^\prime) +\ldots~,
\end{split}
\end{equation}
where the ellipses are $\qq\,$-exact operators.\footnote{Computing this OPE is made easier by choosing $s^\prime=0$, but of course the result is independent of the value of $s^\prime$.} Adopting the conventions from our discussion of the free hypermultiplet, we thus have a noncommutative, but associative, star product defined on the vector space of Higgs branch chiral ring operators given by
\begin{equation}
\OO_i\star\OO_j=\sum_k c_{ij}^{\ph{ij}k}\zeta^{\ell_{ijk}}\OO_k~.
\end{equation}
Additionally this operator algebra determines correlation functions by keeping just the coefficient of the identity operator in the above product. For convenience, let us define a bracket operation for collections of operators of definite degree that strips off the inessential factors of $\zeta$ that appear in the star product,
\begin{equation}
\langle \OO_{i_1}\cdots\OO_{i_n}\rangle \equiv \zeta^{-\frac{1}{2}(\ell_1+\cdots+\ell_n)}{\rm C.T.}\left(\OO_{i_1}\star\cdots\star\OO_{i_n}\right)~,
\end{equation}
Where ${\rm C.T.}$ means taking the constant term. With this definition we have the natural expressions
\begin{equation}
\langle\OO_i\OO_j\OO_k\rangle=c_{ijk}~,\qquad \langle\OO_i\OO_j\rangle=\eta_{ij}~.
\end{equation}

We would like to use the associativity of this algebra (and additional conditions that follow from three-dimensional physics) to bootstrap the structure constants of this algebra for specific theories of interest. However, from our general discussion so far it is not clear how plausible this goal is. In the next section we describe additional properties of the star product that follow from the structure of the superconformal OPE. This allows us to refine our expectations by characterizing the general form of the bootstrap problem for these algebras.

\section{Properties of the protected associative algebra}
\label{sec:algebra_properties}

The star product defined above can be thought of as a noncommutative multiplication operation on the vector space of Higgs (or Coulomb) branch chiral ring operators in a three-dimensional $\NN=4$ SCFT.\footnote{As in Appendix \ref{app:HK_cones}, we will generally denote the chiral ring by $\AA$. When we want to emphasize the connection to the coordinate algebra of the moduli space $\MM$, we may write $\AA=\Cb[\MM]$.} This multiplication is defined by the structure constants appearing in the OPE \eqref{eq:truncated_ope}. While the algebra can be defined for any numerical value of $\zeta\elem\Cb^\star$, we will keep $\zeta$ as an indeterminate parameter, in which case we have a $\Cb[\zeta]$-linear noncommutative algebra on the vector space of power series in $\zeta$ with coefficients in the chiral ring $\AA$:
\begin{equation}
\star:\AA[\zeta]\otimes\AA[\zeta]\mapsto\AA[\zeta]~.
\end{equation}
We will denote this noncommutative algebra by $\AA_{\zeta}$. This algebra has a number of special features that follow from the structure of the underlying superconformal OPE. Below we enumerate these properties in order to define a self-contained algebra problem -- which we will call \emph{the bootstrap problem} -- that the OPE coefficients of our SCFT must solve.

\subsection{Equivariance, selection rules, and symmetry}
\label{subsec:kinematical_properties}

Let us introduce the $\Zb_{\geqslant0}$ grading on $\AA$ by $\suf(2)_H$ charge, or equivalently by conformal dimension,\footnote{We are adopting conventions where the spin one-half representation of $\suf(2)_H$ has unit grading. This allows us to work with integral, instead of half-integral, gradings at the expense of a slight disconnect with the most popular physics conventions.}
\begin{equation}
\AA=\bigoplus_{p\elem\Zb_{\geqslant0}}\AA_p~.
\end{equation}
The star product violates this grading because the twisted translated operators involve lower components of the $\suf(2)_H$ multiplet. Since this violation always comes from multiplication by operators with lower charge than the chiral ring operator, the star product preserves the associated filtration, \ie, 
\begin{equation}
\AA_p\star\AA_q\subset\bigoplus_{k=0}^{p+q}\AA_k[\zeta]~.
\end{equation}
In fact, we have a stronger condition than this filtration. In the twisted translated operators, each unit of violation of $\suf(2)_H$ charge is compensated by a power of $\zeta^{1/2}$. This means that our star product is $\Cb^\star$ equivariant, \ie,
\begin{equation}\label{eq:graded_star_vectorspaces}
\AA_p\star\AA_q\subset \bigoplus_k \zeta^k\AA_{p+q-2k}~.
\end{equation}
This amounts to endowing $\zeta$ with a scaling dimension of two and demanding that $\star$ be a graded multiplication on $\AA[\zeta]$, and indeed $\zeta$ does scale with an effective dimension of two when we act on the supercharges $\q{i}$ with the $\suf(2)_H$ Cartan.

Furthermore, though $\suf(2)_H$ is not a symmetry at the level of $\qq$\,-cohomology, the underlying OPEs and correlation functions of the CFT obey selection rules for $\suf(2)_H$ representations. This means that an operator $\OO_k$ can only appear in $\OO_i\star\OO_j$ if the $\suf(2)_H$ representation of $\OO_k$ in the full theory appears in the tensor product of the representations of $\OO_i$ and $\OO_j$. This truncates the sequence of vector spaces appearing on the right hand side of \eqref{eq:graded_star_vectorspaces} as follows,
\begin{equation}\label{eq:truncated_star_vectorspaces}
\AA_p\star\AA_q\subset \bigoplus_{k=0}^{\min(p,q)} \zeta^k\AA_{p+q-2k}~.
\end{equation}
This means that compared to the most general $\Cb^\star$-equivariant star product, the $\zeta$-expansion of our star product will truncate prematurely.

Finally, we recall that the OPE coefficients in \eqref{eq:covariant_ope} had definite symmetry properties under the exchange of operators, and these are inherited by the structure constants of the star algebra. The result is the requirement that even terms in the $\zeta$-expansion are symmetric under interchange of the two multiplied operators, while odd powers are antisymmetric. A fancy, but succinct, way of expressing this is that the opposite algebra\footnote{The opposite of any algebra is an algebra where the order of multiplication is reversed, so $f\star_{\rm op}g \equiv g\star f$.} with $\zeta$ negated is the same as the original algebra,
\begin{equation}
\AA^{\rm op}_{-\zeta} = \AA_{\zeta}~.
\end{equation}
Alternatively, if we consider $\OO_i\elem\AA_p$ and $\OO_j\elem\AA_q$, then we have
\begin{equation}
\begin{split}
\OO_i\star\OO_j+\OO_j\star\OO_i&\subset~\bigoplus_{k=0,2,\ldots}\zeta^{k}\AA_{p+q-2k}~,\\
\OO_i\star\OO_j-\OO_j\star\OO_i&\subset~\bigoplus_{k=1,3,\ldots}\zeta^{k}\AA_{p+q-2k}~.\\
\end{split}
\end{equation}

\subsection{Leading terms}

Most of the structure constants of our algebra encode unknown two- and three-point functions of Higgs branch operators, and we do not know how to determine them in advance before performing some additional bootstrap analysis. However, the leading and first subleading terms in the $\zeta$ expansion can be understood in terms of the algebraic geometry of the Higgs branch, and this will form the seed for any subsequent analysis.

The leading term in the star product comes directly from the multiplication of Higgs branch \emph{chiral ring} operators in the expansion of the twisted translated operators. These leading terms are thus controlled by the coordinate ring of the Higgs branch of vacua $\Cb[\MM_H]$. In many cases of interest this ring is known. Denoting by $\OO_f$ the operator whose Higgs branch chiral ring is associated to the function $f\elem\Cb[\MM_H]$, we have
\begin{equation}\label{eq:star_leading}
\OO_f\star \OO_g = \OO_{f\cdot g} + O(\zeta)~,
\end{equation}
where we have introduced $\cdot$ as the multiplication in $\Cb[\MM_H]$. We see that the star product can be thought of as a noncommutative deformation of the Higgs branch chiral ring, with $\zeta$ the deformation parameter.

The first subleading term turns out to also be controlled by the Higgs branch. This is more subtle to see, but follows from the formalism of topological descent in Rozansky-Witten theory \cite{Rozansky:1996bq}.\footnote{The following argument regarding the appearance of the poisson bracket in Rozansky-Witten theory here were alluded to in \cite{Bullimore:2015lsa}. A more detailed analysis of this and related constructions is the subject of work in progress by the first author in collaboration with David Ben-Zvi, Mathew Bullimore, Tudor Dimofte, and Andy Neitzke.} In the Rozansky-Witten twist of a four-dimensional $\NN=4$ theory we can build a two-form operator associated to each Higgs branch chiral ring operator (a zero-form operator in the topological theory) $\OO_f$ via descent,
\begin{equation}
\begin{split}
d\OO_f &=~\{Q_{\rm RW},\OO_f^{(1)}\}~,\\
d\OO_f^{(1)} &=~[Q_{\rm RW},\OO_f^{(2)}]~,
\end{split}
\end{equation}
with the property that the integral of $\OO_f^{(2)}$ over a closed two-cycle is a physical operator in the topological theory. This leads to a secondary algebraic operation in the Rozansky-Witten theory defined by
\begin{equation}
(\OO^{(0)}_{f},\OO^{(0)}_{g})=\int_{S^2_0}\OO^{(2)}_f\OO^{(0)}_g~.
\end{equation}
One can verify by explicit calculation that when the theory being twisted is represented as a sigma model onto a hyperk\"ahler manifold, then this secondary algebraic operation gives the poisson bracket of the functions $f$ and $g$.

Now there are two key observations that connect this discussion to our star product. The first is that the equation for the secondary product in the Rozansky-Witten theory is an operator equation up to $Q_{\rm RW}$-exact terms, so even when we are dealing with an SCFT whose Higgs branch is a singular hyperk\"ahler space we can compute the secondary product in a Higgs branch vacuum where locally the theory is a sigma model at low energies. 

The second observation is that this secondary product in the Rozansky-Witten twisted version of an SCFT controls the first subleading term in the $\zeta$-expansion of the star product. Though this is not quite obvious, it becomes clear when one recalls that in the two- and three-point functions of half-BPS operators in a three-dimensional $\NN=4$ SCFT there is a unique superspace structure and the only freedom is in the coefficients appearing in the correlation functions of the superconformal primaries \cite{Ferrara:2001uj}. We then see that (up to an overall universal constant determined by supersymmetry) both the secondary product of the twisted theory and the first subleading term in the star product are measuring the Higgs branch operator of dimension $p+q-2$ appearing in the OPE of Higgs branch operators of dimension $p$ and $q$, so we have\footnote{Note that because of the scaling symmetry of the chiral ring, the poisson bracket is only defined up to an overall constant. In this expression we have implicitly fixed this constant by demanding that the subleading term in the OPE be given by the poisson bracket with unit coefficient.}
\begin{equation}\label{eq:star_leading_and_subleading}
\OO_f\star \OO_g = \OO_{f\cdot g} + \frac{\zeta}{2}\,\OO_{\{f,g\}}+ O(\zeta^2)~.
\end{equation}

\subsection{Reality and positivity}

The associative algebra also inherits reality and positivity properties that follow from unitarity of the underlying SCFT. The first of these is a consequence of CPT symmetry. In Appendix \ref{app:HK_cones} we introduce the $\Cb$-antilinear conjugation operation on the chiral ring $\rho:\AA\mapsto\AA$. This operation is a combination of the CPT operator $\Theta$ and an $\suf(2)_H$ rotation by $\pi$. It is implemented by an anti-unitary operator in the SCFT, so leads to a relation on correlation functions that acts by complex conjugation. At the level of three-point functions, we therefore have\footnote{Note that although the CPT operator will switch the order of the operators on $\Rb_{\rm top}$, it will not change the sign of the explicit factors of $s$ in the expression for the twisted translated operators, so this change of ordering has no effect. Alternatively, we can think of it as changing the total order and also changing the sign of $\zeta$.}
\begin{equation}
\langle\OO_{\rho(i)}\OO_{\rho(j)}\OO_{\rho(k)}\rangle = \overline{\langle\OO_i\OO_j\OO_k\rangle}~.
\end{equation}
Or in terms of structure constants in a fixed basis,
\begin{equation}
\rho_i^{\ph{i}i^\prime}\rho_j^{\ph{j}j^\prime}\rho_k^{\ph{k}k^\prime}c_{i^\prime j^\prime k^\prime}=\overline{c_{ijk}}~.
\end{equation}
This condition can be improved to a true reality condition under certain circumstances. One scenario where this takes place is if there is a basis for $\AA$ in which $\rho$ acts as the identity on some operators. In this case, the three-point function of any such operators will be real by the above equation. More generally, suppose that there is a unitary symmetry $\CC$ in the SCFT that acts identically to $\rho$ in some basis. This is also enough to ensure the reality of three-point functions (and structure constants) in said basis, since we will have
\begin{equation}
\begin{split}
\langle \OO_i\OO_j\OO_k \rangle 		&\xrightarrow{\Theta\circ R_{\pi}} \langle \OO_{\rho(i)} \OO_{\rho(j)} \OO_{\rho(k)} \rangle = \overline{c_{ijk}}~, \\
\ph{\langle \OO_i\OO_j\OO_k \rangle} 	&\xrightarrow{~~\,\,\CC\,~~} \langle \OO_{\rho(i)} \OO_{\rho(j)} \OO_{\rho(k)} \rangle = c_{ijk}~.\\
\end{split}
\end{equation}
In other words, when an anti-unitary and a unitary symmetry act the same way on some observable, then that observable must be real. In all of the examples considered in this paper we will have a basis in which the above relations hold, so the algebra will have a basis with all real structure constants. It is an interesting question whether all three-dimensional $\NN=4$ theories have this property.

Additional constraints arise from the positivity of two-point functions of identical operators in a unitary CFT. In particular, for any complex scalar operator in a CFT we have
\begin{equation}
\label{eq:positive_definite_two_point}
\langle\OO(x)\OO^\dagger(y)\rangle=\frac{n_{\OO}}{|x-y|^{2\Delta_\OO}}~,\qquad n_\OO>0~,
\end{equation}
and this leads to a positivity requirement in the associative algebra. To formulate this requirement, let us introduce the following sesquilinear form on $\AA=\Cb[\MM]$:
\begin{equation}
\theta(f,g) = \langle \OO_{\rho(f)}\OO_{g}\rangle~.
\end{equation}
Recall that $\rho^2=(-1)^{2R}$, so for $f,g\elem\AA_p$ we have
\begin{equation}
\begin{split}
\theta(g,f) &= \langle\OO_{\rho(g)}\OO_f\rangle~,\\
			&= (-1)^{p}\langle\OO_f\OO_{\rho(g)}\rangle~,\\
			&= (-1)^{p}\overline{\langle\OO_{\rho(f)}\OO_{\rho^2(g)}\rangle}~,\\
			&= \overline{\langle\OO_{\rho(f)}\OO_{(g)}\rangle}~,\\
\end{split}
\end{equation}
so this is form is Hermitian, and the positivity requirement in \eqref{eq:positive_definite_two_point} (or more generally the positive definiteness of the norm of the Hilbert space in the SCFT) translates directly into the positive definiteness of this Hermitian form.

These are nontrivial constraints. We will see in our examples that while the reality conditions are easily satisfied, the positivity conditions are very difficult to impose. They may in fact play a key role in making the solution to the bootstrap problem nearly unique.

\subsection{Summary of the bootstrap problem}

We can now succinctly state the algebra problem that we would like to solve. Ideally, its solution will completely, or nearly completely, determine the structure constants of the protected associative algebra. In the rest of the paper we will refer to this as the bootstrap problem. It takes the following form:

Suppose we are given the coordinate algebra of a hyperk\"ahler cone (presumably the moduli space of vacua of a known three-dimensional $\NN=4$ SCFT) as a (graded) poisson algebra with conjugation $(\AA\,,\,\{\cdot,\cdot\}\,,\,\rho)$. The problem is to find a noncommutative star product $\star:\AA[\zeta]\otimes\AA[\zeta]\rightarrow\AA[\zeta]$ such that the following conditions are satisfied. Below, we will always take $f\elem\AA_p$ and $g\elem\AA_q$:
\begin{description}
\item[Associativity]\hfill\\
\vspace{-12pt}
\begin{center}
$(f\star g)\star h = f \star (g\star h)$~,
\end{center}
\item[Equivariance]
$$f\star g = \sum_{k=0}^{\left\lfloor\frac{p+q}{2}\right\rfloor}\zeta^k C^k(f,g)~,\qquad {\rm with}~\quad C^k:\AA_p\otimes\AA_q\mapsto\AA_{p+q-2k}~.$$
\item[Truncation]\hfill\\
\vspace{-12pt}
\begin{center}
$C^k(f,g)=0$ for $k>\min(p,q)$~,
\end{center}
\item[Leading terms]\hfill\\
\vspace{-12pt}
\begin{center}
$C^0(f,g)=f\cdot g$~,\quad $C^1(f,g)=\frac12\{f,g\}$~,
\end{center}
\item[Evenness]\hfill\\
\vspace{-12pt}
\begin{center}
$C^k(f,g)=(-1)^k C^k(g,f)$~,
\end{center}
\item[CPT]\hfill\\
\vspace{-12pt}
\begin{center}
${\rm C.T.}(f\star g\star h) = \overline{{\rm C.T.}(\rho(f)\star\rho(g)\star\rho(h))}$~,
\end{center}
\item[Positivity]\hfill\\
\vspace{-12pt}
\begin{center}
$\theta(f,g)\colonequals C^{p}(\rho(f),g)$ for $f,g\elem\AA_p$ a positive definite Hermitian form.
\end{center}
\end{description}
\vspace{10pt}
\noindent We will see that some of these conditions are much more easily satisfied than others.

\section{Relationship to deformation quantization}
\label{sec:math_phys}

Our bootstrap problem is very similar to the problem of deformation quantization. Indeed, as we will review below, the associativity, equivariance, and leading term conditions precisely define the problem of $\Cb^\star$-equivariant deformation quantization of a graded poisson algebra. There is a significant literature about deformation quantization as a problem in pure mathematics -- see, \eg, \cite{Weinstein,Esposito:formality} and references therein. There are, however, important differences between the bootstrap problem and deformation quantization problem as it is usually formulated. Below we review the conventional formulation of deformation quantization and mention some results about the deformation quantization of hyperk\"ahler cones, which is the case relevant to our goals. We conclude with a comparison of the deformation quantization problem and the bootstrap problem.

\subsection{Deformation quantization}
\label{subsec:def_quant_math}

The starting point in deformation quantization is a commutative poisson algebra $\left(\AA, \{\cdot,\cdot\}\right)$. Often (particularly in physics) this is the algebra of smooth functions on a symplectic manifold, though the case of more general poisson algebras is also well-studied. The goal is to find an associative deformation of the product on $\AA$, usually denoted $\star$, defined on the vector space $\AA\llbracket\hbar\rrbracket$ of formal power series in $\hbar$ with coefficients in $\AA$. Such a product is determined by its action on elements of $\AA$ and extended by linearity to $\AA\llbracket\hbar\rrbracket$. The star product can be written as
\begin{equation}
f\star g = \sum_{k=0}^{\infty} C^k(f,g)\hbar^k~,
\end{equation}
where the $C^k$ are bilinear maps $C^k:\AA\otimes\AA\rightarrow\AA$, and $C^0$ is defined as the original commutative multiplication on $\AA$.\footnote{When $\AA$ is an algebra of smooth functions on some manifold, the $C^k$ are often taken to be bi-differential operators. This is not an essential requirement, and we will not explicitly make such an assumption in our version of the problem.} The requirement that the star product be associative amounts to the following requirement for the $C^k$,
\begin{equation}
\sum_{i+j=n}C^i(C^j(f,g),h)=\sum_{i+j=n}C^i(f,C^j(g,h))~,\qquad f,~g,~h\elem\AA~.
\end{equation}
Without too much trouble one can show that associativity requires that the antisymmetric part of the leading term in the star product defines a poisson bracket -- \ie, it is an antisymmetric bilinear map satisfying the Leibniz rule and Jacobi identities:
\begin{equation}
\{f,g\}_\star\equiv C^1(f,g)-C^1(g,f)~.
\end{equation}
To be a deformation quantization of the original poisson algebra, this poisson bracket should agree with the original one $\{\cdot,\cdot\}$. This situation is sometimes described by saying that deformation quantization is a noncommutative deformation of $\AA$ in the direction of the poisson bracket.

Solutions to the deformation quantization problem are generally organized into large equivalence classes. The origin of such equivalence classes is as follows. Suppose we are given an associative star product. Then consider an arbitrary $\Cb\llbracket\hbar\rrbracket$-linear map $T:\AA\llbracket\hbar\rrbracket\mapsto\AA\llbracket\hbar\rrbracket$ that takes the form
\begin{equation}
\begin{split}
\label{eq:graded_change_of_basis}
T(f)		&=f+\hbar f^{(1)}+\hbar^2 f^{(2)}+\ldots~,\\
T(\hbar)	&=\hbar(1 + a_1 \hbar + a_2 \hbar^2+\ldots)~,
\end{split}
\end{equation}
where $f^{(i)}\elem\AA$ and $a_i\elem\Cb$. Then one can define another associative star product with different structure constants that also satisfies the conditions for deformation quantization by taking
\begin{equation}
\label{eq:equivalent_star}
f\,\tilde{\star}\,g=T^{-1}(T(f)\star T(g))~.
\end{equation}
Because this new product arises from an $\hbar$-dependent change of basis for $\AA\llbracket\hbar\rrbracket$, it can be (and often is) considered an equivalent deformation quantization. In the mathematics literature, this type of change of basis is referred to as a gauge transformation, and the deformation quantization problem is normally treated up to gauge equivalence. In physics, these kinds of ambiguities are familiar as ordering ambiguities that arise in the definition of composite operators.

A more constraining version of deformation quantization is \emph{strict} deformation quantization, in which the series in $\hbar$ appearing in the above discussion are required to converge for some sufficiently small value of $\hbar$.\footnote{In contrast, deformation quantization over $\Cb\llbracket\hbar\rrbracket$ is called \emph{formal}.} These are then honest deformations of the original poisson algebra (not just as an algebra over $\Cb\llbracket\hbar\rrbracket$). Strict deformation quantization is generally a more difficult problem to analyze than formal deformation quantization. We note that in both the formal and the strict cases, the ``gauge group'' associated with transformations of the form given in \eqref{eq:equivalent_star} is infinite-dimensional.

There are many beautiful theorems about the solution of the deformation quantization problem in different contexts -- again, see \cite{Esposito:formality}. Fortunately for us, we will not need to use the full technology developed to address the general problem. This is because of the many simplifications that occur for precisely the case in which we are interested: the quantization of hyperk\"ahler cones.

\subsection{Deformation quantization of hyperk\"ahler cones and classification}
\label{subsec:def_quant_hkcones}

When the poisson algebra of interest has additional structure, one can demand more of the quantization. We are interested in the case where $(\AA\,,\,\{\cdot,\cdot\})$ is the coordinate algebra of a hyperk\"ahler cone. In this case $\AA$ is $\Zb_{\geqslant0}$ graded,
\begin{equation}
\AA=\bigoplus_{p=0}^\infty \AA_p~,\qquad \AA_p\cdot\AA_{q}\subset\AA_{p+q}~,
\end{equation}
and the poisson bracket on $\AA$ has degree $-2$ with respect to this grading,
\begin{equation}
\{\AA_p,\AA_q\}\subset\AA_{p+q-2}~.
\end{equation}
We can then ask for a $\Cb^*$\emph{-equivariant} deformation quantization by demanding that
\begin{equation}
C^k(\AA_p,\AA_q)\subset \AA_{p+q-2k}~.
\end{equation}
This is equivalent to giving $\hbar$ a scaling dimension of two and demanding that $\star$ be a graded multiplication on $\AA\llbracket\hbar\rrbracket$. 

If the star product is $\Cb^\star$-equivariant, only finitely many terms can appear in the $\hbar$ expansion of the product of any two elements of $\AA$. A $\Cb^\star$-equivariant quantization will therefore necessarily give rise to an instance of strict deformation quantization. The equivalence classes of equivariant quantizations are more restricted -- they are induced by maps of the form
\begin{equation}
\begin{split}
\label{eq:graded_equivalent_star}
T(f)		&=~f+\sum_{k=1}^{\lfloor\frac{p}{2}\rfloor} \hbar^k f^{(k)}~,\quad \text{where}~~f\elem\AA^{p}~~\text{and}~~f^{(k)}\elem\AA^{p-2k}~,\\
T(\hbar)	&=~\hbar~.
\end{split}
\end{equation}
This is still an infinite dimensional group of transformations, but crucially, only a finite number of terms can appear in the redefinition of a given operator.

Deformation quantization for hyperk\"ahler cones has been studied extensively in \cite{Braden:2014cb,Braden:2014iea}, building on \cite{Bezrukavnikov:2003cb,Losev:2010cb}. The key result for us is a classification theorem.\footnote{The classification theorem applies most directly to symplectic resolutions. These are singular holomorphic symplectic varieties that can be resolved into smooth holomorphic symplectic manifolds. In three-dimensional superconformal field theories, we often have moduli spaces of vacua that cannot be resolved into symplectic manifolds. In this case, the analogous classification theorem can still be established \cite{Braden:2014cb}.} The classification says that the $\Cb^\star$-equivariant quantizations of $\AA=\Cb[\MM]$ are in one-to-several correspondence with elements of $H^2(\wt\MM,\Cb)$, where $\wt\MM\rightarrow\MM$ is a (universal) smooth symplectic resolution of $\MM$. In the context of quantum field theory, this is the space of FI parameters that resolve the Higgs branch into a smooth variety. The element of $\lambda\elem H^2(\wt\MM,\Cb)$ that corresponds to a given quantization is known as the \emph{period} of the quantization.

The reason the correspondence is one-to-several is that there is a discrete group $W$ -- known as the Namikawa Weyl group -- that acts on $H^2(\wt\MM,\Cb)$ such that different periods that are on the same orbit give equivalent quantizations.\footnote{We should mention here that what we call a ``quantization'' here is actually a more basic object than that treated in the aforementioned papers. There, a quantization is a sheaf of associative algebras on $\MM$ with certain good properties. What we call a quantization is the algebra of global sections of this sheaf.\label{fn:sheafs}} This group is just the Weyl group of the global symmetry that acts as hyperk\"ahler isometries on the \emph{Coulomb} branch. Indeed, the FI parameters that deform the Higgs branch come from turning on scalars in the background vector multiplet that couples to conserved currents for the (topological) Coulomb branch symmetries, so the action of this Weyl group is not surprising. Consequently, the space of inequivalent quantizations of the coordinate ring of a hyperk\"ahler cone, up to gauge transformation, is given by $H^2(\wt\MM,\Cb)/W$.

\subsection{Deformation quantization in SCFT and preferred bases}
\label{subsec:def_quant_scft}

We can now see the relationship between the bootstrap problem of Section \ref{sec:algebra_properties} and traditional deformation. Indeed, upon exchanging $\hbar$ and $\zeta$ many of the points in the previous subsection were replicas of conditions appearing in the bootstrap problem. The leading term and equivariance conditions in the bootstrap problem mean that the algebra we are dealing with is a $\Cb^\star$-equivariant deformation quantization of the chiral ring. Consequently the algebras satisfying these conditions are classified up to gauge equivalence by the theorem mentioned above, and this classification yields a finite-dimensional space of algebras!

We can also make contact with the mathematical literature on the point of the evenness condition. Indeed, there is a notion of an \emph{even quantization} introduced in \cite{Losev:2010cb}, which is a quantization that obeys
\begin{equation}
\AA^{op}_{-\hbar}=\AA_{\hbar}~.
\end{equation}
But this is just our evenness condition!\footnote{This is actually a variation on the notion of an even quantization given in \cite{Losev:2010cb}. The difference is that in \cite{Losev:2010cb} the quantizations being constructed are sheaves of algebras as described in footnote \ref{fn:sheafs}. This makes evenness in the sense defined here a slightly weaker condition than evenness in the sense of \cite{Losev:2010cb}. In particular, with the stronger notion the quantizations with periods related by the Namikawa Weyl group are inequivalent, and as a result the only even quantization is the one with zero period.} By Propositions (3.2) and (3.10) of \cite{Braden:2014cb}, the quantization can only be even if the period of the quantization $\lambda\elem H^2(\wt\MM,\Cb)$ lies on the same Weyl orbit with its negative, $-\lambda = w\cdot\lambda$ for some $w\elem W$. This is a nontrivial condition that will cut down on the space of algebras under consideration.

The remaining conditions -- namely the truncation condition, as well as the reality and positivity conditions -- are not standard in the deformation quantization literature. But this comes as no surprise since these properties are not invariant under the changes of basis described by \eqref{eq:graded_change_of_basis}, and deformation quantization is usually only studied up to such changes of basis. It is fortunate that we have these additional conditions because we care about the precise form of the structure constants of our algebra and the gauge transformations do not correspond to physical equivalences in the SCFT.\footnote{To be more specific, if we were to try to interpret them as such they would come from changes of basis in the chiral ring that mix operators of different scaling dimension. While we are certainly free to do such a thing, we also know that there is a \emph{canonical} basis where we choose operators of definite scaling dimension, and it is in this basis that we are interested in the structure constants of the associative algebra.} Thus what we have are some kind of gauge-fixing conditions. Purely in terms of the algebra problem, it isn't obvious that these conditions should be solvable. If they are solvable, then it isn't obvious that the space of solutions should be finite dimensional. Our bootstrap problem can now be broken down into a few subproblems:
\begin{enumerate}
\item Identify the quantum algebra (up to gauge equivalence). Ideally this will take the form of a generators-and-relations definition of the algebra.
\item Determine the period of the quantization relevant for the theory under consideration.
\item Solve the gauge fixing conditions for the appropriate value of the period. Hopefully the solution will be unique.
\end{enumerate}
In the examples that follow, the first problem is straightforward, while the second two are harder. In the examples of minimal nilpotent orbits, the gauge fixing conditions are irrelevant, while the period can be determined by a localization calculation. For the Kleinian singularities, the period and gauge fixing problems are hard to disentangle, since only some values of the period allow a basis that satisfies the gauge fixing.

\section{Examples: nilpotent orbits}
\label{sec:nilpotent}

The major complication in the bootstrap problem relative to the deformation quantization problem is that the latter works modulo a large group of gauge transformations that are not invariances of the bootstrap problem. For our first class of examples, we will therefore consider cases in which there is no gauge ambiguity even in the deformation quantization problem.

There is a general mechanism for avoiding the ambiguity due to gauge transformations. Suppose that a hyperk\"ahler cone $\MM$ has a nontrivial hyperk\"ahler isometry group 
\begin{equation*}
\label{eq:hk_isometry}
G:\MM\rightarrow\MM~.
\end{equation*}
Then the coordinate ring can be decomposed as a vector space into subspaces transforming in the distinct irreducible representations of $G$:
\begin{equation}
\label{eq:g_rep_decomposition}
\AA=\bigoplus_{\RR}\AA_{\RR}~.
\end{equation}
Now suppose that for any representation $\RR$ of $G$, $\AA_{\RR}$ is supported entirely in a single graded component of $\AA$,
\begin{equation}
\label{eq:rep_grading_overlap}
\forall~\RR~,~~\exists~p\elem\Zb_{\geqslant0}\quad{\rm s.t.}\quad\AA_{\RR}\subset\AA_p~.
\end{equation}
In such a case, a change of basis of the type described by \eqref{eq:graded_change_of_basis} cannot be $G$-equivariant simply because the correction terms involve only terms of strictly lower degree than the operator being corrected. This means that in a $G$- and $\Cb^\star$-equivariant quantization of such a hyperk\"ahler cone, the basis for the quantum algebra will be uniquely determined.

There exists an infinite family of hyperk\"ahler cones whose coordinate algebras satisfy precisely the property \eqref{eq:rep_grading_overlap}, and they all occur as moduli spaces of vacua for known three-dimensional SCFTs \cite{Gaiotto:2008sa}. These are the minimal nilpotent orbits $\OO_{\rm min}(\gf)$ of any simple Lie algebra $\gf$. In the coordinate algebra of minimal nilpotent orbits, the $\Cb^*$ grading and the decomposition according to $\RR$ are related in a very simple way,
\begin{equation}
\label{eq:min_orbit_algebra_decomposition}
\AA_{\circledcirc^n{\rm ad}}=\AA_{2n}~,\qquad \dim{\AA_{2n}}=\dim{\RR_{\circledcirc^n{\rm ad}}}~,
\end{equation}
where $\circledcirc$ is the Cartan product of representations, which simply adds Dynkin labels. In other words, for each $n\elem \Zb$, $\AA_{2n}$ consists of a single element transforming in the representation of $G$ whose Dynkin labels are $n$ times those of the adjoint. The minimal nilpotent orbits can be constructed as algebraic varieties by starting with the polynomial algebra of the Lie algebra in question and then modding out by a certain ideal known as the Joseph ideal \cite{Joseph}. 

The quantization of minimal nilpotent orbits is a fairly well-studied subject (see, \eg, \cite{brylinski,Fronsdal:2009cb} and references therein). For $\gf\neq\slf_n$, the minimal nilpotent orbit of a simple Lie algebra $\gf$ admits a unique $\gf$-equivariant deformation quantization, while for $\gf=\slf_n$ there is a one-parameter family of deformation quantizations \cite{Fronsdal:2009cb}. Furthermore, for $\gf=\slf_{n\geqslant3}$ it was found in \cite{brylinski} that only one value of this parameter is compatible with our evenness condition (called ``parity'' in that work). This result is in accordance with the classification theorem mentioned above, since for $\slf_n$ the minimal nilpotent orbit admits a one-dimensional space of deformations, while for the other simple Lie algebras the minimal nilpotent orbit is rigid. It is only for $\slf_2$ that a generic value of the period satisfies the Weyl group condition, however. 

This leads to the immediate conclusion that \emph{the protected associative algebra of an SCFT whose Higgs branch is the minimal nilpotent orbit of $\gf\neq\slf_2$ is completely determined by the bootstrap}. On the other hand, for $\gf=\slf_2$ we need a single parameter to determine the entire algebra. Fortunately, when we have a Lagrangian realization of these theories we can always compute a single number (in our conventions it will be the two-point function of a generator of the associative algebra) using supersymmetric localization, so the algebra will be exactly determined.

Although these formal results are very satisfactory, it would be better if we had closed form expressions for the structure constants of these (nearly) unique algebras. Here we are in luck, because precisely these algebras appear as (generalized) higher spin algebras, and the structure constants for the classical groups have been worked out by physicists in \cite{Joung:2014qya}. For the special case of $\slf_2$, this is the higher spin algebra $hs[\lambda]$,\footnote{Rather, it is the associative algebra whose commutator algebra is $hs[\lambda]$. We will not be careful about this distinction.} which has a longer history. The structure constants of $hs[\lambda]$ were determined in the original work of \cite{Pope:1989sr,Pope:1990kc}. Below we describe this case in detail and consider various $\NN=4$ theories that realize the algebra for different values of $\lambda$. The other classical groups can also be analyzed using the results of \cite{Joung:2014qya}. We do not consider them in any detail here.

\subsection{\texorpdfstring{$\OO_{\rm min}(\slf_2)$}{Omin(sl2)} and \texorpdfstring{$hs[\lambda]$}{hs[lambda]}}

There is only one nilpotent orbit in $\slf_2$, so the (closure of the) minimal orbit is the same thing as the full nilpotent cone, \ie, the set of two-by-two traceless complex matrices with vanishing determinant. As a complex algebraic variety we therefore have
\begin{equation}
\label{eq:sl2_orbit_as_variety}
\OO_{\rm min}(\slf_2)=\{\Cb^3_{[X,Y,Z]}/\langle XY-Z^2\rangle\}~.
\end{equation}
Alternatively, this space may be thought of as the orbifold $\Cb^2/\Zb_2$. The poisson bracket and conjugation operations are determined entirely by their action on the generators $X$, $Y$, and $Z$,
\begin{alignat}{4}
\label{eq:sl2_poisson_and_conjugation}
&\{Z,X\}=X~,&\qquad &\{Z,Y\}=-Y~,&\qquad &\{Y,X\}=2Z~,\\
&\rho(X)=Y~,&\qquad &\rho(Y)=X~,&\qquad &\rho(Z)=-Z~.\nn
\end{alignat}
We see that $X$, $Y$, and $Z$ are the moment maps for a hamiltonian $SU(2)$ action on $\OO_{\rm min}(\slf_2)$, with $Z$ corresponding to the Cartan and $X$ and $Y$ the raising and lowering operators, respectively. The algebra \eqref{eq:sl2_orbit_as_variety} decomposes as described in \eqref{eq:min_orbit_algebra_decomposition}, with one multiplet in each integer spin representation of $\slf(2)$. In particular, we can define
\begin{equation}\label{A_1 basis}
\OO^{j}_{m}\colonequals
\begin{cases}
~~Z^{j-m}X^{m} ~,\qquad~\, m>0~,\\
~~Z^{j+m}Y^{-m}~,\qquad m<0~,
\end{cases}
\end{equation}
and the $\OO^{j}_{m}$ with $|m|\leqslant j$ transform in the spin $j$ representation. 

Any star product for these operators is restricted by $\slf_2$ selection rules to take the form
\begin{equation}
\OO^{i}_{m}\star\OO^{j}_{n}=\sum_{k=0}^{i+j} \zeta^k g^{ij}_{k}(m,n) \OO^{i+j-k}_{m+n}~,
\end{equation}
where the structure constants are constrained by $\slf_2$ covariance to take the form
\begin{equation}
g^{ij}_{k}(m,n)=\frac{K(i,m)K(j,n)}{K(i+j-k,m+n)}C^{i,j,i+j-k}_{m,n,m+n}\,\chi^{ij}_{k}~,
\end{equation}
where $C^{i,j,i+j-k}_{m,n,m+n}$ are Clebsch-Gordan coefficients and the $K(i,m)$ are normalization factors related to our choice of basis in \eqref{A_1 basis},
\begin{equation}
K(i,m)=\sqrt{(i+m)!(i-m)!}~.
\end{equation}
The problem of determining the structure constants $\chi^{ij}_{k}$ that make this star product associative was solved in \cite{Pope:1990kc}. The solution is parameterized by a single numerical coefficient $\mu$, and takes the form
\begin{equation}
\chi^{ij}_{k}(\mu)~=~\left(\frac{-1}{4}\right)^k\phi^{ij}_{k}(\mu)\sqrt{\frac{(2i+2j-k+1)!}{k!(2i-k)!(2j-k)!(2i+2j-2k+1)}}~,
\end{equation}
where
\begin{equation}
\phi^{ij}_{k}(\mu)~\equiv~\pFq{4}{3}{-\tfrac12-2s~~,~~\tfrac32+2s~~,~~-\tfrac{k}{2}~~,~~-\tfrac{k}{2}+\tfrac12~~}{~~\tfrac12-i~~,~~\tfrac12-j~~,~~i+j-k+\tfrac32~~}{~~1}~,
\end{equation}
with $\mu\equiv s(s+1)$\footnote{The structure constants are invariant under $s\leftrightarrow -1-s$, making them well-defined functions of $\mu$.}. In most of the literature the structure constants are expressed in a different form that doesn't make the Clebsch-Gordan coefficients manifest. In particular, one has
\begin{equation}
g^{ij}_{k}(m,n)=\left(\frac{-1}{4}\right)^k\frac{\phi^{ij}_{k}(\mu)}{(k!)}N^{ij}_{k}(m,n;\mu)~,
\end{equation}
where\footnote{The reader should beware that these structure constants are related to those in the higher spin literature by a shift in many of the arguments.}
\begin{equation}
N^{ij}_{k}(m,n)=\sum_{a=0}^k(-1)^a \binom{k}{a}\frac{(i+m)!(i-m)!(j+n)!(j-n)!}{(i+m+a-k)!(j-n+a-k)!(i-m-a)!(j+n-a)!}~.
\end{equation}
The two-dimensional higher spin algebra $hs[\lambda]$ is defined as the commutator algebra of this associative algebra with $\mu = \frac{1}{4}(\lambda^2-1)$. 

It is straightforward to verify that these structure constants obey most of the conditions enumerated in section \ref{sec:algebra_properties}. In particular, the truncation condition -- which would normally be a nonstandard additional condition to impose in deformation quantization -- is here guaranteed by $\slf_2$ covariance, since all of the operators in the coordinate ring are in the same $\slf_2$ and $\suf(2)_H$ representations. The CPT and positivity properties are not completely automatic; they impose minor constraints on the possible values of $\lambda$. The CPT constraint requires $\mu$ to be real, while positivity for this algebra requires that 
\begin{equation}
(-1)^{j-m}\,{\rm C.T.}\left(\OO^{j}_{-m}\star\OO^{j}_{m}\right)>0~,\qquad \forall~j,~m~.
\end{equation}
It is easy to check that this requirement amounts to $\mu\leqslant0$. This can be contrasted with the case of higher spin algebras, where one normally requires $\mu\geqslant-\frac14$ so that there exist unitary representations of the algebra. It is interesting that there is a small overlap between our allowed values of $\mu$ and those of interest for higher spin studies.

The only freedom in determining the structure constants of this algebra is in the choice of $\mu$. As it happens, there are several $\NN=4$ SCFTs with $\Cb^2/\Zb_2$ as their Higgs branch of vacua, and they will realize the $hs[\lambda]$ as their protected associative algebra for different values of $\lambda$. For ease of comparison, here we display a few star products (in our conventions) as a function of $\mu$:
\begin{equation}
\begin{split}
Z\star Z &= Z^2 + \tfrac13\mu\zeta^2~,\\
Z\star X &= ZX + \tfrac12 X \zeta~,\\
Z\star Y &= ZY - \tfrac12 Y \zeta~,\\
X\star Y &= Z^2 - Z \zeta - \tfrac23 \mu \zeta^2~.
\end{split}
\end{equation}
Below we consider three different examples of theories that realize this algebra for two different values of $\mu$.

\subsubsection*{$\Zb_2$ gauge theory}

The simplest theory that has the minimal nilpotent orbit of $\slf_2$ as its Higgs branch is the $\mathbb Z_2$ gauge theory of the free hypermultiplet $(Q,\widetilde Q)$. This is also a subsector of the $U(1)_{2}\times U(1)_{-2}$ ABJM theory studied in \cite{Chester:2014mea}. The Higgs branch chiral ring is the $\Zb_2$-invariant restriction of the Higgs branch chiral ring of the free hypermultiplet. We take it's generators to be given by\footnote{To alleviate the burden of too many indices, we use different naming conventions for the free hypermultiplet scalars than in Section \ref{sec:superalgebra}. The relation is $\wt Q \leftrightarrow q^1_1$, $Q\leftrightarrow q^1_2$.}
\begin{equation}
\label{eq:higgs_branch_generators_z2_gauge}
Z = \tfrac14 Q\widetilde{Q}~, \qquad X = \tfrac14 Q^2~, \qquad Y = \tfrac14 \widetilde{Q}^2~.
\end{equation}
We can directly compute star products using the Moyal-Weyl-Groenewold product \eqref{Moyal-product}. For example, the star products of these generators of the chiral ring are given by
\begin{equation}
\begin{aligned}
Z\star Z &= Z^2 - \tfrac{1}{16}\zeta^2~, \\
Z\star X &= ZX + \tfrac{1}{2}X\zeta~, \\
Z\star Y &= ZY - \tfrac{1}{2}Y\zeta~, \\
X\star Y &= Z^2 - Z\zeta + \tfrac{1}{8}\zeta^2~.
\end{aligned}
\end{equation}
Comparing with the structure constants above, we see that this is an instance of the algebra with $\mu=-\frac{3}{16}$, or $\lambda=\frac12$. Incidentally, the fact that $hs[\frac12]$ can be formulated in terms of a Moyal product is well-known -- see, \eg, \cite{Gaberdiel:2011wb,Kraus:2012uf} for a review.

\subsubsection*{SQED with $N_f=2$}

The next theory is the $\NN=4$ SCFT obtained as the IR fixed point of a $U(1)$ gauge theory with two charged hypermultiplets $(Q_i,\wt Q_i)$ with $i=1,2$ -- see Appendix \ref{app:localization}. In this theory we take the gauge invariant operators that generate the Higgs branch chiral ring to be
\begin{equation}
\label{eq:higgs_branch_generators_A1_gauge}
Z = \tfrac12\left(Q_1\tilde Q_1 + Q_2\tilde Q_2\right)~, \qquad X = Q_1 Q_2~,\qquad Y= \tilde Q_1\tilde Q_2~,
\end{equation}
and there is an $F$-term relation $Q_1\tilde Q_1-Q_2\tilde Q_2 = 0$. For this theory we can compute the two-point function of the $Z$ operator, which is the moment map operator for the Cartan subalgebra of the $SU(2)$ hyperk\"ahler isometry. We perform this computation in Appendix \ref{app:localization}, and we find that with an appropriate normalization of operators, we have
\begin{equation}
\begin{aligned}
Z\star Z &= Z^2 - \tfrac{1}{12}\zeta^2~, \\
Z\star X &= ZX + \tfrac{1}{2}X\zeta~, \\
Z\star Y &= ZY - \tfrac{1}{2}Y\zeta~, \\
X\star Y &= Z^2 - Z\zeta + \tfrac{1}{6}\zeta^2~.
\end{aligned}
\end{equation}
This is the higher spin algebra with $\mu=-\frac14$, \ie, $\lambda=0$. It may be of interest to note that along with $\lambda=\frac12$, this is the other value for $\lambda$ that had special properties in the investigation of \cite{Pope:1989sr}. In particular, for precisely $\lambda=0$ one can define an infinite extension of the higher spin algebra to a $W_{\infty}$ algebra without including any negative spins. This is also the ``free fermion'' value for $\lambda$ \cite{Kraus:2011ds}. We are not certain whether this coincidence has deeper significance.
	
\subsubsection*{$U(2)_{2}\times U(1)_{-2}$ ABJ theory}

Finally, the $U(2)_{2}\times U(1)_{-2}$ ABJ theory, which is believed to describe the IR fixed point of $\NN=8$ supersymmetric Yang-Mills theory with $SO(3)$ gauge group, has the same ``Higgs branch chiral ring''.\footnote{Here we use scare quotes because the enhanced supersymmetry of the theory makes the distinction of Higgs versus Coulomb branch chiral ring nonstandard. Our terminology is appropriate if we treat the theory as a special case of an $\NN=4$ theory.} The $\suf(2)$ flavor symmetry of the Higgs branch is actually an $R$-symmetry of the extended supersymmetry of the theory, and the moment map two-point function is related by supersymmetry to the central charge $c_{T}$. This was computed using supersymmetric localization in \cite{Chester:2014fya}. The conventions for normalizing operators used in that work were different than for us -- we have set the coefficient of the leading term in the star product to be one, while they normalized their operators to have unit two-point functions. This means that we have
\begin{equation}
\lambda_{2i,2j,2i+2j-2k}^{\rm CLPY}=g^{ij}_{k}(i,-j)\sqrt{\frac{g^{kk}_{2k}(k,-k)}{g^{ii}_{2i}(i,-i) g^{jj}_{2j}(j,-j)}}~.
\end{equation}
With this change of conventions, we find that in this theory the protected associative algebra is exactly the same as that for the quiver gauge theory described above, \ie, 
\begin{equation}
\mu=-\frac{1}{4}~,\qquad \lambda=0~.
\end{equation}
It is then straightforward to reproduce the values shown in Table 4 of \cite{Chester:2014mea}.

\vspace{10pt}

It may be interesting to note that while the unitarity bounds for this algebra only require $\mu < 0$, we have found in all three examples that we have $-\frac14\leqslant\mu <0$. The relevance is that in the context of higher spin gravity, one restricts attention to $\mu\geqslant-\frac14$ due to the absence of unitary representations of the higher spin algebra for smaller values \cite{Gaberdiel:2011wb}. This may hint at an important role for representations of the protected associative algebra in the analysis of our bootstrap problem.

\section{Examples: affine quiver theories}
\label{sec:kleinian}

Our second class of examples are two-complex-dimensional hyperk\"ahler cones -- the Kleinian singularities,
\begin{equation}
\MM_{\Gamma}=\Cb^2/\Gamma~,\qquad \Gamma\in\left\{A_{n\geqslant1},D_{n\geqslant4},E_{6,7,8}\right\}~.
\end{equation}
The special case of $\Gamma=A_1\equiv\Zb_2$ is the same as the minimal nilpotent orbit of $\slf(2)$, but for other choices of discrete group these are not nilpotent orbits. These singularities are realized as the Higgs branches of $\NN=4$ quiver gauge theories whose quivers are in the shape of the affine quiver diagrams of type $\Gamma$, with the ranks of each unitary gauge group equal to the Dynkin index of the corresponding node. For example, the $A_n$ cases are realized by considering a circular quiver of Abelian gauge groups of $n+1$ nodes. 

These hyperk\"ahler cones can be realized as algebraic varieties in $\Cb^3$, \ie, $\Cb^2/\Gamma = \{(X,Y,Z)\in \Cb^3\  |\  \phi_\Gamma(X,Y,Z)=0  \}$ where $\phi_\Gamma$ is a polynomial in the variables $X,Y,Z$ determined by $\Gamma.$ Concretely,
\begin{equation}
\begin{aligned}
&\Gamma = A_n:\qquad  &&\phi_{A_n}=XY - Z^{n+1}~, \\
&\Gamma = D_n:\qquad  &&\phi_{D_n}=X^{n-1}  + X Y^2 + Z^2~,\\
&\Gamma = E_6:\qquad  &&\phi_{E_6}=X^2 + Y^3 + Z^4~,\\
&\Gamma = E_7:\qquad  &&\phi_{E_7}=X^2 + Y^3 + YZ^3~,\\
&\Gamma = E_8:\qquad  &&\phi_{E_8}=X^2 + Y^3 + Z^5~.
\end{aligned}
\end{equation}
Such a description arises naturally when $\MM_\Gamma$ is realized as a hyperk\"ahler quotient. All of these spaces can also be realized in simple discrete gauge theories of a single hypermultiplet. The poisson bracket for each of these examples can be written succinctly in terms of the defining polynomial,
\begin{equation}\label{poissonbracket}
\{X,Y \} = \frac{\partial \phi_\Gamma}{\partial Z}~,\qquad 
\{Y,Z \} = \frac{\partial \phi_\Gamma}{\partial X}~,\qquad
\{Z,X \} = \frac{\partial \phi_\Gamma}{\partial Y}~.
\end{equation}

Compared to the $D_n$ and $E_{6,7,8}$ cases, deformation quantization for the $A_n$ singularities is substantially simplified thanks to the presence of a $U(1)$ flavor symmetry, which imposes additional selection rules and thus restricts the number of operators that can appear in any given star product. We will therefore focus on these examples. It would be interesting to attack the more challenging $D_n$ and $E_n$ theories in the future.

\begin{table}[ht]
\renewcommand{\arraystretch}{1.2}
\begin{center}
\begin{tabular}{c|c|c}
 & $\Delta$ & $\phantom{+}q_F$ \\
 \hline
 $Z$ & $2$ & $\phantom{+}0$ \\
 $X$ & $n+1$ & $+1$ \\
 $Y$ & $n+1$ & $-1$ \\
\end{tabular}
\caption{$\Cb^*$ grading $\Delta$ and $U(1)_F$ charges $q_F$ of the generators of the $A_n$ singularity.\label{Ansingularityfields}}
\end{center}
\end{table}

We will henceforth be focusing on the $A$ series of Kleinian singularities. Our conventions for the $\Cb^\star$ grading and $U(1)$ flavor charges of the generators of the coordinate algebras are summarized in Table \ref{Ansingularityfields}. Note that $Z$ plays the role of a moment map for the flavor symmetry, and it will correspond to the $U(1)$ moment map operator for any SCFT realizing one of these singularities as its moduli space.

We can define a complete basis for $\AA=\Cb[\MM_{A_n}]$ as follows,
\begin{equation}\label{A_n basis}
\OO^{i,j}\colonequals
\begin{cases}
~~X^{j} Z^{i-j}~,\qquad~\, j\geqslant0~,\\
~~Y^{-j} Z^{i+j}~,\qquad j<0~,
\end{cases}
\end{equation}
where $j\in\Zb$ and $i\geqslant|j|$. Here $\Delta_{ij}\colonequals\Delta(\OO^{i,j}) = 2i +(n-1)|j| $ and $q_F(\OO^{i,j}) = j$.  The goal is to define a star product on the vector space spanned by these operators that obeys all of the conditions outlined in our formulation of the bootstrap problem in Section \ref{sec:algebra_properties}.

\subsection{A note on strategy}
\label{subsec:bootstrap_strategy}

Before considering individual examples, let us briefly comment on the strategies that we can employ in trying to solve the bootstrap problem for these algebras.

The first -- and perhaps most obvious -- strategy is to simply write down the most general possible star product that quantizes the (known) chiral ring and obeys the appropriate truncation and evenness properties. In these examples it should also respect $U(1)_F$ charge conservation. This product will then take the form
\begin{equation}
\label{starproductAn}
\OO^{i,j}\star\OO^{k,l}=\sum_{m=0}^{\text{min}(\Delta_{ij},\Delta_{kl})} \zeta^m \ c^{(i,j),(k,l)}_m\  \OO^{\frac{\Delta_{ij}+\Delta_{kl}}{2}-m-\frac{n-1}{2}|j+l|,j+l}~,
\end{equation}
where the truncation condition is implemented in the range of the sum on the right hand side, and evenness requires that
\begin{equation}
c^{(i,j),(k,l)}_m = (-1)^m c^{(k,l),(i,j)}_m~.
\end{equation}
Operators $\OO^{i,j}$ appearing on the right hand side of \eqref{starproductAn} not satisfying $i\geqslant|j|$ are set to zero. Without loss of generality, the composite operators in \eqref{A_n basis} can be normalized such that the structure constants $c^{(i,j),(k,l)}_0$ have value one:
\begin{equation}\label{normalizationcomposites}
c^{(i,j),(k,l)}_0 = \phantom{+}1~.
\end{equation}
Additionally, we can impose normalization conditions such that the poisson bracket appears in the leading term with the desired normalization,
\begin{equation}\label{normalizationPoisson}
c^{(1,0),(1,1)}_1 = \frac{1}{2} ~, \qquad  c^{(1,0),(1,-1)}_1 = -\frac{1}{2} ~, \qquad  c^{(1,1),(1,-1)}_1 = -\frac{n+1}{2} ~.
\end{equation}
This completely fixes the normalizations of all of our operators.

The star product will not be associative for general values of the structure constants. In true bootstrap fashion, we can then impose associativity conditions for every triple of operators and see how the structure constants are constrained. In practice, this involves ordering all triples of operators according to their total dimension and imposing the associativity conditions in order. It is difficult to build the positivity conditions into this process, so they must be checked after the fact. It is not at all obvious, but nevertheless turns out to be true in our examples, that the solutions to the resulting associativity problem are completely determined by a finite number of free coefficients. We will return to the examples after considering an alternative perspective for solving this problem.

A second strategy is available in examples where we know how to solve the \emph{deformation quantization problem} but not the \emph{bootstrap problem}. Indeed, for the Kleinian singularities this is the case, and the relevant noncommutative algebras are known as spherical subalgebras of symplectic reflection algebras \cite{EGGO,Gor06}. These algebras can be understood very concretely as finitely generated noncommutative algebras with certain relations imposed. In this case the associativity conditions are already solved, and the remaining problem is to find an appropriate \emph{basis} for the noncommutative algebra where the solution to deformation quantization is upgraded to a solution of the bootstrap problem.

For all of the $A$-series Kleinian examples, the relevant noncommutative algebra is generated by noncommutative variables $\hat X$, $\hat Y$, and $\hat Z$ that obey commutation relations of the form
\begin{equation}\label{noncommC2/An}
[\hat X,\hat Y] = -\zeta P(\hat Z)~, \qquad [\hat Z,\hat X ] = \zeta \hat X~, \qquad [\hat Z,\hat Y ] = -\zeta\hat Y~,
\end{equation}
where $P(\hat Z)$ is a polynomial in $\hat Z$ of degree $n$ (with appropriate powers of $\zeta$ inserted) with leading term given by $P(\hat Z) = (n+1) \hat Z^n + \ldots$. This algebra has a center generated by \cite{Smith:1990}
\begin{equation}
\hat \Omega = \frac{1}{2}(\hat X \hat Y + \hat Y \hat X) - \frac{1}{2}(Q(\hat Z + \zeta) + Q(\hat Z))~,
\end{equation}
where $Q(\hat Z)$ is a polynomial in $\hat Z$ of degree $n+1$ such that
\begin{equation}\label{defQ}
Q(\hat Z + \zeta) - Q(\hat Z) = \zeta P(\hat Z)~.
\end{equation}
The condition \eqref{defQ} uniquely determines the coefficients of $Q$ in terms of the coefficients of $P$ except for the lowest order term, \ie, the coefficient of $\zeta^{n+1}$. The non-commutative algebra generated by $\hat X, \hat Y, \hat Z$, with commutation relations \eqref{noncommC2/An}, quotiented by the two-sided ideal generated by $\hat \Omega$ is the associative algebra that solves the $\Cb^\star$-equivariant deformation quantization problem for the Kleinian singularities, and the coefficients in $Q$ (alternatively, the coefficients in $P$) parameterize the \emph{period} of these quantizations.

The evenness condition in the bootstrap problem implies that if we take the opposite algebra and change the sign of $\zeta$ then the algebra should be unchanged. This already imposes constraints on the parameters appearing in $P$ and $Q$. In particular, the polynomials $P(\hat Z)$ and $Q(\hat Z+\zeta) +Q(\hat Z)$, being invariant under passing to the opposite algebra, must only contain terms with even powers of $\zeta$. This is the incarnation for these examples of the requirement from Section \ref{sec:math_phys} that the period live on the same Namikawa Weyl orbit with its negative, and it reduces the number of parameters determining the algebra to $\lfloor\frac{n+1}{2}\rfloor$.

Now our second step is to start with such an associative algebra and determine a basis for the vector space of operators that solves the bootstrap problem. It is helpful to first note that any monomial in $\hat X, \hat Y, \hat Z$ can always be expressed as a sum of terms of the form
\begin{equation}\label{A_n basis noncomm}
\hat\OO^{i,j}\colonequals
\begin{cases}
~~\hat X^{j} \hat Z^{i-j}~,\qquad~\, j\geqslant0~,\\
~~\hat Y^{-j} \hat Z^{i+j}~,\qquad j\leqslant0~.
\end{cases}
\end{equation}
This will be our canonical ordering.\footnote{There is nothing special about this choice of ordering. It is just one choice of many that could be made.} We can then write down the most general basis of operators of the form
\begin{equation}\label{mapintononcommAn}
\OO^{i,j} \rightarrow \hat\OO^{i,j} + \sum_{k=|j|}^{i-1} \zeta^{i-k} \ \alpha^{i,j}_k \ \hat\OO^{k,j}~,
\end{equation}
with \emph{a priori} arbitrary coefficients $\alpha^{i,j}_k$. An inverse map can be obtained by mapping the highest dimensional non-commutative operator back one by one. We then demand that in this basis, the evenness and truncation conditions hold.

It is straightforward to show that these conditions will hold for all products if they do so for the following simple products:
\begin{equation}
\OO^{i,j} \star \OO^{1,0}~,~\OO^{1,0} \star \OO^{i,j} \qquad \text{and } \qquad \OO^{i,j} \star \OO^{1,\pm 1}~,~\OO^{1,\pm1} \star \OO^{i,j}~.
\end{equation}
The second strategy is therefore to constrain the coefficients appearing in \eqref{mapintononcommAn} using the conditions just introduced. We found this strategy to be somewhat more computationally efficient. We will see that in addition to the parameters determining the period of the quantization, only a finite number of free coefficients associated to the choice of basis survive.

\subsection{Revisiting the Kleinian singularity \texorpdfstring{$\MM_{A_1} = \Cb^2/A_1$}{M(A1)=C2/A1}}

In the previous section, the solution to the deformation quantization of $\OO_{\rm min}(\slf_2)$ was presented and seen to solve the bootstrap problem for $\mu<0$. In some sense, the approach described there was that of our first strategy above. It may be illuminating to reconsider this case from the point of view of the second strategy outlined above. For this case evenness requires the polynomial $P(\hat Z)$ to be given by $P(\hat Z) = 2 \hat Z$ after suitably normalizing the operators, so we have the commutation relations
\begin{equation}\label{noncommC2/A1}
[\hat X,\hat Y] = -2\zeta \hat Z~, \qquad [\hat Z,\hat X ] = \zeta \hat X~, \qquad [\hat Z,\hat Y ] = -\zeta\hat Y~,
\end{equation}
which one readily recognizes as the standard commutation relations of an $\slf_2$ algebra. The polynomial $Q(\hat Z)$ takes the form $Q(\hat Z) = \hat Z \hat Z - \zeta \hat Z +\kappa \zeta^2$ and thus the relation $\hat\Omega=0$ reads
\begin{equation}
\hat \Omega =  0 \Longleftrightarrow  \frac{1}{2}(\hat X \hat Y + \hat Y \hat X) - \hat Z\hat Z = \kappa \zeta^2~.
\end{equation}
The upshot is clear: we are setting the quadratic Casimir operator $\hat C_2 = \frac{1}{2}(\hat X \hat Y + \hat Y \hat X) - \hat Z\hat Z$ of $\slf(2)$ equal to $\kappa \zeta^2$. The full non-commutative algebra can then be understood as the central quotient of the universal enveloping algebra of $\slf_2$: $\UU(\slf_2)/\langle\hat C_2 -  \kappa \zeta^2\rangle$. We have rediscovered the fact that this quantum algebra will be the higher spin algebra $hs[\lambda]$, where our $\kappa$ is the same as $\mu$. As we explained in Section \ref{sec:nilpotent}, the $SU(2)$ symmetry of this algebra renders the choice of basis unique, so all structure constants follow from the value of $\mu$.

\subsection{The Kleinian singularity \texorpdfstring{$\MM_{A_2} = \Cb^2/A_2$}{M(A2)=C2/A2}}

This is our first example of an algebra for which the gauge invariance associated to changes of basis becomes a real problem. We have pursued both of the above strategies to analyze this example with similar successes in both cases. Here we will focus on the second strategy.

As in the $A_1$ case, there is a single free coefficient $\kappa$ that determines the polynomials $P$ and $Q$ -- though in this case there are three potential parameters and the evenness condition eliminates two of them. The noncommutative algebra has commutation relations
\begin{equation}\label{noncommC2/A3}
[\hat X,\hat Y] = -3\zeta \hat Z^2 + \kappa\zeta^3 ~, \qquad [\hat Z,\hat X ] = \zeta \hat X~, \qquad [\hat Z,\hat Y ] = -\zeta\hat Y~,
\end{equation}
and we quotient by the two-sided ideal generated by
\begin{equation}
\hat\Omega=\frac12\left(\hat X\hat Y+\hat Y\hat X\right)-\hat Z^3+\frac{2\kappa-1}{2} \zeta^2 \hat Z~.
\end{equation}
We find a one-parameter family of bases that satisfy the evenness and truncation constraints, with the parameter appearing in the definition of the $Z^2$ operator,
\begin{equation}
Z^2 \equiv \hat Z^2 + \alpha \zeta^2~.
\end{equation}
With normalizations \eqref{normalizationcomposites} and \eqref{normalizationPoisson}, we find the following expressions for the star products of the Higgs branch generators,
\begin{equation}\label{starproductC2/Z3}
\begin{aligned}
Z \star Z &= Z^2 - \alpha \zeta^2~,\\
Z \star X &= ZX +\frac{1}{2} \zeta X~,\\
Z \star Y &= ZY -\frac{1}{2} \zeta Y~,\\
X \star Y &= Z^3 - \frac{3}{2} \zeta Z^2 - \frac{3\alpha + \kappa}{4\alpha}\zeta^2  Z + \frac{3\alpha+\kappa}{2}\zeta^3~.
\end{aligned}
\end{equation}
Note that $\alpha$ appears directly as the constant term in the $Z\star Z$. Since $Z$ is the moment map operator in these theories, we will be able to compute $\alpha$ using supersymmetric localization when we have a path integral realization of the theory.

\begin{figure}
    \centering
    \begin{subfigure}[b]{0.45\textwidth}
    	\includegraphics[scale=.53]{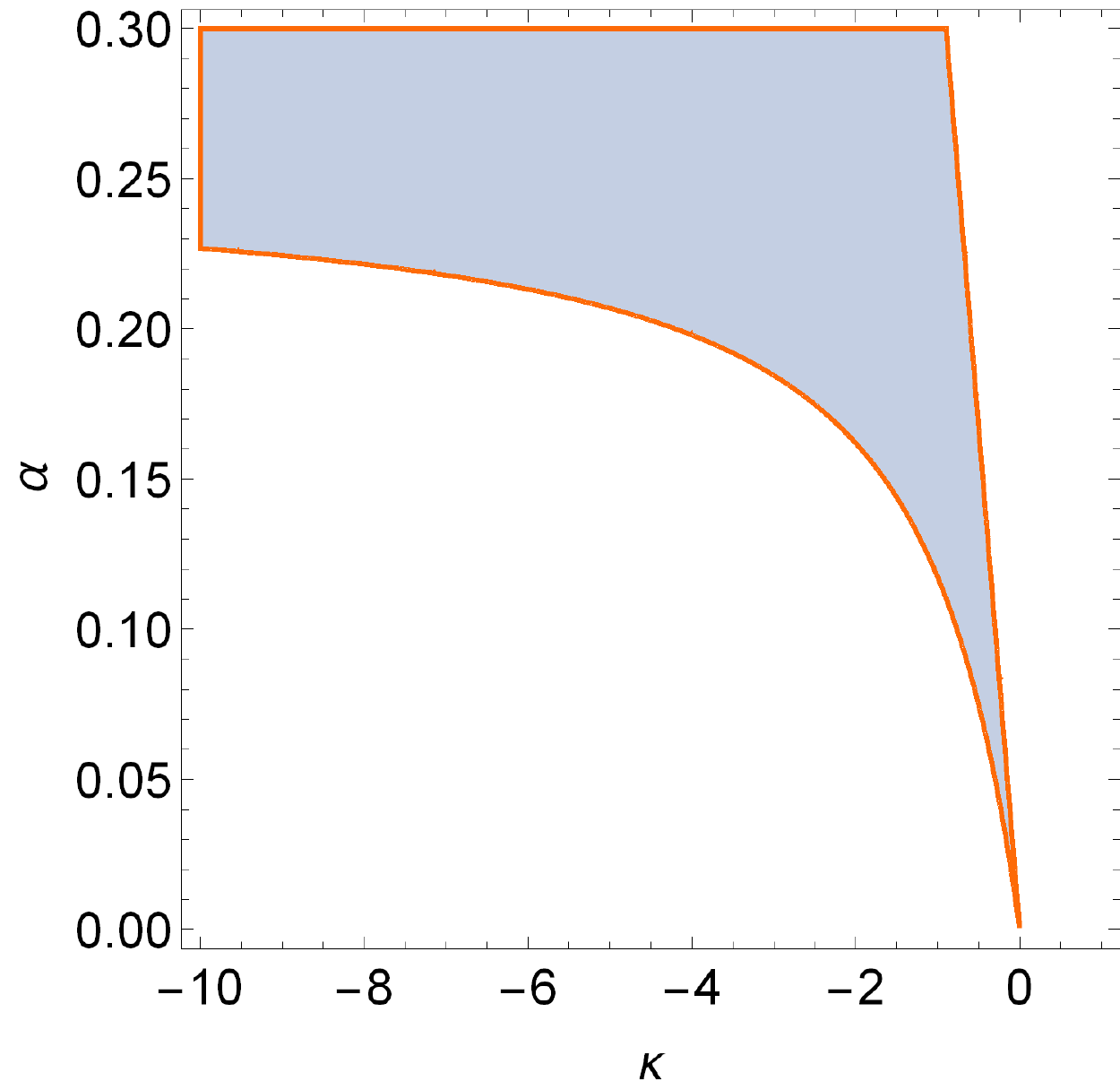}
		\label{fig:a2_sequence_1}
    \end{subfigure}
    ~
    \begin{subfigure}[b]{0.45\textwidth}
    	\includegraphics[scale=.53]{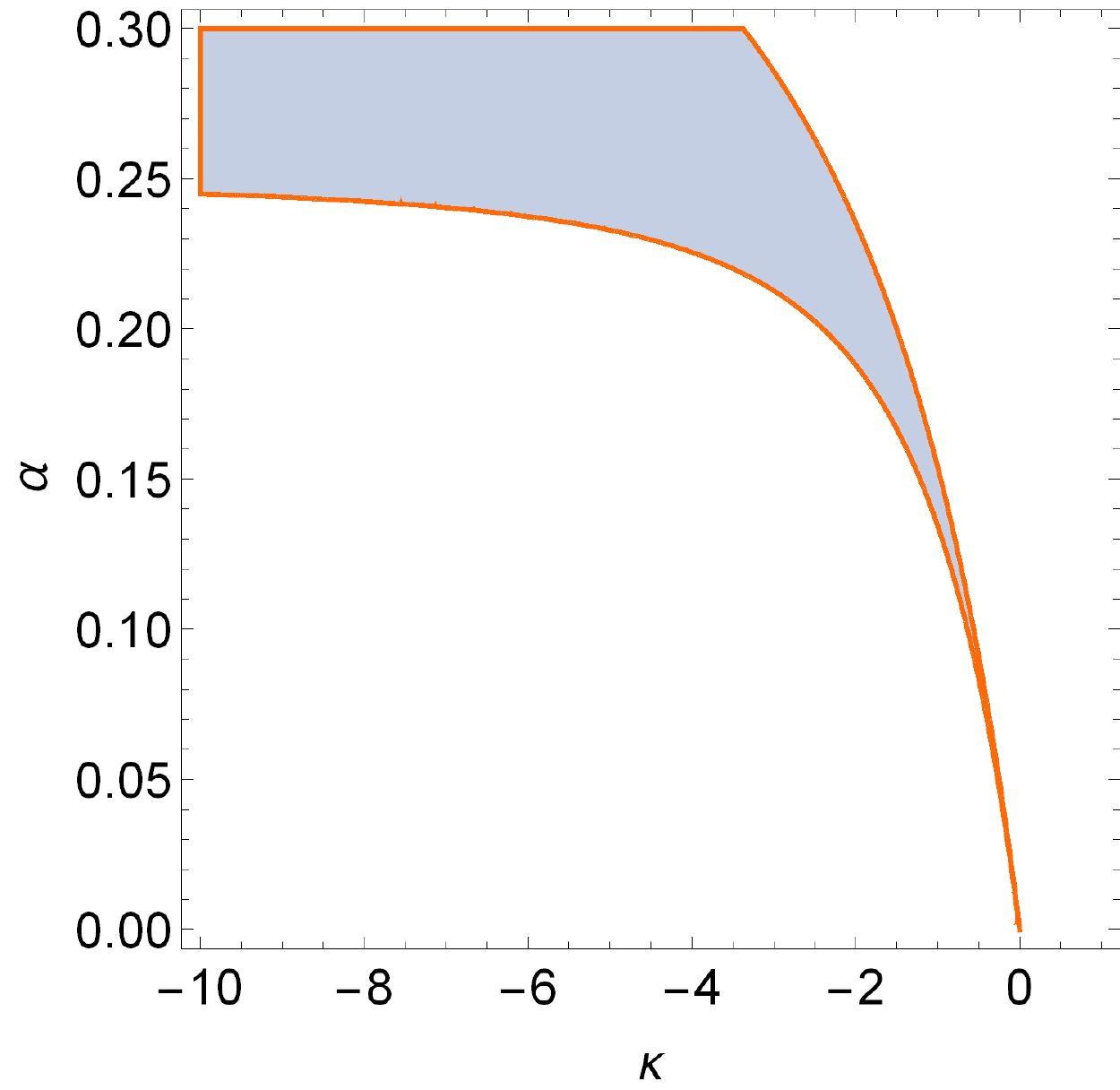}
		\label{fig:a2_sequence_2}
    \end{subfigure}

\vspace{8pt}
    \centering
    \begin{subfigure}[b]{0.45\textwidth}
    	\includegraphics[scale=.53]{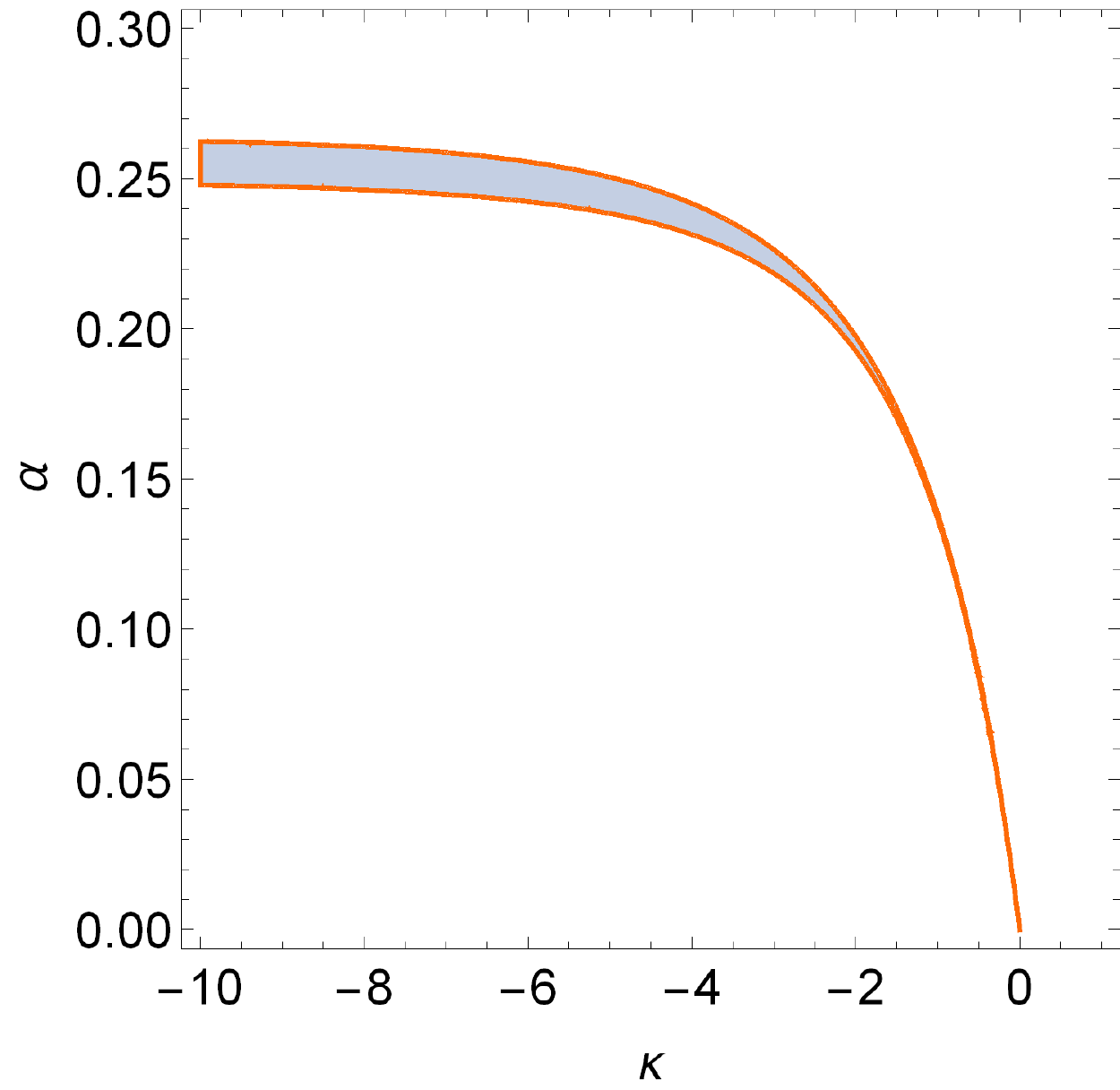}
		\label{fig:a2_sequence_3}
    \end{subfigure}
    ~
    \begin{subfigure}[b]{0.45\textwidth}
    	\includegraphics[scale=.53]{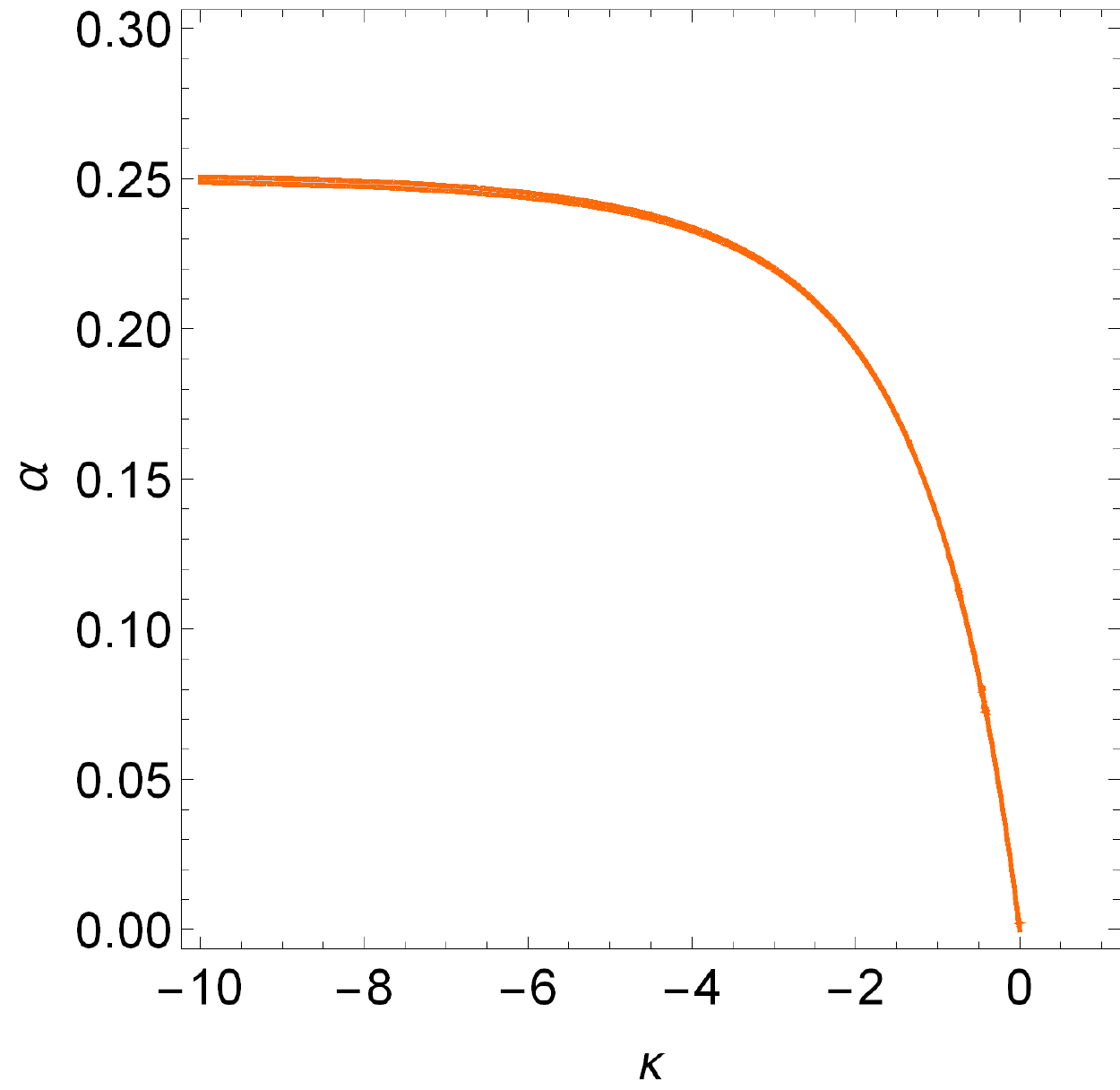}
		\label{fig:a2_sequence_4}
    \end{subfigure}
    \caption{Regions in the $(\kappa, \alpha)$ plane that are not excluded by the constraints of unitarity for operators of dimension less than or equal to two (top left), three (top right), four (bottom left), and five (bottom right).\protect\footnotemark}\label{fig:a2_sequence}
\end{figure}

\begin{figure}[t!]
\centering
\includegraphics[scale=.45]{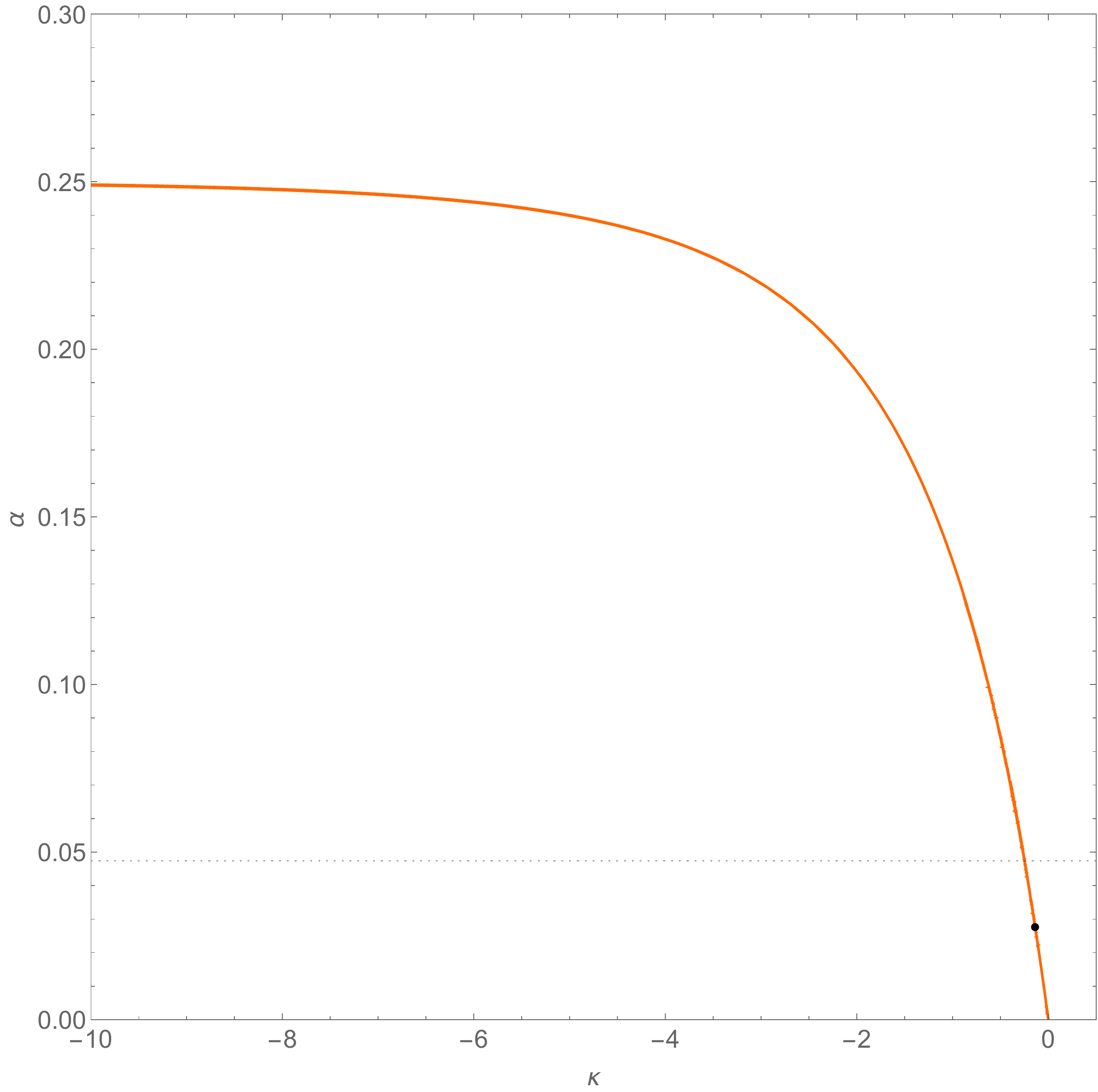}
\caption{Allowed values of $(\kappa,\alpha)$ after imposing unitarity bounds for the operators listed in the text. The horizontal dotted black line is at $\alpha = (18-\frac{144}{\pi^2})/72$, which is the value of $\alpha$ in the $A_2$ affine quiver gauge theory. The black dot corresponds to the $\Zb_3$ gauging of a free hypermultiplet.}
\label{fig:unitarityboundsZ3}
\end{figure}

By imposing the truncation and evenness constraints we have reduced the infinite-dimensional space of gauge-equivalent algebras for a given period to a one-dimensional space of algebras for a given period. Our remaining constraint is that of unitarity, which requires that the constant terms in certain star products have definite sign. In particular, for this algebra we have $\rho(X)=Y$, $\rho(Y)=-X$, and $\rho(Z)=-Z$, so our constraints are all given by
\begin{equation}
{\rm C.T.}\left((-1)^{i+j} X^i Z^j \star Y^i Z^j\right) ~>~0~.
\end{equation}
We computed the constant terms appearing in the following star products:
\begin{equation}
\begin{aligned}
&Z^i \star Z^i \qquad  &&\text{for} \quad i=1,2,\ldots,6~, \\
&XZ^i \star YZ^i \qquad  &&\text{for} \quad i=0,1,2,3,4~,\\
&X^2Z^i \star Y^2Z^i \qquad  &&\text{for} \quad i=0,1,2,3~,\\
&X^3Z^i \star Y^3Z^i \qquad  &&\text{for} \quad i=0,1~, \\
&X^4 \star Y^4~.&&
\end{aligned}
\end{equation}
By requiring \footnotetext{To facilitate the rendering of these plots, we have superposed the curve of figure \ref{fig:unitarityboundsZ3} for small $\alpha$.} that each constant term have the correct sign we find that many values of $(\kappa,\alpha)$ are excluded by unitarity. The exclusion plots of this type are displayed in Figs. \ref{fig:a2_sequence} and \ref{fig:unitarityboundsZ3}. The most salient feature of these plots is that the allowed region in the $(\alpha,\kappa)$ plane appears to collapse to a one dimensional region that includes points for all values of $\kappa\leqslant0$. In this sense, the addition of unitarity appears to give us a perfect gauge fixing condition for these noncommutative algebras (for an appropriate range of $\kappa$).

\begin{figure}[t!]
\centering
\raisebox{-0.5\height}{{
\includegraphicsif{scale=.30}{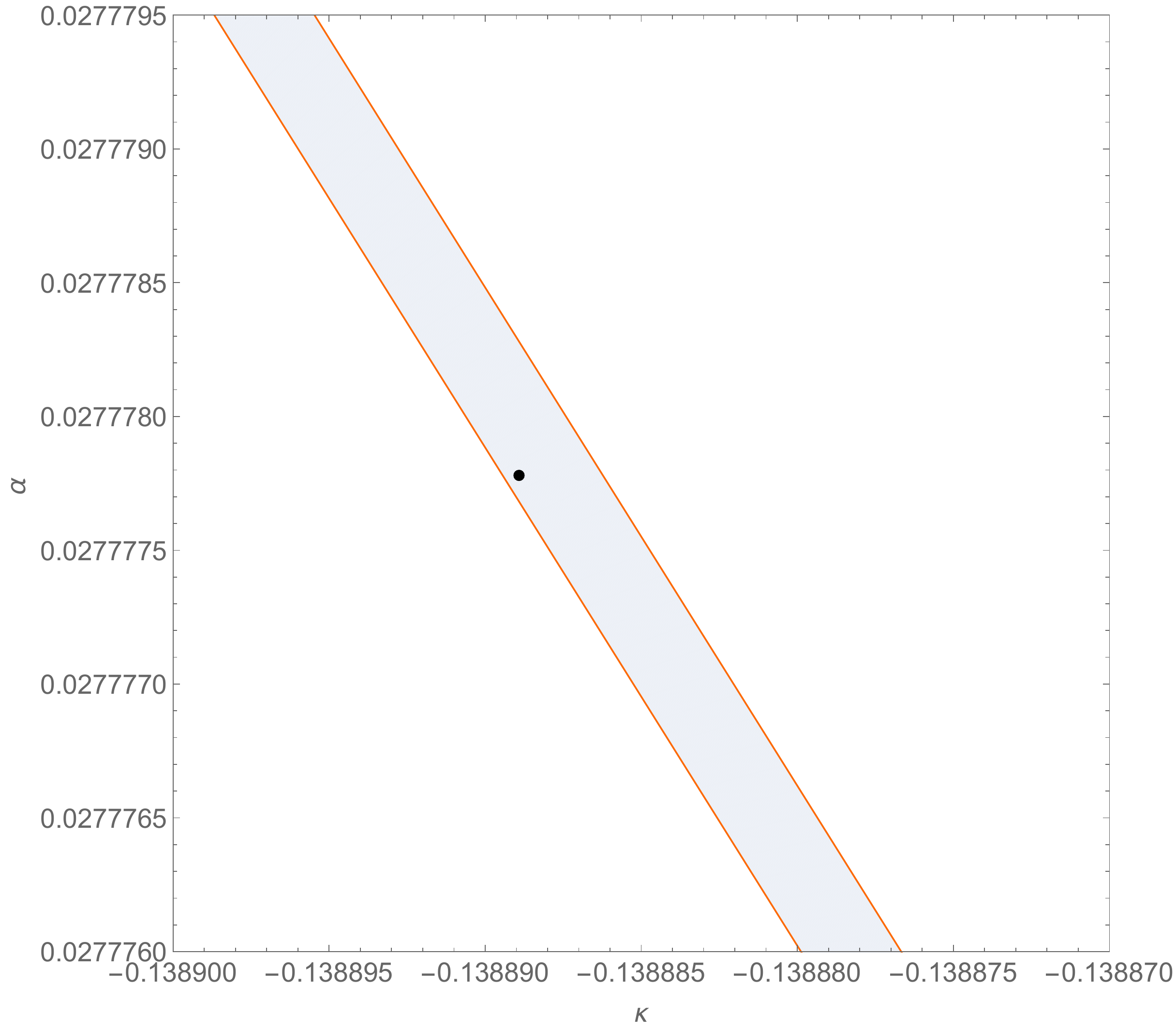}}}
\quad
\raisebox{-0.5\height}{{
\includegraphicsif{scale=.30}{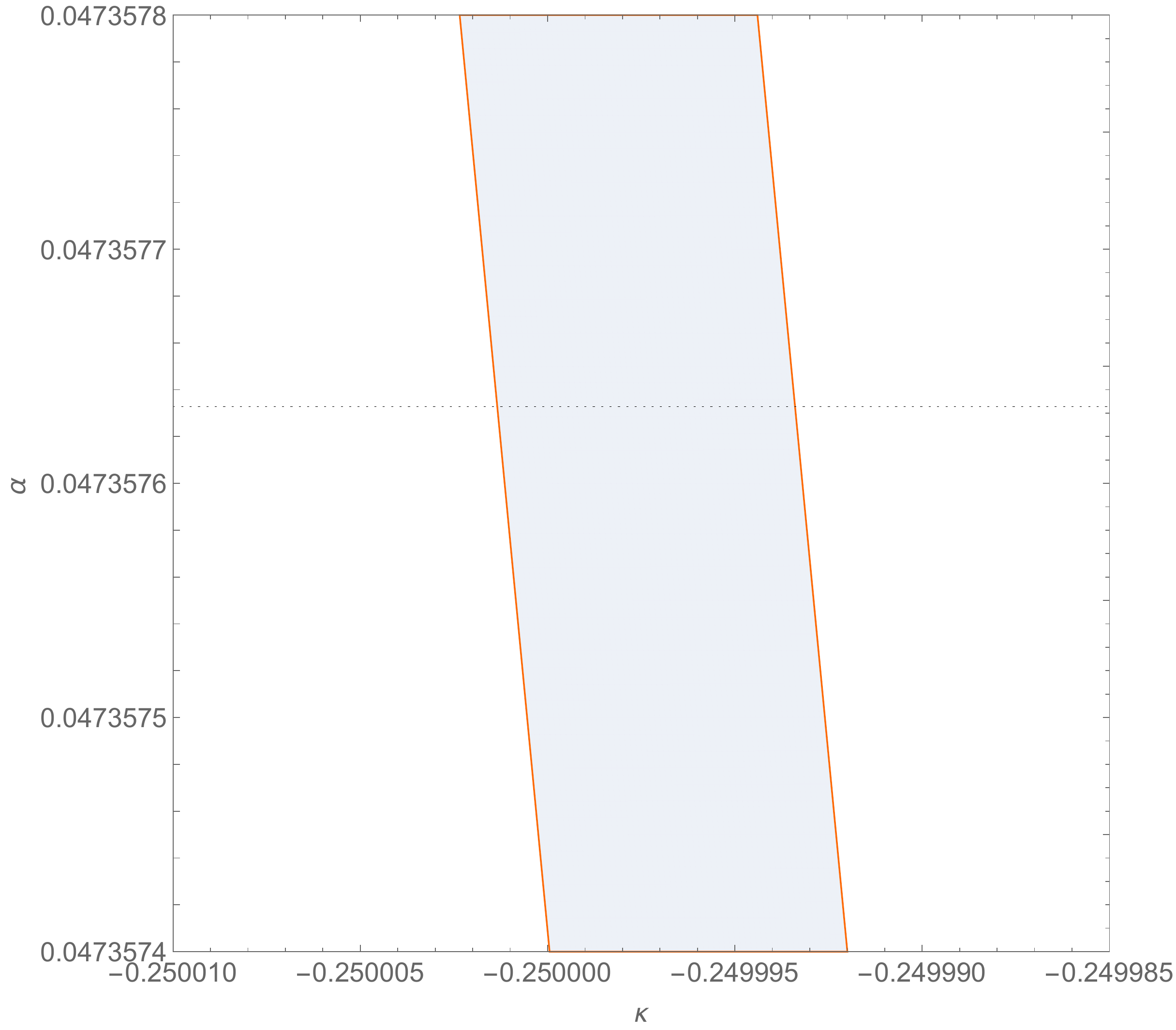}}}
\caption{Close up views of the unitarity bounds for coefficients $\kappa$ and $\alpha$ near the points of interest for the free (left) and affine quiver (right) theories. The allowed values lie in the shaded region. The black dot in the left plot indicates the location of the $\Zb_3$ gauge theory of the free hypermultiplet. The dotted black line denotes value of $\alpha$ for the affine $A_2$ quiver gauge theory.}
\label{fig:unitarityboundsZ3_zoom}
\end{figure}

Let us now consider two specific theories that have $\Cb^2/{\Zb_3}$ as their Higgs branch. The first one is the $\Zb_3$ gauge theory built from a free hypermultiplet. Its Higgs branch chiral ring is obtained by restricting oneself to the $\Zb_3$ invariant subsector of the Higgs branch chiral ring operators of the free hypermultiplet described in subsection \ref{subsec:free_hyper}.  We define
\begin{equation}
Z = \frac{1}{6} Q\wt Q~, \qquad X = \frac{1}{6^{\frac{3}{2}}}Q^3~, \qquad Y = \frac{1}{6^{\frac{3}{2}}}\wt Q^3~,
\end{equation}
which indeed satisfy $XY=Z^3$. Star products can be computed using the Moyal-Weyl-Groenewold star product \eqref{Moyal-product}. One straightforwardly obtains the following example star products:
\begin{equation}
\begin{aligned}
Z\star Z &= Z^2 - \frac{1}{36}\zeta^2 \\
Z\star X &= ZX +\frac{1}{2} \zeta X \\
Z\star Y &= ZY -\frac{1}{2} \zeta Y \\
X\star Y &= Z^3 - \frac{3}{2} \zeta Z^2 + \frac{1}{2} \zeta^2 Z - \frac{1}{36} \zeta^3~.
\end{aligned}
\end{equation}
We find in terms of our parameterization that this algebra has $\alpha=\frac{1}{36}$ and $\kappa =-\frac{5}{36}.$ These values are indicated in Figure \ref{fig:unitarityboundsZ3} by the black dot. Zooming in around these values, one finds the left plot in Figure \ref{fig:unitarityboundsZ3_zoom}.

The other theory we would like to consider is the IR fixed point of the $\NN=4$ quiver gauge theory whose quiver is the $A_2$ affine Dynkin diagram with $U(1)$ gauge groups at each node. In appendix \ref{app:localization} we compute the two-point coefficient $\tau$ for canonically normalized flavor symmetry currents in this theory using supersymmetric localization \cite{Closset:2012vg}. In terms of the normalizations used in the appendix, we have
\begin{equation}
Z\star Z = -\frac{\tau}{2} \zeta^2 + Z^2~,
\end{equation}
while  since $X$ and $Y$ have charge $\pm 3$ in those conventions, we also have
\begin{equation}
Z\star X = 3\zeta X + ZX~, \qquad Z\star Y = -3\zeta Y + ZY~.
\end{equation}
Rescaling $Z\rightarrow \frac{Z}{6}$ brings us back to the conventions of this section and we then have
\begin{equation}
Z\star Z =  -\frac{\tau}{72} \zeta^2 + Z^2~,
\end{equation}
from which we can read off $\alpha = -\frac{\tau}{72}$. In appendix \ref{app:localization}, $\tau$ was computed to be $\tau = 18 - \frac{144}{\pi^2}$ and thus
\begin{equation}
\alpha=\frac{\frac{144}{\pi^2}-18}{72}~.
\end{equation}
Given this value, the coefficient $\kappa$ is strongly constrained by unitarity, as can be seen in Figure \ref{fig:unitarityboundsZ3} and in particular in the right plot of Figure \ref{fig:unitarityboundsZ3_zoom}. It must take its value in the range
\begin{equation}
-.2500015~<~\kappa~<~-.2499965~.
\end{equation}
It is hard to ignore the presence of the rational value $-\frac14$ within this range. Assuming that this is the true value, this seems to be a strong indication that the value of $\kappa$ should be exactly computable. This is an interesting direction for future study.


\begin{figure}[t!]
    \centering
    \begin{subfigure}[b]{0.45\textwidth}
    	\includegraphics[scale=.53]{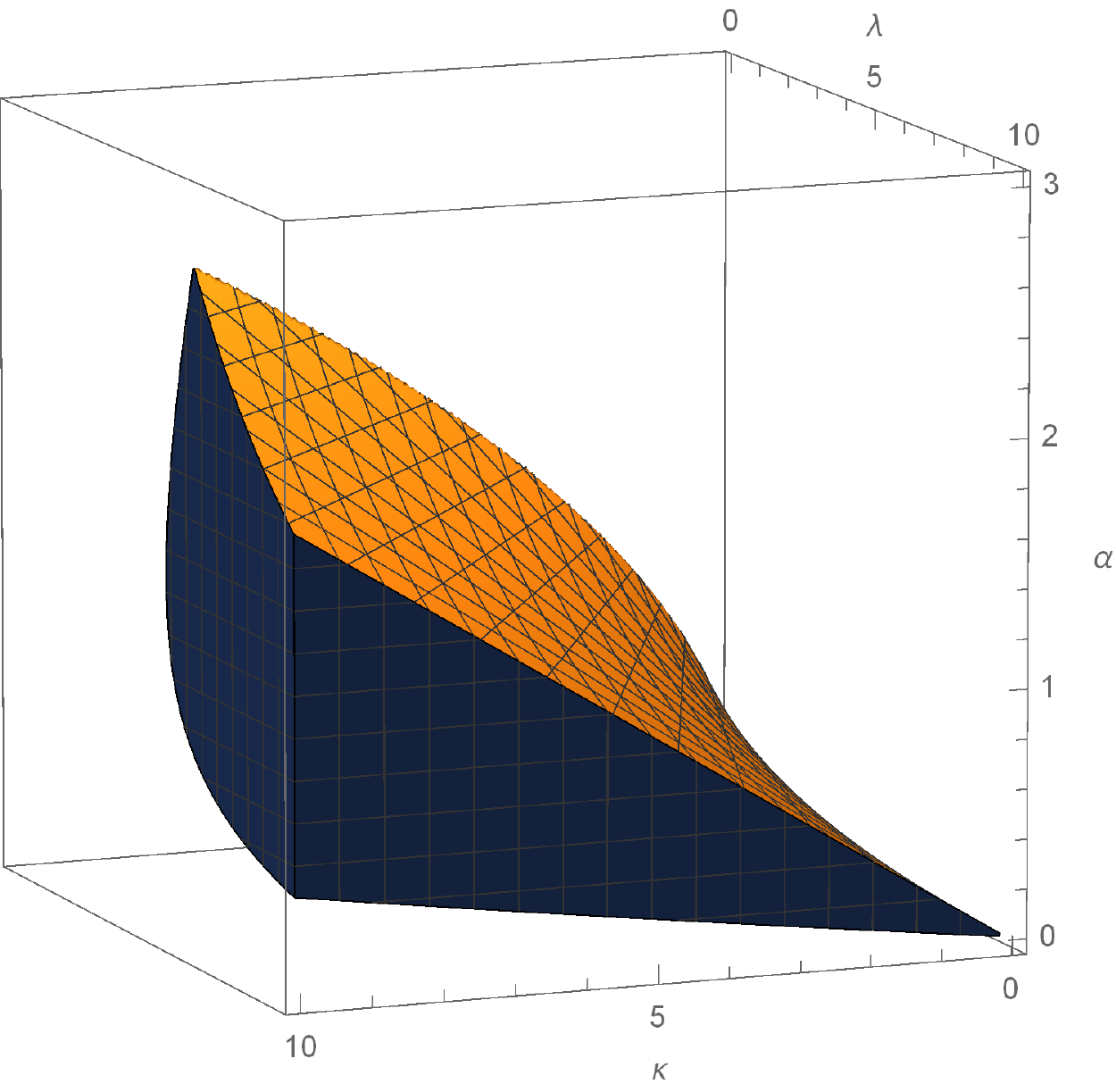}
		\label{fig:a3zoom_1}
    \end{subfigure}
    ~
    \begin{subfigure}[b]{0.45\textwidth}
    	\includegraphics[scale=.53]{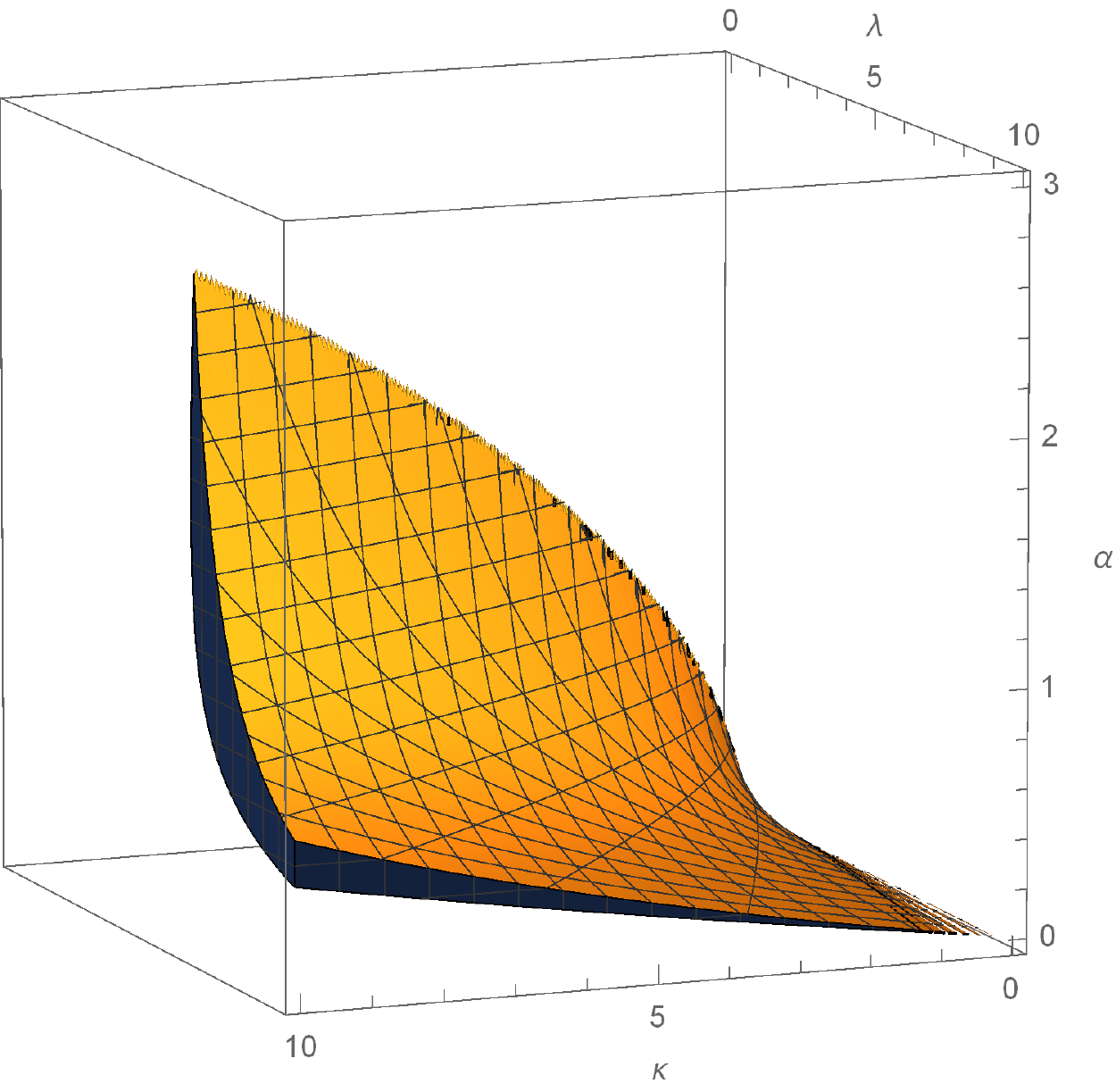}
		\label{fig:a3zoom_2}
    \end{subfigure}

\vspace{8pt}
    \centering
    \begin{subfigure}[b]{0.45\textwidth}
    	\includegraphics[scale=.53]{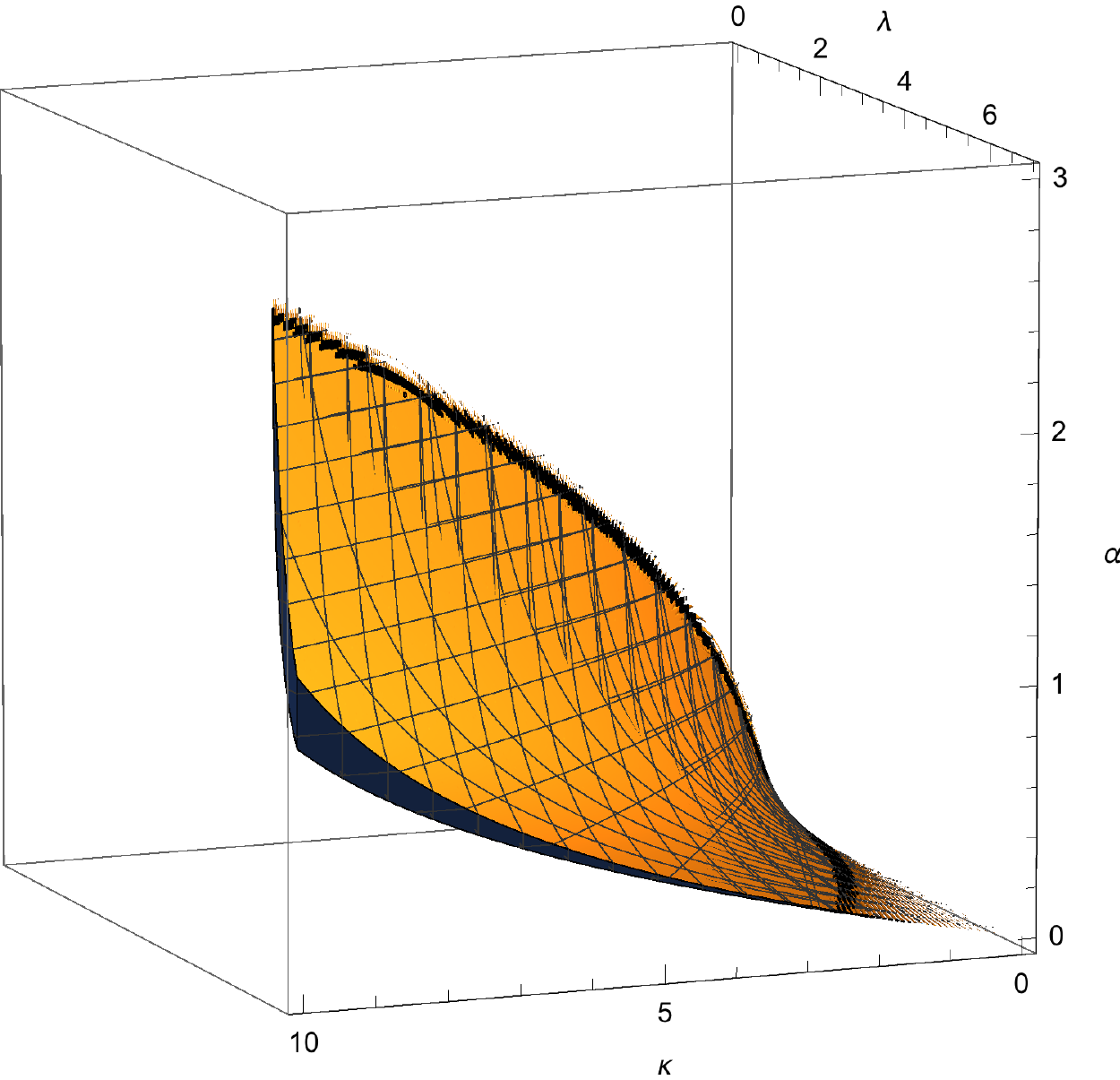}
		\label{fig:a3zoom_3}
    \end{subfigure}
    ~
    \begin{subfigure}[b]{0.45\textwidth}
    	\includegraphics[scale=.53]{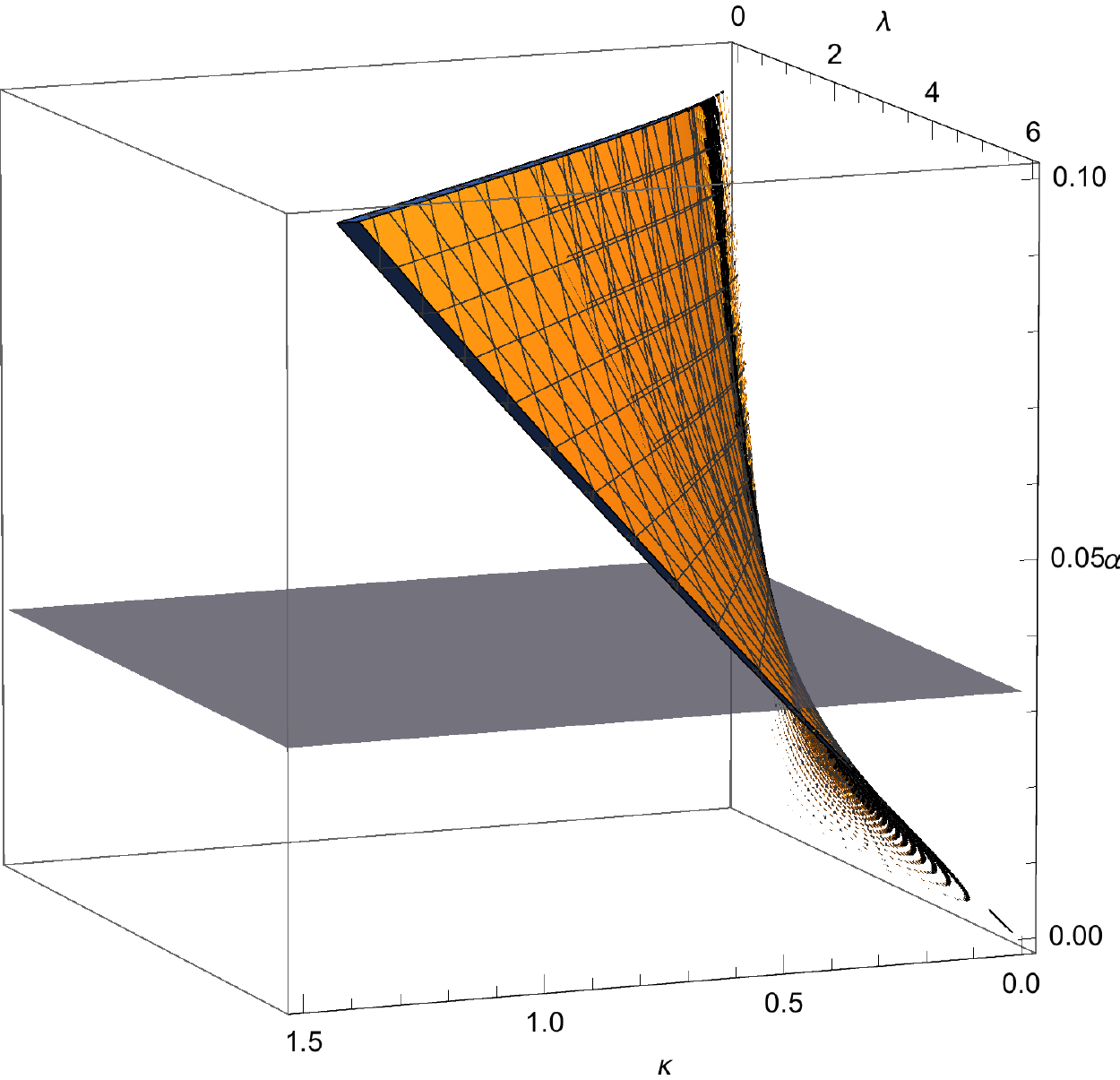}
		\label{fig:a3zoom_4}
    \end{subfigure}
    \caption{Constraints imposed by unitarity on the parameters of the $A_3$ star product. The allowed regions satisfy unitarity bounds for operators of dimension up to two (upper left), three (upper right), and four (lower left). In the lower right we show at higher resolution the intersection of the allowed region with the plane representing the value of $\alpha$ relevant to the affine $A_3$ quiver gauge theory.}\label{fig:a3zoom}
\end{figure}

\subsection{The Kleinian singularity \texorpdfstring{$\MM_{A_3} = \Cb^2/A_3$}{M(A3)=C2/A3}}

The analysis for $\Cb^2/A_3$ is analogous to the $A_2$ case. Here we find that the allowed even, truncated star products depend on three free coefficients -- two associated with the space of even periods of the corresponding quantum algebra, and one additional free coefficient associated with the choice of basis. 

The quantum algebra is defined by
\begin{equation}\label{noncommC2/Z4}
[\hat Z,\hat X ] = \zeta \hat X~, \qquad [\hat Z,\hat Y ] = -\zeta\hat Y~,\qquad [\hat X,\hat Y] = -4\zeta \hat Z^3 -2(\lambda-1)Z\zeta^3 ~,
\end{equation}
where we quotient by the two-sided ideal generated by
\begin{equation}
\hat\Omega=\frac12\left(\hat X\hat Y+\hat Y\hat X\right)-\hat Z^4 - \lambda \hat Z^2 \zeta^2 - \kappa \zeta^4~.
\end{equation}
The additional coefficient associated to the choice of basis again appears in the definition of the $Z^2$ operator,
\begin{equation}
Z^2 \equiv \hat Z^2 + \alpha \zeta^2~.
\end{equation}
In the normalizations of \eqref{normalizationcomposites} and \eqref{normalizationPoisson}, we then find for the $\star$-products of generators
\begin{equation}\label{starproductC2/Z4}
\begin{aligned}
Z \star Z &= Z^2 - \alpha \zeta^2 \\
Z \star X &= ZX +\frac{1}{2} \zeta X \\
Z \star Y &= ZY -\frac{1}{2} \zeta Y \\
X \star Y &= Z^4 -2 \zeta Z^3 - f(\alpha, \kappa,\lambda) \zeta^2 Z^2 +  \frac{-2\kappa + \alpha + \lambda\alpha}{5\alpha}\zeta^3 Z - 2\frac{-2\kappa + \alpha + \lambda\alpha}{5} \zeta^4 ~,
\end{aligned}
\end{equation}
where
\begin{equation}
f(\alpha, \kappa,\lambda) = \frac{-2\kappa(\lambda-4\alpha) + (1+\lambda)(-5+6\lambda-14\alpha)\alpha}{7(\kappa+\alpha(2-3\lambda+5\alpha))}~,
\end{equation}
All the other $\star$-products we computed can be expressed in terms of the same three coefficients. 

We have derived unitarity bounds on these parameters associated to the two-point functions of operators of dimension less than or equal to five. In Fig. \ref{fig:a3zoom} we display the allowed region in parameter space after imposing unitarity bounds for various numbers of dimensions. The main observation to make is that as the number of unitarity bounds imposed is increased, the allowed region appears to collapse to a codimension one subspace of the naive parameter space, with at most one point allowed for each value of $\kappa$ and $\lambda$. We take this as experimental support of our gauge fixing conjecture.

As in the previous example, we can realize a point in this space with the $\Zb_4$ gauge theory of a free hypermultiplet. The Higgs branch chiral ring is given by the $\Zb_4$ singlet sector of the free hypermultiplet Higgs branch chiral ring. Its generators can be taken to be
\begin{equation}
Z = \frac{1}{8}Q \wt Q~, \qquad X = \frac{1}{64}Q^4~, \qquad Y = \frac{1}{64}\wt Q^4~,
\end{equation}
and satisfy $XY=Z^4$. Star products can be computed in this theory using the Moyal-Weyl-Groenewold star product \eqref{Moyal-product}, and read
\begin{equation}
\begin{aligned}
Z\star Z &= Z^2 - \frac{1}{64}\zeta^2 \\
Z\star X &= ZX +\frac{1}{2} \zeta X \\
Z\star Y &= ZY -\frac{1}{2} \zeta Y \\
X\star Y &= Z^4 - 2 \zeta Z^3 + \frac{9}{8} \zeta^2 Z^2 - \frac{3}{16} \zeta^3 Z + \frac{3}{512} \zeta^4~.
\end{aligned}
\end{equation}
From this we can identify the parameters of \eqref{starproductC2/Z4} in this theory as $\kappa = \frac{105}{4096},\lambda=\frac{43}{32}$ and $\alpha = \frac{1}{64}$.

\begin{figure}
    \centering
    \begin{subfigure}[b]{0.45\textwidth}
    	\includegraphics[scale=.53]{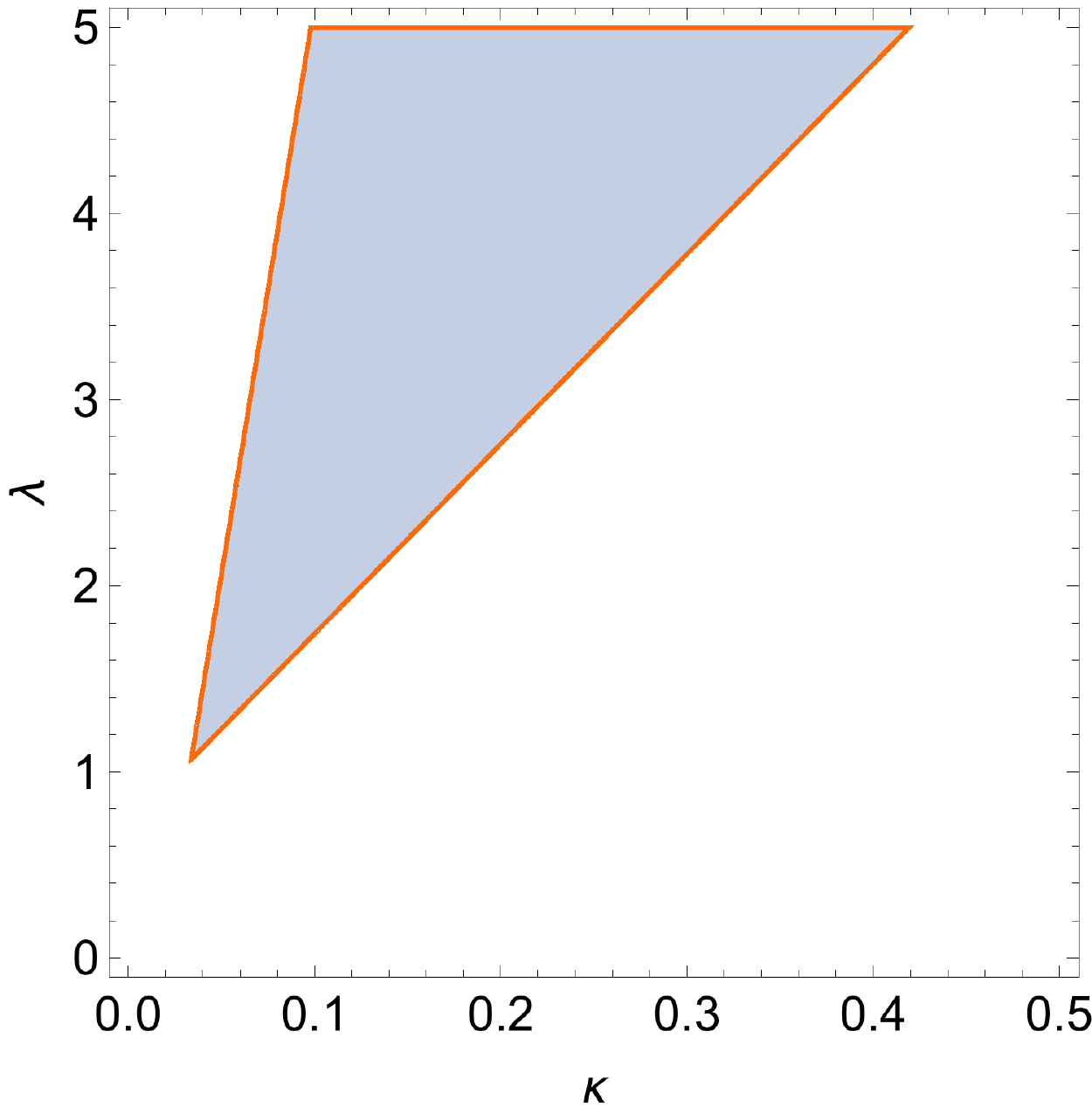}
		\label{fig:a4slice_1}
    \end{subfigure}
    ~
    \begin{subfigure}[b]{0.45\textwidth}
    	\includegraphics[scale=.53]{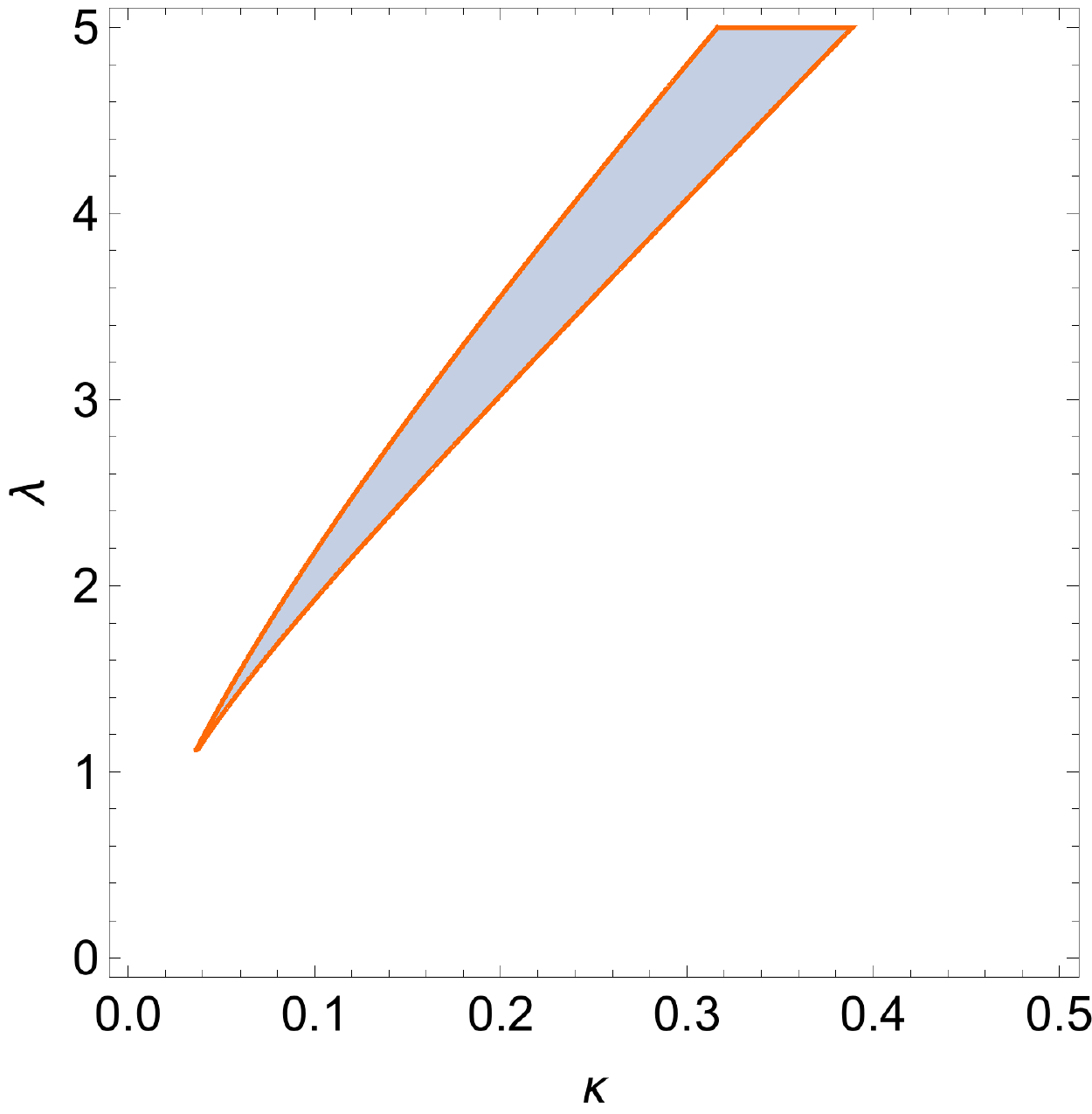}
		\label{fig:a4slice_2}
    \end{subfigure}

	\vspace{8pt}
    \centering
    \begin{subfigure}[b]{0.45\textwidth}
    	\includegraphics[scale=.53]{./figures/Z4slice_dim3.pdf}
		\label{fig:a4slice_3}
    \end{subfigure}
    ~
    \begin{subfigure}[b]{0.45\textwidth}
    	\includegraphics[scale=.53]{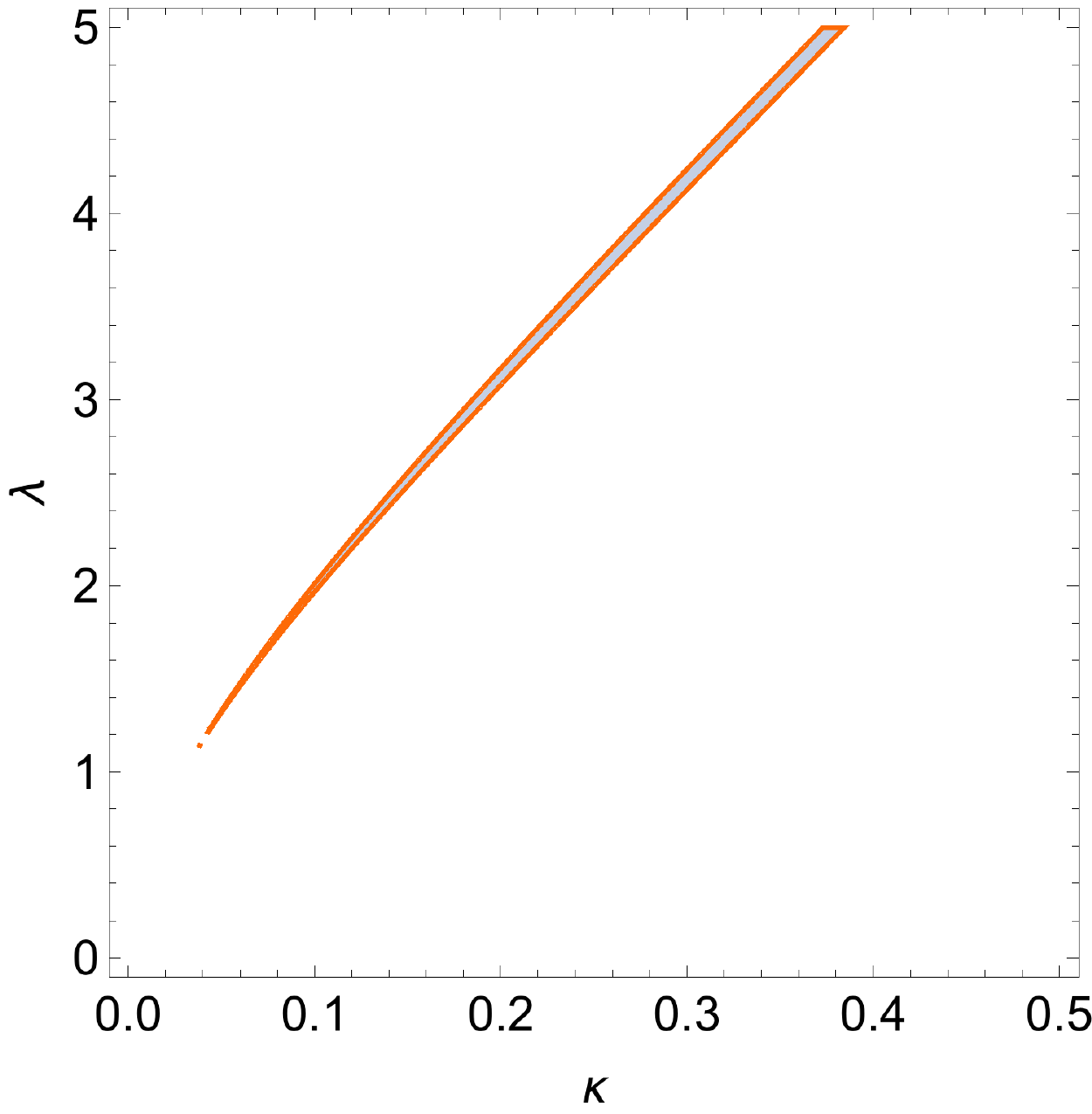}
		\label{fig:a4slice_4}
    \end{subfigure}
    \caption{Allowed values of the period parameters for the $A_3$ theory when $\alpha$ is set equal to its value for the $A_3$ affine quiver gauge theory. The bounds correspond to requiring the correct sign for two-point functions of operators up to dimension two (top left), three (top right), four (bottom left), and five (bottom right).}\label{fig:a4_slices}
\end{figure}

Another point should be realized by the quiver gauge theory with the $A_3$ affine Dynkin diagram as its quiver with $U(1)$ nodes. In the $A_3$ gauge theory, $\alpha$ is given equal to $\frac{\tau}{128}$, where $\tau$ was computed in Appendix \ref{app:localization} using supersymmetric localization. The constraints of unitarity for algebras with $\alpha$ equal to this value are shown in Fig. \ref{fig:a4_slices}. We see that the two-dimensional parameter space collapses to a single curve as more unitarity bounds are included. We do not know how to compute where on this curve the actual algebra should sit.

\subsection*{Comment on $\Cb^2/A_4$}

Finally, let us mention the case of $\Cb^2/A_4$ without going into detail. Here, like in the $A_3$ case, the space of even periods is two-dimensional, and we might hope that again there would be a one dimensional space of even, truncating bases which could further be reduced by unitarity to a unique basis for each allowed quantum algebra.

Alas, it is not so simple, and our preliminary investigations indicated that for these algebras there is a two-dimensional space of even, truncating bases for each choice of even period. However, for an assortment of values of the even period we then found that unitarity conditions restrict the basis parameters to a bounded domain. This gives us hope that ultimately unitarity will prove sufficient to collapse the allowed basis for each allowed period to a single point, in agreement with Conjecture \ref{conj:main}. However, it is clear that there are plenty of sharp tests of this conjecture within reach. We hope to return to some of them in the near future.

\section{Future directions}
\label{sec:prospects}

There are a number of interesting directions for extending the present work. Let us mention several of them here.

\subsection*{Localization} It is natural to ask whether the algebras studied here are accessible by supersymmetric localization in theories that have Lagrangian descriptions. It is not immediately obvious that such an approach should work, because the supercharges used to define the protected associative algebra involve superconformal generators, which are absent in the ultraviolet description of any three-dimensional gauge theory. What's more, the transcendentality of the structure constants described in our Kleinian singularity examples make it implausible that the correlators encoded in our algebra can be computed in any simple way in the UV theory. A possible exception to these objections is for theories, such as the ABJM theories, that are described with actions that are classically superconformally invariant. It would be very interesting to see whether the cohomology described in this paper can be investigated directly in those theories using the classical Lagrangian.

\subsection*{Omega deformation}
A more promising direction for accessing the structure introduced in this paper with path integral techniques involves the omega deformation. Indeed, in \cite{Yagi:2014toa,Bullimore:2015lsa} it has been observed that turning on an omega deformation in three-dimensional $\NN=4$ gauge theories leads to a quantization of the chiral ring. There is a puzzle involved in understanding this coincidence more deeply. The omega deformation is a true \emph{deformation} of the theory, so the exact relationship between observables in the deformed theory and the original theory are obscured. Nevertheless, the similarity between the two constructions may hint at a deeper connection between the omega background and observables at the superconformal fixed point \cite{Beem_wip}.

\subsection*{Local quantization on the moduli space}
Our approach in the examples to solving for the protected associative algebra (in the physical basis) was rather inelegant. This is much different from the approach used in the mathematics literature to prove theorems about these quantizations. There one works locally on $\MM$ (or on a larger space fibered over the space of resolutions of $\MM$) and builds a sheaf of noncommutative algebras that are easy to define locally. The algebra of interest to us is then roughly the algebra built out of global sections of this sheaf. It would be interesting to know whether the details of our problem -- in particular the properties required to define the physical basis -- can be understood in such a framework.

\subsection*{Inclusion of conformal defects}
The cohomological construction that gives rise to the protected associative algebra can be extended to configurations where conformal defect operators are present in such configurations that the relevant supercharge is still a symmetry. This allows for the inclusion of BPS line operators lying on $\Rb_{\rm top}$ and/or BPS boundary conditions or domain walls filling the plane transverse to $\Rb_{\rm top}$ and intersecting it at the origin. On general grounds, the algebra supported on the line defects will still be a deformation quantization of the chiral ring, but the evenness constraint will no longer hold. Boundaries will need to furnish bi-modules and modules, respectively, for the quantized algebra. Most of these general comments also hold in the case where the quantization in question arises due to an omega deformation \cite{BDG_to_appear}. It would be interesting to see if in the superconformal construction there are additional strong constraints analogous to those defining the physical basis in the present work.

\subsection*{Extensions of the algebra} In a similar vein, the associative algebra we have described can be realized in a configuration in which the three-dimensional $\NN=4$ theory is supported on the worldvolume of a conformal defect in a higher-dimensional SCFT. The only case where we are certain this can be arranged is in the case of a four-dimensional $\NN=4$ SCFT with a conformal boundary or domain wall. Again one can see without much work what sort of algebraic structure must arise from such a configuration. In particular, the half-BPS chiral ring operators of the bulk $\NN=4$ theory will be able to interact with the quantized algebra on the boundary, and they will necessarily form a subalgebra that is central to the full noncommutative algebra.

\vspace{10pt}
More generally, we have found in this paper that the properties of superconformal field theories suggest a new viewpoint from which to consider an already well-studied subject in mathematics -- namely the quantization of symplectic singularities. At least some of the conditions we have studied seem rather natural from the point of view of the hyperk\"ahler structure of these singularities. Indeed, the construction of twisted translated operators in the three-dimensional SCFT looks like it could be understood as a nontrivial fibration of spacetime over the twistor sphere of the Higgs branch. It seems likely that developing the theory surrounding the physical bases for such quantum algebras could lead to a great deal of interesting mathematics as well as a additional exact results for interesting superconformal field theories.

\acknowledgments

The authors would like to thank Dionysios Anninos, Mathew Bullimore, Clay C\'ordova, Kevin Costello, Tudor Dimofte, Pavel Etingof, Maxime Gabella, Davide Gaiotto, Amihay Hanany, Eric Perlmutter, David Poland, Nathan Seiberg, David Simmons-Duffin, Andrew Tolland, Ben Webster, and Edward Witten for helpful conversations and useful suggestions. We are grateful to the Aspen Center for Physics (partially supported by NSF grant \#1066293) for hospitality during the summer workshop ``From scattering amplitudes to the conformal bootstrap''.
C.B. gratefully acknowledges support from the Frank and Peggy Taplin Fellowship at the IAS.
C.B. is also supported in part by the NSF through grant PHY-1314311. 
W.P. is supported in part by the DOE grant DOE-SC0010008. 
The work of L.R. is supported in part by NSF Grant PHY-1316617.

\appendix


\section{Hyperk\"ahler geometry and chiral rings}
\label{app:HK_cones}

This purpose of this appendix is to provide a collection of relevant facts about hyperk\"ahler cones and their connections to three-dimensional superconformal field theory. Most of the discussion here appears in \cite{Gaiotto:2008nz}, with additional mathematically precise exposition available in \cite{Swann:1991}. The present appendix is included for the reader's convenience and to emphasize the specific structures that play a role in the present work. In addition we collect the relevant data for the $A$-type Kleinian singularities to set up the bootstrap problem described in the text.

\subsection*{Hyperk\"ahler manifolds and cones}
\label{subapp:hyperkahler}

A hyperk\"ahler manifold $\MM$ is a manifold possessing three inequivalent complex structures -- denoted $\JJ^{1,2,3}$ -- obeying $\JJ^1\JJ^2=\JJ^3$ such that $\MM$ is k\"ahler with respect to all three complex structures. We denote the corresponding k\"ahler forms $\omega_{1,2,3}$. In a given complex structure, say $\JJ^3$, one has k\"ahler form $\omega_3$ and the other k\"ahler forms combine into a holomorphic symplectic form $\Omega^{(2,0)}\colonequals\omega_1+i\omega_2$. For this reason, from the point of view of complex geometry hyperk\"ahler manifolds are often treated essentially as holomorphic symplectic manifolds.

The moduli spaces of three-dimensional $\NN=4$ SCFTs are singular hyperk\"ahler spaces. As additional structure they admit a dilatation (homothety) a non-holomorphic $SU(2)_R$ isometry rotates the complex structures as a triple.\footnote{The $SU(2)_H$ and $SU(2)_C$ symmetries of an SCFT act as isometries on the Higgs and Coulomb branches of vacua, respectively. Here we are just dealing with a single hyperk\"ahler cone, and we will call whichever is the relevant symmetry $SU(2)_R$.} The dilatation symmetry makes $\MM$ into a cone, with the $SU(2)_R$ isometry acting on the base.

In general a hyperk\"ahler manifold can look much different as a holomorphic symplectic manifold depending on the choice of which complex structure one considers. However for a hyperk\"ahler cone, because the complex structures are all rotated into one another under the action of $SU(2)_R$, things look the same regardless of the choice of complex structure. Thus it will not be essential to specify in \emph{which} complex structure we are considering our cones. Any choice will be equally good.

\subsection*{Coordinate algebras}
\label{subapp:HK_rings}

Our interest in hyperk\"ahler cones is mainly related to their algebraic structure. In particular, all of the examples considered in the present paper -- and to the best of our knowledge all hyperk\"ahler cones that occur as the moduli spaces of three-dimensional $\NN=4$ SCFTs -- are affine algebraic varieties. We can then focus on the coordinate algebra $\AA\equiv\Cb[\MM]$ of the hyperk\"ahler cone, which will be a finitely generated, reduced, commutative $\Cb$-algebra. On a generic hyperk\"ahler manifold, the coordinate algebra might depend on a choice of complex structure, but as we just mentioned all complex structures are equivalent on a hyperk\"ahler cone, and one may meaningfully discuss \emph{the} coordinate algebra.

Two totally standard features of the coordinate algebra of a hyperk\"ahler cone are the following:
\begin{enumerate}
\item[$\bullet$]{{\bf $\Zb_{\geqslant0}$ Grading} -- A choice of complex structure picks out a Cartan subalgebra $U(1)_R\subset SU(2)_R$ that leaves the complex structure fixed. The dilatation isometry on $\MM$ is then complexified by that cartan to give a holomorphic $\Cb^\star$ action on the coordinate algebra of $\MM$. The elements of the coordinate algebra are half-integrally graded under this action -- but we will use conventions in this paper such that the grading is defined with an extra factor of two so that the algebra is graded by non-negative integers,
\begin{equation}
\AA=\bigoplus_{p\elem\Zb_{\geqslant0}}\AA_p~,
\end{equation}
where $\AA_0$ is just the constant functions. Additionally, $\AA_2$ comprises moment maps for hyperk\"ahler isometries of $\MM$.
}
\item[$\bullet$]{{\bf Poisson bracket} -- The holomorphic symplectic two-form has charge two (in our conventions) with respect to the $\Cb^\star$ action. This turns the coordinate algebra into a Poisson algebra with Poisson bracket of degree negative two, which in local coordinates takes the form
\begin{equation}
\{f,g\}=(\Omega^{-1})^{ij}\partial_if\partial_jg~.
\end{equation}
Note that the holomorphic two-form doesn't have a preferred overall normalization, so neither does this poisson bracket. In the text we generally choose a convenient normalization for the poisson bracket as a way of fixing the normalizations of chiral ring operators in $\NN=4$ SCFTs.
}
\end{enumerate}
The above conditions only involve knowledge of $\MM$ as a holomorphic symplectic manifold. There is an additional structure that can be described at the level of the coordinate algebra that encodes more of the hyperk\"ahler structure of $\MM$.
\begin{enumerate}
\item[$\bullet$]{{\bf Conjugation} -- The coordinate algebra admits a privileged order-four $\Cb$-antilinear endomorphism. We can think of this as follows. Every holomorphic function lives in a finite-dimensional $SU(2)_R$ multiplet of functions on $\MM$. A holomorphic function is the highest weight state in its multiplet with respect to the Cartan decomposition introduced earlier. The lowest weight state in same multiplet is anti-holomorphic. Let us introduce the notation $\sigma(f)$ to denote the anti-holomorphic function obtained thusly. Then there is a complex \emph{anti}-linear map endomorphism of the space of holomorphic functions that can be defined as
\begin{eqnarray}
\rho:~~\AA~~&\rightarrow&~~\AA~,\nn\\
\rho(f)~&=&~~\overline{(\sigma(f))}~.
\end{eqnarray}
This conjugation operation has the property that $\rho^2=(-1)^{2R}$, so it is order at most four, though in some examples it will be of order two and so an involution. It is natural to adopt a basis for the coordinate algebra that diagonalizes the action of $\rho^2$.}
\end{enumerate}

Using the information about this conjugation operation an $\MM$ as a holomorphic symplectic variety, one can almost completely reconstruct $\MM$ as a hyperk\"ahler manifold via its twistor space \cite{Gaiotto:2008nz}. Although we will not make use of any twistor space technology in the present paper, there are many hints that the algebraic construction we are discussing is closely connected to it.

\subsection*{Translation to physics}

The properties of hyperk\"ahler coordinate algebras are closely connected with the properties of half-BPS local operators in three-dimensional $\NN=4$ SCFTs. This is because the half-BPS operators are in one to one correspondence with the elements of the coordinate ring. In the text, we often label half-BPS operators by the holomorphic function $f:\MM\to\Cb$ that describes the expectation value of that operator at each point in the moduli space. Chiral ring multiplication is then just the multiplication of holomorphic functions in the coordinate algebra,
\begin{equation}
\OO_f\OO_g=\OO_{fg}~.
\end{equation}
The grading of the coordinate algebra corresponds in the obvious way to the grading of the chiral ring by $R$-charge. We further have argued in Section \ref{sec:algebra_properties} that the poisson bracket is reflected in certain singular terms in the OPE between Higgs (or Coulomb) branch operators. We will not repeat the argument here.

The conjugation operation on the coordinate algebra lifts to an anti-unitary symmetry of an $\NN=4$ SCFT. This symmetry is simply the composition of the CPT operator, $\Theta$, and an $SU(2)_R$ rotation by $\pi$, $R_{\pi}$ that sends $(\omega_1,\omega_2,\omega_3)\to(\omega_1,-\omega_2,-\omega_3)$. We then have
\begin{equation}
[(\Theta\circ R_{\pi}),\OO_{f}] = \OO_{\rho(f)}~.
\end{equation}
Of course, one could make other choices for a rotation by $\pi$ if so desired. This would amount to improving the action of $\rho$ with the action of any unitary $\Zb_2$ symmetry, or in geometric language, any $\Cb$-linear involution of $\MM$.

\subsection*{Examples}
\label{subapp:HK_examples}

To make the above discussion completely clear, we start with the simplest example of a hyperk\"ahler cone, $\Cb^2$. This is the Higgs branch of the theory of a free hypermultiplet. Taken in a complex structure with complex coordinates $(z,w)$, we have
\begin{eqnarray}
\Omega^{(2,0)}&=&dw\wedge dz~,\nonumber\\
\omega^{(1,1)}&=&\frac{i}{2}\left(dw\wedge d\bar{w}+dz\wedge d{\bar z}\right)~,\\
\Omega^{(0,2)}&=&d\bar{w}\wedge d{\bar z}~,\nonumber
\end{eqnarray}
The coordinate algebra is just $\Cb[z,w]$, and we take $z$ and $w$ to each have scaling dimension one. The Poisson bracket induced by the holomorphic symplectic form is given by
\begin{equation}
\{w,z\}=1.
\end{equation}
The real structure and the corresponding anti-linear conjugation take the following form,
\begin{equation}
\begin{split}
\sigma(z)&=\ph{-}\overline{w}~~~\Longrightarrow~~~\rho(z)=\ph{-}w~,\\
\sigma(w)&=-\overline{z}~~~\Longrightarrow~~~\rho(w)=-z~.
\end{split}
\end{equation}

\vspace{15pt}
\noindent The next simplest example is a quotient of the first example, the cone $\Cb^2/\Zb_2$. We can realize this space as the algebraic variety defined by $\{XY=Z^2\}\subset\Cb^3$. If we want, we may think of the generators of the coordinate ring in terms of the coordinates $z$ and $w$ on the covering space as
\begin{equation}
X=z^2~,\quad Y=w^2~,\quad Z = zw~.
\end{equation}
The holomorphic symplectic form is now given by
\begin{equation}
\Omega^{(2,0)}=\frac{dX\wedge dY}{-2Z}~,
\end{equation}
which encodes the following Poisson brackets for the generators of the coordinate ring,
\begin{equation}
\begin{split}
\{Y,X\}&=2Z~,\\
\{Z,X\}&=\ph{2}X~,\\
\{Z,Y\}&=-Y~.
\end{split}
\end{equation}
Conjugation now acts according to
\begin{equation}
\begin{split}
\rho(Z)&=-Z~,\\
\rho(X)&=\ph{-}Y~,\\
\rho(Y)&=\ph{-}X~.
\end{split}
\end{equation}

\vspace{15pt}
\noindent We include one more example that appears in the paper and that illustrates a subtle phenomenon that will be important. This is the cone $\Cb^2/\Zb_3$. We can again realize this as an algebraic variety in $\Cb^3$ defined by the homogeneous equation $XY=Z^3$. In terms of generators on the $\Cb^2$ cover, we have
\begin{equation}
X= z^3~,\quad Y= w^3~,\quad Z= zw ~.
\end{equation}
The holomorphic symplectic form is given by
\begin{equation}
\Omega^{(2,0)}=\frac{dX\wedge dY}{-3Z^2}~,
\end{equation}
the Poisson brackets of the generators are
\begin{equation}\label{eq:c2z3_poissons}
\begin{split}
\{Y,X\}&=3Z^2~,\\
\{Z,X\}&=\ph{3}X~,\\
\{Z,Y\}&=-Y~.
\end{split}
\end{equation}
Conjugation now acts according to
\begin{equation}\label{eq:c2z3_conjugations}
\begin{split}
\rho(Z)&=-{Z}~,\\
\rho(X)&=\ph{-}{Y}~,\\
\rho(Y)&=-{X}~.
\end{split}
\end{equation}
The thing to notice here is that there is a meaningful correspondence between signs in the poisson bracket and signs appearing in the conjugation operation. If we changed the poisson bracket to have have a minus sign in front of $X$ and a plus sign in front of $Y$ on the right hand side of \eqref{eq:c2z3_poissons} without making a similar change in \eqref{eq:c2z3_conjugations}, it would be a meaningful change to the hyperk\"ahler data of this manifold that could not be cancelled by any redefinition of variables.\footnote{We would like to particularly thank Davide Gaiotto and Andy Neitzke for discussions on this topic.} This means that one needs to be very careful in defining the conjugation operation -- indeed, the unitarity bounds that we derive and study in Section \ref{sec:kleinian} depend in a detailed manner on these choices.

\section{Moment map norms from localization}
\label{app:localization}

In this appendix we describe the calculation of two-point functions for moment map operators -- that is, for half-BPS operators of dimension one. These necessarily have conserved currents for global flavor symmetries as descendants. Their expectation values on the Higgs branch are those of the moment map for the Hamiltonian flavor symmetry action.

These two-point functions can be computed because they are related to the canonically normalized current two-point functions by supersymmetry, and these current two-point functions can be computed using supersymmetric localization \cite{Closset:2012vg}. We proceed as follows. First we consider the $\NN=4$ theory as a special case of an $\NN=2$ theory by choosing an embedding of $OSp(2|4,\Rb)\hookrightarrow OSp(4|4,\Rb)$. The $SO(2)_r \cong U(1)_r$ symmetry of the $\NN=2$ theory depends on the chosen embedding and is an abelian subgroup of the $\NN=4$ $R$-symmetry -- its commutant in $SU(2)_C\times SU(2)_H$ is realized as additional flavor symmetries. We then define a supersymmetric coupling of the $\NN=2$ theory to the round three-sphere background using an $r$-symmetry that is an arbitrary mixture of the true $r$-symmetry with an abelian subgroup of the flavor symmetry. The current two-point function is then the second derivative of this free energy with respect to the mixing coefficient, evaluated when the mixing coefficient vanishes.

Let us first establish conventions and normalizations. In an $\NN=2$ SCFT with abelian flavor currents $j_\mu^{(a)}$, the current two-point function is fixed by conformal invariance to take the form
\begin{equation}
\left\langle j_\mu^{(a)}(x) j_\nu^{(b)}(0) \right\rangle = \frac{\tau^{ab}}{16\pi^2} \left(\delta_{\mu\nu} \partial^2 - \partial_\mu \partial_\nu \right)\frac{1}{x^2}~,
\end{equation}
where $\tau^{ab}$ is a positive definite matrix. The normalization is chosen such that in the theory of a single free chiral multiplet $Q$ of unit flavor charge, we have
\begin{equation}
\begin{split}
\langle Q(x)Q^\dagger(0)\rangle &= \frac{1}{|x|}~,\\ 
\int_{S^2_0} \left(\hat{n}\cdot \vec{j}(x)\right)Q(0) &= i Q(0)~, \\
\left\langle j_\mu(x)j_\nu(0)\right\rangle &= \frac{1}{16\pi^2} \left(\delta_{\mu\nu} \partial^2 - \partial_\mu \partial_\nu \right)\frac{1}{x^2}~,
\end{split}
\end{equation}
so that $\tau_{\rm chiral}=1$.

To relate the coefficient $\tau$ to the moment map two-point function coefficient, we consider a single free $\NN=4$ hypermultiplet. It decomposes in two free chiral multiplets $(Q,\wt Q)$ -- to which we assign $U(1)$ flavor symmetry charges $+1$ and $-1$ respectively -- so $\tau_{\rm hyper} = 2$.\footnote{The full flavor symmetry of the free hypermultiplet is $USp(2)$. We focus on its (rescaled) Cartan component here.} Let $\mu^{(ab)}(x)$ denote the $SU(2)_H$ triplet moment map operator for this $U(1)$ flavor symmetry, which we normalize according to
\begin{equation}
\mu^{(12)}(x) Q(0) \sim \frac{Q(0)}{|x|} + \ldots~, \qquad \mu^{(12)}(x) \wt Q(0) \sim -\frac{\wt Q(0)}{|x|}+ \ldots~.
\end{equation}
Concretely, our normalizations amount to defining $\mu^{11} = 2Q\wt{Q}$, $\mu^{(12)} = (QQ^\dagger - \wt Q \wt Q^\dagger)$, and $\mu^{22}=-2Q^\dagger\wt{Q}^\dagger$. It is straightforward to then compute
\begin{equation}
\langle\mu^{(12)}(x) \mu^{(12)}(0)\rangle = \frac{2}{|x|^2}~, \qquad \langle\mu^{11}(x) (\mu^{11}(0))^\dagger \rangle = \frac{4}{|x|^2}~, 
\end{equation}
where $\mu^{11}, \mu^{22}$ complete the $SU(2)$ triplet and satisfy $(\mu^{11})^\dagger = -\mu^{22}$. Superconformal symmetry then guarantees that with these normalizations for the moment map operators in terms of the flavor charges, there is a general relation
\begin{equation}
\left\langle\mu^{(12),a}(x) \mu^{(12),b}(0)\right\rangle = \frac{\tau^{ab}}{|x|^2}~, \qquad \left\langle\mu^{11,a}(x) \left(\mu^{11,b}(0)\right)^\dagger \right\rangle = \frac{2\tau^{ab}}{|x|^2}~. 
\end{equation}

Now consider an $\NN=2$ supersymmetric Lagrangian theory and denote the UV $U(1)_r$ charges of the bottom components of its chiral multiplets $Q_i$ as $r_0(Q_i)$. Let us further denote its abelian flavor symmetry charges as $F_a$. The $U(1)_r$ charges of the infrared conformal field theory, which equal the conformal dimensions, are generally obtained by mixing with flavor symmetries, $r(Q_i)(t) = r_0(Q_i) + \sum t_a F_a(Q_i).$ Let $t_*$ denote the values of the mixing parameters corresponding to the infrared $r$-charges. The three-sphere partition function $Z_{S^3}(t)$ can be computed exactly via supersymmetric localization \cite{Jafferis:2010un,Hama:2010av,Hama:2011ea} and is a function of the mixing parameters; its free energy $F(t) = -\log Z_{S^3}(t)$ satisfies
\begin{equation}
\frac{\partial}{\partial t_a}  \operatorname{Re} F(t) \Big|_{t=t_*} = 0~, \qquad \frac{\partial^2}{\partial t_a \partial t_b} \operatorname{Re} F(t)\Big|_{t=t_*} = -\frac{\pi^2}{2} \tau_{ab}~. 
\end{equation}
Here the first equation encodes the use of F-maximization \cite{Jafferis:2010un} to determine the infrared $r$-charges, \ie, to fix the mixing parameters $t_*$. In our $\NN=4$ examples there is no mixing and we will always have $t_*=0$. When the first equation is satisfied, the second one simplifies to
\begin{equation}\label{eq:tau_without_prefactors}
\tau_{ab} = \frac{2}{\pi^2}\operatorname{Re}\frac{1}{Z(t_*)}\frac{\partial^2}{\partial t_a \partial t_b} Z(t)\Big|_{t=t_*}~.
\end{equation}

\paragraph{Free chiral multiplet.} Let us consider the simplest example of a single free, massless chiral multiplet. The theory is conformal, with the complex scalar having $r_0=\frac{1}{2}$. Mixing with the $U(1)$ flavor symmetry is parameterized as $r(t) = r_0 + t$ -- obviously we will have $t_* =0$. The three-sphere partition function obtained via localization is given by
\begin{equation}
Z_{S^3}(t) \sim e^{\ell\left(1-r(t) \right)} = e^{\ell\left(\frac{1}{2} - t \right)}~.
\end{equation}
The function $\ell(z)$ was introduced in \cite{Jafferis:2010un} and satisfies the useful properties
\begin{equation}
\ell^\prime(z) = -\pi z \cot(\pi z)~,\qquad \ell\left(\frac{1}{2}-i z \right) + \ell\left(\frac{1}{2}+i z \right) = - \log \left(2\cosh(\pi z)\right)~.
\end{equation}
We have neglected prefactors in the partition function since they play no role in determining the flavor current two-point function coefficient $\tau$. Indeed, we have
\begin{equation}
\tau = \frac{2}{\pi^2} \operatorname{Re} \frac{1}{Z(0)} \frac{\partial^2 Z(t)}{\partial t^2}\Big|_{t=0} =  \frac{2}{\pi^2} \operatorname{Re} \frac{1}{e^{\ell(1/2)}} \frac{\pi^2}{2}e^{\ell(1/2)} = 1~,
\end{equation}
as expected.

\begin{figure}[ht]
\centering
\begin{tikzpicture}
\draw (-1.5,0) circle [radius=.48]; \node at (-1.5,0) {$U(1)$};
\draw (1.5,0) circle [radius=.48]; \node at (1.5,0) {$U(1)$};
\draw  (-1.08,.2)--(1.08,.2);
\draw  (1.08,-.2)--(-1.08,-.2);
\end{tikzpicture}
\caption{Quiver diagram for the $A_1$ gauge theory.\label{fig:A1_quiver}}
\end{figure}
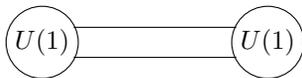

\paragraph{$A_1$ theory.} The first nontrivial example relevant for the text is the $\NN=4$ superconformal field theory whose Higgs branch is the $A_1$ singularity. Microscopically it is a $U(1)$ gauge theory with two charged hypermultiplets. The symmetry rotating the two hypermultiplets constitutes an $SU(2)$ flavor symmetry. In light of generalizing the computation to other Kleinian singularities, it will be convenient to think of the theory as a $U(1)^2/U(1)$ gauge theory with two bifundamental $\NN=4$ hypermultiplets, which can be encoded in the quiver in Fig. \ref{fig:A1_quiver}
The corresponding $\NN=2$ chiral multiplets $Q_i, \wt Q^i$ for $i=1,2$ have their $U(1)^2\times U(1)_F$ charges given in Table \ref{U(1)^2charges}.
\begin{table}[ht]
\centering
\begin{tabular}{c|cccc}
& $U(1)_1$ & $U(1)_2$  & $U(1)_F$\\ \hline
$Q_1$ & $\phantom{+}1$ & $-1$ & $\ph{-}1$\\
$Q_2$ & $\phantom{+}1$ & $-1$ &  $-1$\\
$\wt Q^1$ & $-1$ & $\phantom{+}1$ & $-1$ \\
$\wt Q^2$ & $-1$ &$\phantom{+}1$ & $\ph{-}1$\\
\end{tabular}
\caption{Charges under the $U(1)^2$ gauge symmetry and $U(1)_F$ flavor symmetry. Note that the fields are uncharged under the diagonal $U(1).$\label{U(1)^2charges}}
\end{table}
We have omitted the free, neutral chiral multiplets arising from the $\NN=4$ vector multiplets as they are irrelevant for the computation of the current two-point function coefficient.

We now need to compute the partition function $Z_{S^3}(t)$, where $t$ parametrizes the mixing of the $U(1)_F$ symmetry with the $U(1)_r$ charges of the chiral multiplets. Concretely, the bottom components of the chiral superfields $Q_1$ and $\wt Q^2$ are given $r$-charge $r_f(t) = \frac{1}{2} + t$, while the bottom components of the chiral superfields $Q_2$ and $\wt Q^1$ are given $r$-charge $r_a(t) = \frac{1}{2} - t$. Thanks to $\NN=4$ supersymmetry, the infrared $r$-charges are specified by $t_* = 0$. We find
\begin{equation}
\begin{split}
Z_{S^3}(t)	&\sim \int d\sigma_1 d\sigma_2 \  \delta(\sigma_1 + \sigma_2) \ e^{\ell\left(1-r_f(t) \pm i (\sigma_1-\sigma_2) \right)+\ell\left(1-r_a(t) \pm i (\sigma_1-\sigma_2) \right) } \\
			&\sim \int dx  \ e^{\ell\left(1-r_f(t) \pm i x \right) + \ell\left(1-r_a(t) \pm ix \right)}~.
\end{split}
\end{equation}
The delta function implements the modding out of the diagonal $U(1)$ and we used shorthand notation to denote a product over all signs independently in the integrand. We also omitted overall factors: they will not be important, since the coefficient $\tau$ can be computed using \eqref{eq:tau_without_prefactors},
\begin{equation}
\tau = \frac{2}{\pi^2} \operatorname{Re} \frac{1}{Z(0)} \frac{\partial^2 Z(t)}{\partial t^2}\Big|_{t=0}  = \frac{8}{3}~.
\end{equation}

\paragraph{$A_2$ theory.} This theory has a microscopic description as a quiver gauge theory with quiver given by the affine Dynkin diagram of $A_2,$ whose nodes are $U(1)$ gauge groups. 
In $\NN=2$ language, this means that we have six chiral matter multiplets $Q_i, \tilde Q_i$ for $i=1,2,3$ whose bottom components have $U(1)^3$ charges as in Table \ref{U(1)^3charges}.
\begin{table}[ht]
\centering
\begin{tabular}{c|ccc}
 & $U(1)_1$ & $U(1)_2$ & $U(1)_3$ \\ \hline
 $Q_1$ & $\phantom{+}1$ & $-1$ & $\phantom{+}0$ \\
 $Q_2$ & $\phantom{+}0$ & $\phantom{+}1$ & $-1$ \\
 $Q_3$ & $-1$ & $\phantom{+}0$ & $\phantom{+}1$ \\
 $\wt Q_1$ & $-1$ & $\phantom{+}1$ & $\phantom{+}0$ \\
 $\wt Q_2$ & $\phantom{+}0$ & $-1$ & $\phantom{+}1$ \\
 $\wt Q_3$ & $\phantom{+}1$ & $\phantom{+}0$ & $-1$  
\end{tabular}
\caption{Charges of matter fields in the $A_2$ quiver theory. All fields are uncharged under the diagonal $U(1).$\label{U(1)^3charges}}
\end{table}
The fields are uncharged under the diagonal $U(1)$ gauge symmetry, which should be modded out. Furthermore, the theory has a $U(1)_F$ flavor symmetry which assigns charge $+1$ to the $Q_i$ and charge $-1$ to the $\wt Q_i$. The vector multiplets decompose into a $U(1)^3 / U(1)$ neutral chiral multiplet, which will not play a role, and $U(1)^3 / U(1)$ $\NN=2$ vector multiplets.

Mixing in the flavor symmetry, the bottom components of the chiral superfields $Q_i$ are assigned $r$-charge $r(t) = \frac{1}{2} + t$, while the bottom components of the chiral superfields $\wt Q_i$ are assigned $r$-charge $\wt r(t) = \frac{1}{2} - t$. Again one has $t_* =0.$ The partition function $Z_{S^3}(t)$ as a function of the parameter $t$ is then computed by the following integral
\begin{equation}
\begin{split}
Z(t)\sim \int d\sigma_1 d\sigma_2 d\sigma_3\ \delta(\sigma_1 + \sigma_2 + \sigma_3)\ &e^{\ell\left(1-r(t) + i (\sigma_1-\sigma_2) \right) + \ell\left(1-r(t) + i (\sigma_2-\sigma_3) \right) + \ell\left(1-r(t) + i (\sigma_3-\sigma_1) \right)}\\
\times\,&e^{\ell\left(1-\wt r(t) - i (\sigma_1-\sigma_2) \right) + \ell\left(1-\wt r(t) - i (\sigma_2-\sigma_3) \right) + \ell\left(1-\wt r(t) - i (\sigma_3-\sigma_1) \right)}~.
\end{split}
\end{equation}
Again, the delta function removes the diagonal $U(1)$ and we omitted unimportant overall constants. Changing variables to $x_1 = \sigma_1 - \sigma_2,$ $x_2 = \sigma_2 - \sigma_3$ and $x_3 = \sigma_1+\sigma_2+\sigma_3$, we find
\begin{equation}
\begin{split}
Z(t)\sim \int dx_1 dx_2 \ & e^{\ell\left(1-r(t) + i x_1 \right) + \ell\left(1-r(t) + i x_2 \right) + \ell\left(1-r(t) + i (-x_1-x_2) \right)}\\
\times &e^{\ell\left(1-\wt r(t) - i x_1 \right) + \ell\left(1-\wt r(t) - i x_2 \right) + \ell\left(1-\wt r(t) - i (-x_1-x_2) \right)}~.
\end{split}
\end{equation}
From this integral representation we can compute the current two-point function analytically, and we find
\begin{equation}
\tau =  \frac{2}{\pi^2} \operatorname{Re} \frac{1}{Z(0)} \frac{\partial^2 Z(t)}{\partial t^2}\Big|_{t=0} = 18 - \frac{144}{\pi^2}~.
\end{equation}

\paragraph{$A_3$ theory.} Finally, we consider theory with quiver given by the $A_3$ affine Dynkin diagram.
Following the same steps as in the previous examples, we arrive at an integral representation for the current two-point function that can be performed analytically. Ultimately, we obtain the following expression for the current two-point function:
\begin{equation}\label{tauforA3}
\tau=\restr{\frac{512(\pi^2-6)}{51\pi^2+18\,\partial_{a}\left(\frac{1}{\Gamma(a)}\,\pFq{3}{2}{1,2,2}{3,a}{-1}\right)-36\,\partial_{b}\left(\frac{1}{\Gamma(b)}\,\pFq{2}{1}{1,2}{b}{-1}\right)-36}}{a\to2,~b\to3}
\end{equation}
Since this is a somewhat inscrutable expression, we note that numerically this is $\tau \approx 4.1821109\ldots$

\bibliography{./auxiliary/biblio}
\bibliographystyle{./auxiliary/JHEP}

\end{document}